\documentclass{aa}

\usepackage{graphicx}
\usepackage{txfonts}
\usepackage{lscape}
\usepackage{xcolor}
\usepackage{float}
\usepackage[flushleft]{threeparttable}
\begin{document} 

   \title{Barium stars as tracers of $s$-process nucleosynthesis in AGB stars}

   \subtitle{II. Using machine learning techniques on 169 stars}
   
   \author{J.W. den Hartogh\inst{1,2}
   \and 
   A. Yag\"ue L\'opez\inst{1,2,3}
   \and
   B. Cseh\inst{1,2,4}
   \and
   M. Pignatari\inst{1,2,5,11,12} 
   \and
   B. Vil\'agos\inst{1,2,4,6}
   \and 
   M. P. Roriz\inst{7}
   \and
   C. B. Pereira\inst{7}
   \and
   N. A. Drake\inst{8,9}
   \and
   S. Junqueira\inst{7}
   \and
   M. Lugaro\inst{1,2,6,10}
   }
   %1: konkoly
   %2: MTA
   %3: LANL
   %4: Monash
   %5: Hull
   %6: ELTE
   %7: rio de janeiro
   %8: saint petersburg
   %9: itajuba
   %lendulet somewhere here or in acknowledgements? here

   \institute{Konkoly Observatory, Research Centre for Astronomy and Earth Sciences (CSFK), ELKH, H-1121, Budapest, Konkoly Thege M. \'ut 15–17., Hungary
   \and
   {CSFK, MTA Centre of Excellence, Budapest, Konkoly Thege Miklós út 15-17., H-1121, Hungary}
   \and
   {Computer, Computational and Statistical Sciences (CCS) Division, Center for Theoretical Astrophysics, Los Alamos National Laboratory, Los Alamos, NM, 87545, USA}
    \and
    {MTA-ELTE Lend{\"u}let "Momentum" Milky Way Research Group, Hungary}
   \and
   {E.~A.~Milne Centre for Astrophysics, University of Hull, HU6 7RX, United Kingdom}
   \and
   {ELTE E\"{o}tv\"{o}s Lor\'and University, Institute of Physics, Budapest 1117, P\'azm\'any P\'eter s\'et\'any 1/A, Hungary}
    \and
    {Observat\'orio Nacional/MCTI, Rua General Jos\'e Cristino, 77 Sao Cristovao, Rio de Janeiro, Brazil}
    \and
    {Laboratory of Observational Astrophysics, Saint Petersburg State University, Universitetski pr. 28, 198504, Saint Petersburg, Russia}
    \and
    {Laboratório Nacional de Astrof\'isica/MCTI, Rua dos Estados Unidos 154, Bairro das Na\c c\~oes, 37504-364, Itajub\'a, Brazil}
    \and
    {School of Physics and Astronomy, Monash University, VIC 3800, Australia} 
    \and
    {NuGrid Collaboration, \url{http://nugridstars.org}}
    \and
    {Joint Institute for Nuclear Astrophysics - Center for the Evolution of the Elements}
}

   \date{Received May 28, 2022; accepted November 28, 2022}
 
  \abstract
  {
  Barium (Ba) stars are characterized by a high abundance of heavy elements made by the \textit{slow} neutron capture process (\textit{s}-process). Such an observed peculiar signature is due to mass transfer from a stellar companion, bound in a binary stellar system with the Ba star observed today. The signature is created when the stellar companion was an Asymptotic Giant Branch (AGB) star.
  }
  %useful to disentangle the different physical features of AGB stars}
   {We aim to analyse the abundance pattern of 169 Ba stars, using machine learning techniques and the AGB final surface abundances predicted by \textsc{Fruity} and Monash stellar models.}
   {We developed machine learning algorithms that use the abundance pattern of Ba stars as input to classify the initial mass and metallicity of its companion star using stellar model predictions. We use two algorithms: the first exploits neural networks to recognise patterns and the second is a nearest-neighbour algorithm, which focuses on finding the AGB model that predicts final surface abundances closest to the observed Ba star values. In the second algorithm we include the error bars and observational uncertainties to find the best fit model. The classification process is based on the abundances of Fe, Rb, Sr, Zr, Ru, Nd, Ce, Sm, and Eu. We selected these elements by systematically removing $s$-process elements from our AGB model abundance distributions, and identifying those whose removal has the biggest positive effect on the classification. We excluded Nb, Y, Mo, and La. Our final classification combines the output of both algorithms to identify for each Ba star companion an initial mass and metallicity range. }
   {With our analysis tools we identify the main properties for 166 of the 169 Ba stars in the stellar sample. The classifications based on both stellar sets of AGB final abundances show similar distributions, with an average initial mass of M=2.23M$_{\odot}$ and 2.34M$_{\odot}$ and an average [Fe/H]=-0.21 and -0.11, respectively. We investigated why the removal of Nb, Y, Mo, and La improves our classification and identified 43 stars for which the exclusion had the biggest effect. We show that these stars have statistically significant different abundances for these elements compared to the other Ba stars in our sample.
   We discuss the possible reasons for these differences in the abundance patterns.} 
   {}


   \maketitle

%------------------------------------------------------------------

\section{Introduction}

Barium (Ba) stars are part of binary systems \citep{McClure80,McClure83} and their companions are white dwarfs that went through the asymptotic giant branch (AGB) phase \citep[see][for reviews on AGB stars]{falk_ARAA,2006_Straniero_Gallino_Cristallo,2014karakas}. The spectra of Ba stars show enhanced abundances of \textit{slow} neutron-capture (\textit{s} process) elements heavier than Fe as compared to solar \citep{bidelman51}. It is commonly accepted that the overabundance of these s-process elements is due to mass transfer from the (former AGB) companion star.

There are two neutron source reactions in AGB stars that can provide neutrons for the s process. The first is the $^{13}$C($\alpha$,n)$^{16}$O reaction, which produces the neutrons in the $^{13}$C-rich region ($^{13}$C-pocket) between the He- and H-burning shells in the AGB star. The second neutron source is $^{22}$Ne($\alpha$,n)$^{25}$Mg, which produces only a small amount of neutrons compared to the $^{13}$C-pocket and is activated in the thermal pulses, which are recurrent convective episodes of He-burning. The $^{22}$Ne neutron source produces higher neutron densities than the $^{13}$C neutron source, which allows for the activation of branching points along the s-process path \citep[e.g.,][]{bisterzo:15}. The $s$-process abundances made in the He intershell are brought to the surface by the Third dredge-up events (TDUs) \citep[e.g.,][]{gallino:98}. The characterisation of the s-process signature in AGB stars is commonly done by comparing the s-process production at the first s-process peak (neutron magic number N = 50, including Sr, Y and Zr) and at the second s-process peaks (neutron magic number N = 82, including Ba, La and Ce), either by deriving an average production or by using individual elements from those two peaks \citep[][]{busso:01,cseh2018,lugaro2020}.

Homogeneous observations of Ba stars used to be limited, until the study of \citet{deCastro} was published. In that work the high-precision spectra of 182 Ba star candidates were presented, as observed and analysed in a self-consistent manner. Error bars for the ratios of the first and second peak s-process elements were calculated by \citet{cseh2018}. Those authors also found that for each specific metallicity a spread of about a factor of 3 is present in the ratios. This range needs to be explained by variation in features like initial mass and internal mixing. 

The complete s-process pattern of stellar evolution models carries the signatures of those features \citep[see][for an extensive review]{2014karakas}. Therefore, it should possible to infer the initial AGB mass and metallicity, and other physical features by comparing the individual abundance patterns of Ba stars to single AGB models. However, this pattern cannot be directly compared with the s-process patterns of AGB models. The reason for this is that the material in the envelope of the Ba star is a mixture of its original envelope and the accreted material from the AGB star donor. To account for this mixture, we must modify the s-process patterns in the AGB final surface abundances by a dilution factor. 

\citet[henceforth `Paper 1']{2022Cseh} started the process of matching individual spectra with single models applying a simplified method of matching the models to the determined [Ce/Fe] abundances. These authors investigated the 28 of the Ba stars out of the sample of \citet{deCastro} for which it was possible to better determine the masses \citep{jorissen19} using Gaia DR2 data \citep{gaia2018}. \citet{jorissen19} estimated the masses of the companions of the Ba stars, the white dwarfs, by using assumptions of the inclination of the binary system and the mass ratio of the two binary stars. Using an initial-final mass relation \citep{El-Badry18} it was then possible to estimate the initial mass of the companion, and thus the initial mass of the AGB models. Of the 28 Ba stars, the authors of Paper 1 were able to find good matches with the models for 21 Ba stars, and also independently confirmed the initial AGB mass estimate from \citet{jorissen19} for those stars. The authors of Paper 1 also reported on some problems. For instance, for 16 of those 21 cases the observed abundances of Nb, Ru, Mo and/or Nd, Sm were higher than predicted. Furthermore, 3 stars were matched with AGB models with a lower mass than those of \citet{jorissen19}.  Finally, 4 stars showed higher first \textit{s}-process peak abundance values than in the AGB models. 

Comparing all the Ba stars to all the AGB models of \textsc{\textsc{fruity}} and Monash one-by-one without any information on the initial AGB mass is a time consuming task, and prone to human bias. After all, the full set of 169 Ba stars has to be compared to the thousands of diluted AGB final surface abundances. The use of modern machine learning (ML) techniques allows us to automatise this classification and perform it faster, while also removing the human bias in the process. These techniques are increasingly common to use, as data sets in astronomy keep increasing in size. Another recent example is \citet{2021Karinkuzhi}, in which ML techniques are compared to the manual analysis to identify s-process enrichment in Ba stars.

The aim of this Paper is to identify the main features of the old AGB star companion of 169 Ba stars, by comparing observations to theoretical AGB models.
The structure of the Paper is as follows: in Section \ref{ref:datasets} we describe our data sets of Ba star spectra and sets of AGB final surface abundances. In Section \ref{sec:algorithms} we present our full pipeline of algorithms and the set of elements we include in our pipeline. In Section \ref{sec:28stars}, we compare the results of our classification methods to the ones classified in Paper 1. Then, in Section \ref{sec:full-set-classification} we apply the classification to the full set of 169 stars, draw statistics and show the peculiar cases. In Section \ref{sec:disc} we focus on understanding why the exclusion of certain elements leads to better classifications. We present our conclusions in Section \ref{sec:conclusions}.

%------------------------------------------------------------------
\section{Data sets}
\label{ref:datasets}

\subsection{Abundances of Ba stars}
\label{sec:abundances}
For the verification and validation of our classification algorithms, we use the same set of 28 Ba stars from Paper 1. These 28 stars are the overlap of the samples of \citet{deCastro, roriz21, Roriz21_heavy} and \citet{jorissen19}. As a result, we have for those stars a large set of self-consistently derived abundances, independent estimates of the initial mass of the AGB star and of the mass of the Ba star observed today in the binary system. The mass estimates provide an extra constraint for the models, which sped up the classification during the manual classification of Paper 1 by reducing the pool of possible models. Because we do not have the mass estimates for all 169 Ba stars, we do not include them in our classification algorithms. Instead, the initial masses of the AGB stars are a result of our classification algorithms.

After verifying and validating the algorithms, we use them to classify the full set of Ba stars. All stars were observed with FEROS \citep[Fiberfed Extended Range Optical Spectrograph, R = 48 000,][]{1999Kaufer}, allowing to derive a large set of elemental abundances around the $s$-process peaks, as well as light elements \citep[see][for discussions on those]{deCastro,roriz21,Roriz21_heavy,2022Cseh}. A total of 182 stars were analysed by \citet{deCastro}, from which 169 stars were found to be a Ba star. This is our whole set with $s$-process elemental abundances available for Rb, Sr, Y, Zr, Nb, Mo, Ru, La, Ce, Nd, Sm, and Eu for most of the stars.

For those elements whose abundances were derived using three or more lines, we include the observational error bars from the spectral measurements \citep[for details, see][]{Roriz21_heavy}.
For some elements the abundances were derived from less than three spectral lines. In these cases (except Rb) the error bars would be at the order of at least the representative cases in Tables 2, 3, and 4 in \citet{Roriz21_heavy} and we adopted an artificial error bar of 0.5 dex to account for the fact that the measurement was uncertain. When Rb is not available, we add an artificial Rb abundance of 1 $\pm$ 1 dex.

Note that 7 stars (HD 749, HD 5424, HD 12392, HD 116869, HD 123396, HD 210709, HD 223938) are also published by \cite{AB2006}.  Overall we found that their abundances are in agreement with our values (mostly within the error bars, where available) when compared to \cite{AB2006} and \cite{allen:07} data using the same ionisation states for the elements. The largest discrepancy is between the Ru abundances for HD 12392: \cite{Roriz21_heavy} reported 0.74 $\pm$ 0.26 dex, while \cite{allen:07} derived 1.41 dex for the same star. 

\subsection{AGB final surface abundances}
\label{sec:agbyields}
The AGB final surface abundances included in our comparison are from two extensive sets of AGB stellar nucleosynthesis models: \textsc{fruity}\footnote{http://fruity.oa-teramo.inaf.it/} \citep{2009cristallo,2011cristallo,2013piersanti,2015cristallo} and Monash \citep{2012lugaro,2014fishlock,2016karakas,2018karakas}. We have included all initial stellar masses M$_{\rm{ini}}$ (ranging from M$_{\rm{ini}}$ 1.25 to 8 M$_{\odot}$) and all initial metallicities Z$_{\rm{ini}}$ (ranging from 0.0028 to 0.03) of both sets of AGB final surface abundances. Furthermore, we include the slow rotating models with an initial rotational velocity at or below 60 km/s, even though evidence may suggest that those models rotate too fast during the AGB phase \citep[see][]{2019denhartogh}. Finally, we note that Monash and \textsc{fruity} models use different methods and assumptions to inject protons in the He-intershell to create the $^{13}$C-pocket. 
Monash inserts an artificial mixing profile, which is the same in each thermal pulse, while \textsc{fruity} allows for the creation of the profile by free parameters based on physical mixing processes. 

The AGB final surface abundances cannot directly be compared to the abundances derived from the spectra of the Ba stars. As explained in the Introduction, a dilution factor has to be included. The diluted abundances are given by
\begin{equation}
\left[\frac{\rm{X}}{\rm{Fe}}\right]_{\rm{Ba}}=\rm{log}_{10}\left[ (1-\delta) \cdot 10^{[X/Fe]_{\rm{ini}}} + \delta \cdot 10^{[X/Fe]_{\rm{AGB}}} \right]
\label{eq:delta}
\end{equation}
where [X/Fe]$_{\rm{ini}}$ is the initial abundance of element X, [X/Fe]$_{\rm{AGB}}$ is the final surface abundance of the AGB model, and $\delta$ is the effect of dilution, which we treat as a free parameter and relates to the dilution factor, \textit{dil}, as $\delta$ = 1/dil. We limit $\delta$ to a maximum of 0.9, like in Paper 1 because higher values imply that the Ba star envelope is (almost) entirely composed of AGB material. We assume the initial abundances follow a scaled solar distribution, which makes [X/Fe]$_{\rm{ini}}$ = 0 in Eq. (\ref{eq:delta}). We dilute each model within our sets of AGB final surface abundances with $\delta$ = [0, 0.9] with increments of 0.002, creating a grid of diluted AGB final surface abundances. We remove from the analysis diluted abundance distributions if they are identical within the first three decimals to other abundance distributions derived from the same AGB model. AGB final surface abundances where the maximum [X/Fe] value of all the included elements is below 0.2 are also excluded as those are considered to not be enriched enough in \textit{s}-process nucleosynthesis.

\subsection{Correlations between elements}
\label{sec:corr_el}
\begin{figure}
    \centering
    \includegraphics[width=0.9\linewidth]{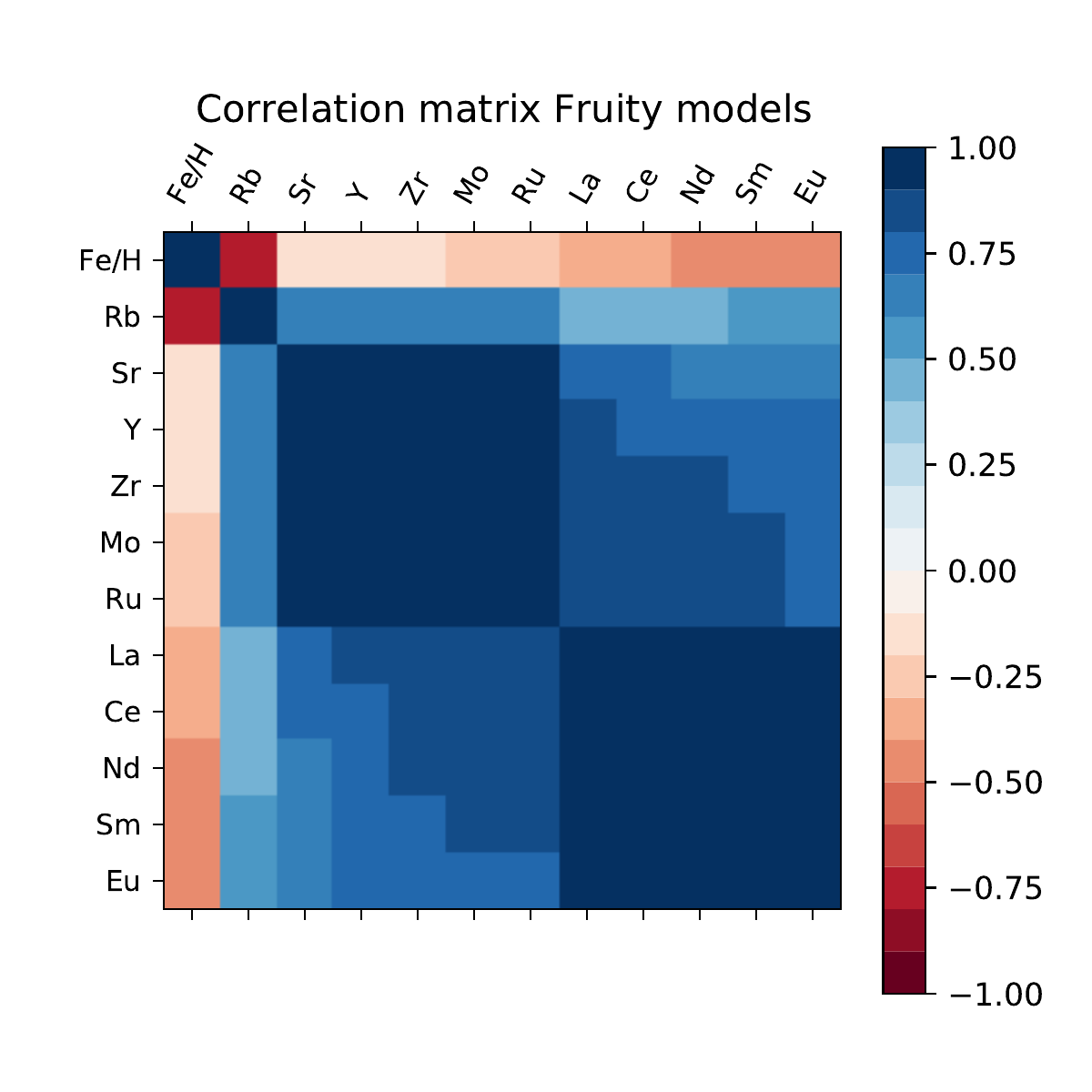}
    \vspace{-10mm}\\
    \includegraphics[width=0.9\linewidth]{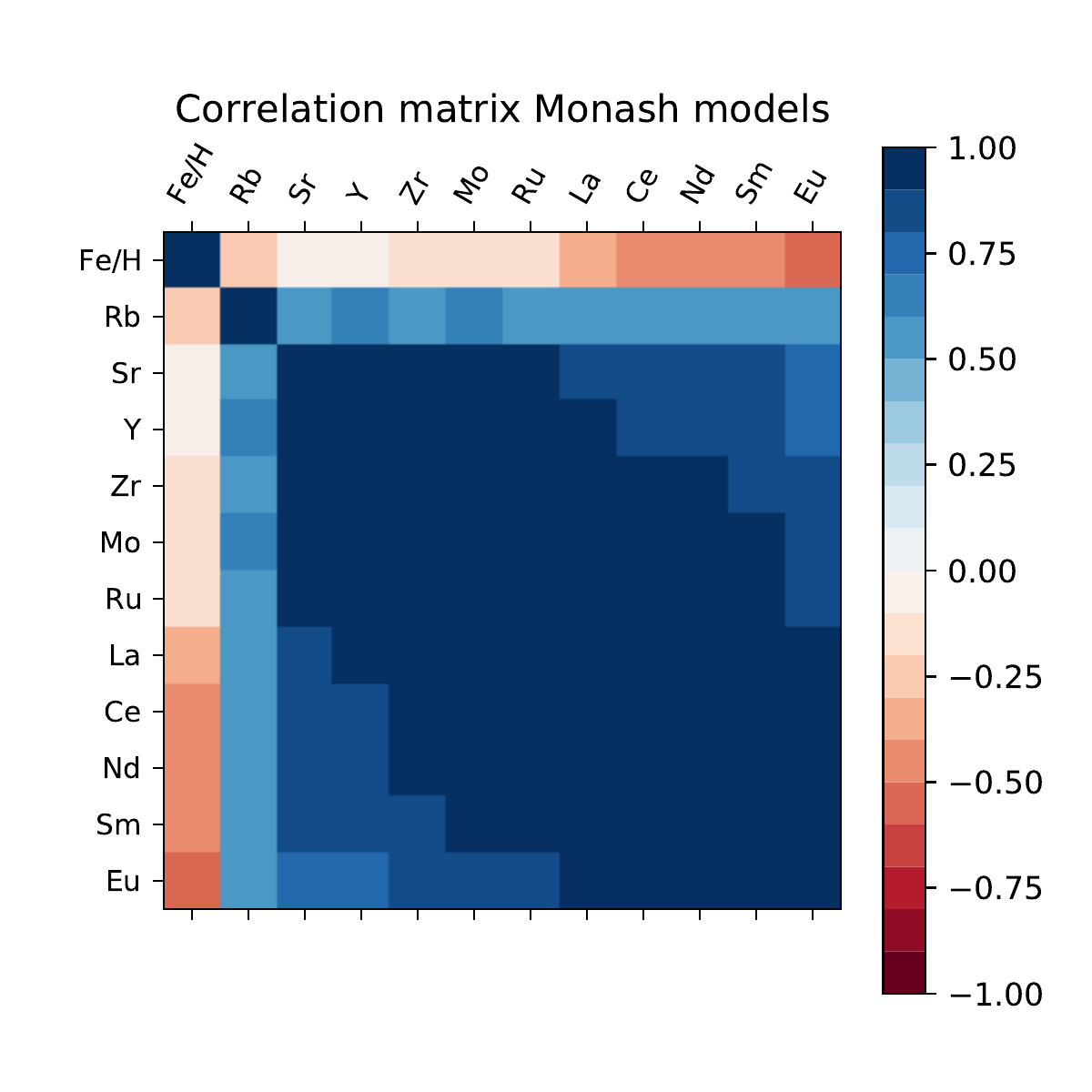}  
    \vspace{-10mm}\\
    \includegraphics[width=0.9\linewidth]{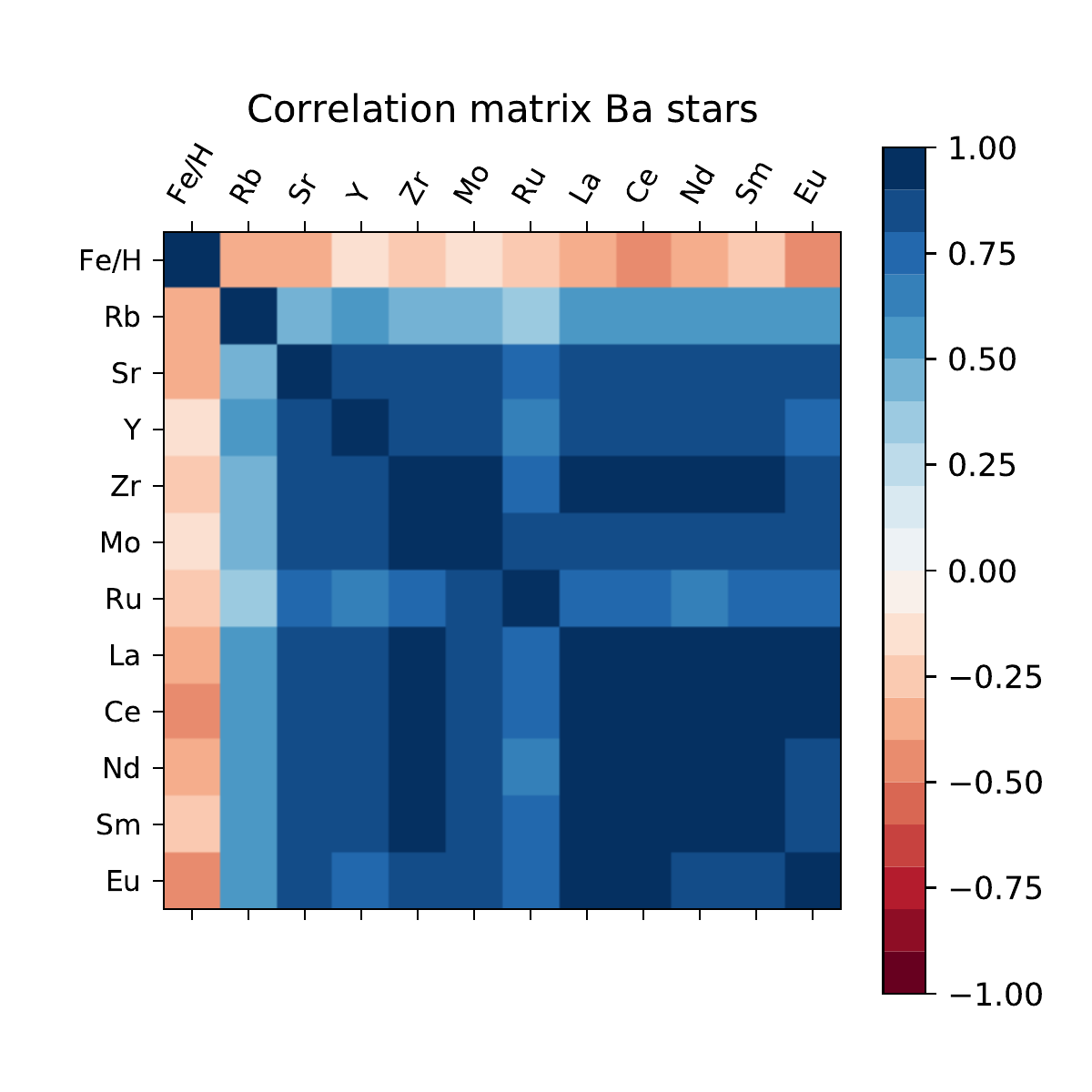}    
    \caption{Correlation matrix demonstrating the linear relationship between pairs of elements in the AGB final surface abundances of \textsc{fruity} (top panel), of Monash (middle panel), and in the observed Ba star sample (bottom panel, excluding the error bars on the observations).}
    \label{fig:corrmat}
\end{figure}

We perform an initial comparison of the correlation coefficients between the different elements within one data set and we perform this comparison for the Ba stars and the two sets of AGB final surface abundances. We include only Fe and elements heavier than Fe in this comparison, and we do not include the error bars when analysing the observed data. The resulting correlation matrices are shown in Fig. \ref{fig:corrmat}.  

In all three panels we see that [Fe/H] is negatively correlated with the other elements (red shaded boxes), and that all the other elements are positively correlated with each other (blue shaded boxes). Strong positive correlations are found in all panels, but the patterns are different. As we mentioned in Section \ref{sec:agbyields}, the profiles and subsequent $^{13}$C-pockets in the \textsc{fruity} models are less homogeneous than in the Monash models. Therefore, the s-process patterns in the AGB final surface abundances of \textsc{fruity} are also less homogeneous than the s-process patterns in the AGB final surface abundances of Monash. 

In the  AGB final surface abundances of \textsc{fruity} (top panel) we find strong positive correlations (darkest blue, above 0.9) between all the first peak elements and again between all the second peak elements. The AGB final surface abundances of Monash (middle panel) show strong correlations (above 0.9) between all the first peak elements, all the second peak elements, and also between the two peaks. In contrast, the correlations between the observed abundances show weaker correlations, only the second peak elements are strongly correlated with each other and with Zr. A first result thus is that there seems to be a difference between the Ba star observations and our theoretical understanding of the nucleosynthesis in the first s-process peak. However, we do not know if this difference is due to observational uncertainties or theoretical limitations.

\section{Algorithms}
\label{sec:algorithms}

We use two different classification algorithms. The first is an artificial neural network (ANN) ensemble (Section~\ref{sec:ann}), while the second is a nearest-neighbour classification algorithm (Section~\ref{sec:nearest}). The motivation for using two algorithms is to supplement their respective shortcomings: the nearest-neighbour algorithm finds the closest fitting model without regards for any particular pattern, while the ANN ensemble is more sensitive to the pattern and therefore might fail to find a match for an observed set of abundances if the pattern in that set does not match the patterns within the AGB final surface abundances. We start this section with a visualisation of the classification problem, then we introduce the classification algorithms in Sections \ref{sec:ann} and \ref{sec:nearest} and how we combine their results in Section \ref{sec:combining}. At the end of the section we introduce the set of elements we use in our classification.

\subsection{Visualisation of classification challenge}
\label{sec:visualisation}
\begin{figure}
    \centering
    \includegraphics[width=\linewidth]{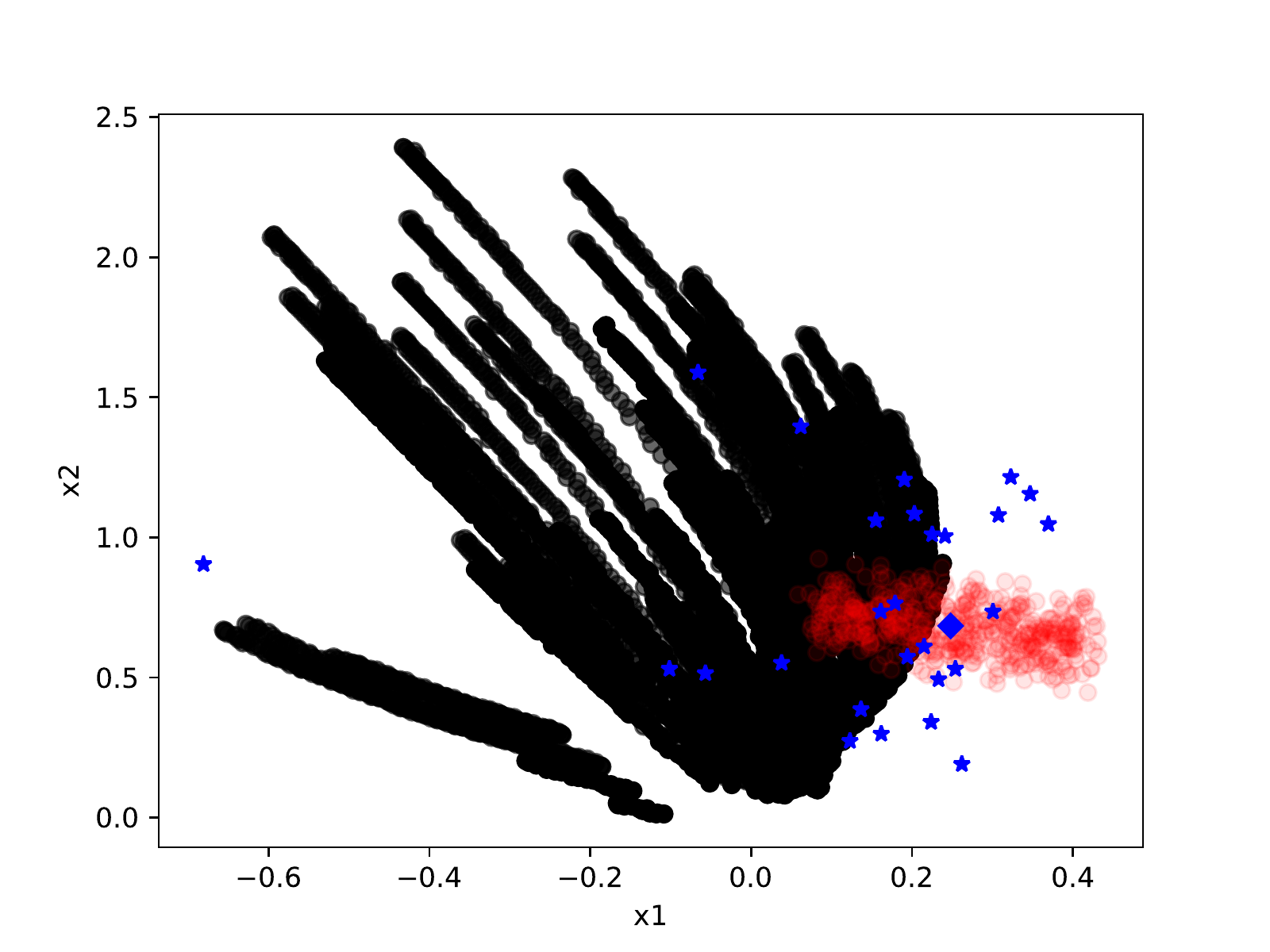}
    \caption{Visualisation of the diluted AGB final surface abundances set of Monash (black circles, dilution increments of 0.01 instead of 0.002 to improve the visibility of the figure) and the subset of 28 Ba stars (blue stars). Principal component analysis (see text for explanation) was used to reduce the dimensions of the AGB final surface abundances sets and derived abundances to only 2 linearly uncorrelated dimensions, x1 and x2, that do not have a physical meaning. For one star, BD~$-$14$^{\circ}$2678 (blue diamond), we also plotted the results of Monte-Carlo sampling of the error bars on the abundances (red circles, showing 500 samples), for details see Section \ref{sec:nearest}.}
    \label{fig:2Dfigure}
\end{figure}

We visualise our classification problem in Figure \ref{fig:2Dfigure}, where we show the set of AGB final surface abundances of Monash and the set of Ba star observations after reducing them into 2 dimensions. This is done using principal component analysis, which is a procedure that converts observations of correlated features into an often smaller set of linearly uncorrelated features. For one star, we also show the results of Monte-Carlo sampling the error bars on the observed abundances, which we will explain in detail in Section \ref{sec:nearest}. From this figure we can identify two potential issues: (1) several observed Ba stars fall outside the parameter space covered by the set of AGB final surface abundances and (2) some of the diluted models become indistinguishable from each other. The same issues are present when using the set of AGB final surface abundances of \textsc{fruity} instead of the Monash set.

The first issue is probably due to the observed spectra showing abundance trends that the models cannot match. This may be caused by present limitations in the theoretical AGB nucleosynthesis models and/or by observational uncertainties. Our classification algorithms can thus identify abundance pattern of Ba stars that might be difficult to classify, which allows us to further specify the features that are present in the AGB final surface abundances but not in the Ba stars and vice versa. 

The second issue can be improved by either grouping similar AGB final surface abundances and giving this group one name or label at the start or at the end of the classification pipeline, or by using a technique to classify the observed spectra that does not require the AGB final surface abundances to be named.

\subsection{Artificial Neural Networks (ANN)}
\label{sec:ann}
We use Tensorflow \citep{tensorflow} to create our ANN ensembles, one for each set of AGB final surface abundances. The training, cross-validation and testing sets are 56\%, 24\% %(coming from a 30\% split of 80\%) 
and 20\% respectively of either diluted set. The sets are well-balanced and randomly shuffled before a new network is trained for a certain set of AGB final surface abundances. For the training of these networks, we group all the models with the same initial mass and metallicity in one short label. The reason being that the models with same mass and metallicity only differ in parameters such as $^{13}$C-pocket size and (low) rotational velocity. These variations in the abundances are often already accounted for by the dilution process. 

We trained our network with a randomly selected 56\% of the diluted set of the AGB final surface abundances as introduced in Section \ref{sec:agbyields}. Therefore, the values for the initial mass of 28 stars (as presented in \citet{jorissen19} are not included in our algorithms.

The typical accuracy of a single ANN is around 95\% for both the \textsc{fruity} and Monash ANN. We use for the input layer a number of nodes equal to the number of elements we use to classify, plus all the possible unique pairs of elements\footnote{This adds up to a total number of $n(n-1)/2 + n$ nodes, where $n$ is the number of elements used in the classification}. We also include a dropout rate of 30\%, meaning the probability of training a node in a particular training iteration, to reduce the change of overfitting. We do not use the mass and metallicity as input variables.
The number and size of hidden layers is different for each set of AGB final surface abundances: for the \textsc{fruity} set we use a single hidden layer of size 100 times the output layer, which has 65 nodes (one per \textsc{fruity} label). For the Monash set we use two hidden layers of size 10 times the output layer, which has 69 nodes. Every hidden layer has a dropout rate of 30\%. A \textsc{fruity} ANN is trained with 4 epochs with a learning rate of 0.001 while a Monash ANN is trained with 2 epochs and a learning rate of 0.0015. For both cases we use Tensorflow RMSprop optimization algorithm.

Our ensembles are a collection of 20 ANNs for each set of AGB final surface abundances. We return the predictions from all the networks, so that we may have more than one classification for each observation. This is not the standard method when including ensembles of neural networks: usually when ensembles are used, the ensemble output layer is calculated as the median of each individual ANN output layer node-wise. However, this procedure may hide relevant statistical information from the ensemble that can be useful when obtaining a range of masses and metallicities for our classification. Therefore, we instead take the classification of each ANN in the ensemble. When all ANNs agree on the classification, there is only one match from the whole ensemble to compare to the outcome of the nearest-neighbour algorithm, leading to smaller ranges in the overlapping classification. When the ANNs do not agree, there will be a collections of matches from the ensemble to compare to the outcome of the nearest-neighbour algorithm, leading to larger ranges in the overlapping classification.

When classifying the observations with the ensembles, a value of 1 dex (with an error of 1 dex) is used for a missing abundance\footnote{We tested other values too: 0 - 0.5 - 1.0 - 1.5 - 2.0 dex and found that the classifications of these 7 stars do not strongly depend on this value.}. This is the case in 7 of the 169 Ba stars. This approach is necessary due to the ANNs requiring all the inputs to classify. The arbitrary value and error bar do not pose a problem for the classification given the dropout regularization used during the training, as long as the substituted values are less than 30\% of the total.

\subsection{Nearest-neighbour and goodness of fit}
\label{sec:nearest}
This algorithm takes each set of AGB final surface abundances and calculates the goodness of fit (GoF) of all diluted AGB final surface abundances. It returns the one with the highest GoF and the corresponding dilution. 

One common way of calculating the GoF of AGB final surface abundances in the literature for a given observed Ba spectra is through the $\chi^2$-test. However, as stated by \cite{2015Abate} and \cite{2021stancliffe}, using $\chi^2$ for the GoF is `naive' because the observed abundances of the different s-process elements cannot be correctly modelled as independent normal distributions. In addition, in a traditional $\chi^2$ test an observed value of 0 indicates no detection, while in [X/Fe] notation it simply indicates that the abundance over iron scales like solar.

To fix these two shortcomings we have decided to use a modified $\chi^2$-test. First, we divide by the 1$\sigma$ uncertainty value instead of the observed value,
\begin{equation}\label{eq:chi_sq_our}
    \chi^2_m = \sum{\frac{(X_i-O_i)^2}{E_i}},
\end{equation}
where $X_i$ and $O_i$ are the predicted and observed abundances for element $i$, and $E_i$ is the observational uncertainty. The index $i$ only represents the elements considered in the classification (see Section~\ref{sec:elements}). By dividing by the observational uncertainty, this test weights by how accurate an observation is. The more the predicted and observed abundances agree, the smaller the $\chi^2_m$ value is. The GoF is then calculated from a tail distribution of $\chi^2_m$ values (see Eq. (\ref{eq:chi_sq_our})) calculated for random samples of the observed abundances (see below). The tail distribution is $1 - CDF$, where $CDF$ is the cumulative distribution function of all calculated $\chi^2_m$ values. The GoF of an AGB model for a given star is the value of this observation-derived tail distribution at where $\chi^2_m$ equals to that of the given, diluted AGB model. Thus, the smaller the value for $\chi^2_m$, the higher the value for the GoF. 

To create the tail distribution we use a Monte-Carlo method with which we modify the observed values $O_i$ to generate 10$^5$ possible predictions $X_i$. We take into account that the abundances depend on the atmospheric parameters by randomly sample the atmospheric parameters from normal distributions and propagate these uncertainties into the $O_i$ to generate $X_i$. We use these generated $X_i$ in Eq. (\ref{eq:chi_sq_our}) to calculate our tail distribution to which we compare our predicted models.

The dependency of the abundance determinations with the physical parameters is taken as an average of Tables 8, 9 and 10 from \cite{deCastro} and Tables 2, 3 and 4 from \citet{Roriz21_heavy}. The final distributions for the abundances are normalized such that the median is equal to $O_i$ and the standard deviation is equal to $E_i$. Finally, we reject all classifications with a GoF below 50\% and return the 5 models with the highest GoF values.

\subsection{Combining the classifications into one}
\label{sec:combining}

With our algorithms we classify our set of Ba stars, leading to a set of four files (two per set of AGB final surface abundances, using our two algorithms) with the results for all Ba stars. These four different classification sets are then combined into one for each star for each set of predicted AGB abundances. This includes a mass range, a metallicity range, and the minimum of the GoF values from the shortest distance algorithm. In order to define the mass and metallicity ranges, we collect all the masses and metallicities from the two classifications of each set of AGB abundances. Then we identify the overlapping values within the mass and metallicity ranges defined independently. We do this for \textsc{Fruity} and Monash separately by selecting the overlap between the classification from the nearest-neighbour algorithm and of the ensemble of neural networks for each star using either the \textsc{Fruity} or Monash set. Thus if the nearest-neighbour classification using the Monash set results in a mass range of 2.0-3.0 M$_{\odot}$ and the ensemble of neural networks using the Monash set in a range of 2.5-3.0 M$_{\odot}$, then the table of Monash classifications will list 2.5-3.0 M$_{\odot}$ as overlapping mass range.

For some stars the overlapping range is only one value, e.g., 2.5 M$_{\odot}$. This is because our sets of AGB final surface abundances are discrete and often have gaps of 0.5 M$_{\odot}$ between consecutive masses. To correct for this, we broaden the final ranges slightly: we add (subtract) 0.25 M$_{\odot}$ to the maximum (minimum) of our determined mass range, and multiply (divide) by 1.7 the maximum (minimum) of our determined metallicity range. These factors are based on the values for the initial mass and metallicity that are present in our sets of AGB final surface abundances and are chosen to close the gaps between the initial values of the mass and metallicities that are present in the sets. After applying these factors we have our final classifications.

All algorithms as well as the 4 different sets of classifications can be found online\footnote{\url{https://github.com/AndresYague/Ba_star_classification}}.

\subsection{Elements used in classification}
\label{sec:elements}

The abundances of 21 elements are measured from the spectra of the Ba stars sample considered in our analysis. However, in this paper we focus on the s-process classification and we thus eliminate all elements lighter than Fe. That leaves us with 13 elements. Finally, we also exclude Nb since this element is often shown to have an observed abundance much higher than theoretical s-process predictions \citep[][]{2022Cseh}.

As mentioned in Sections \ref{sec:corr_el} and \ref{sec:visualisation}, there might be trends in the observed data that are not present in the AGB final surface abundances and vice versa. We aim to match all Ba stars with a s-process pattern and we thus want to exclude the elements that decrease the GoF of our classifications. To determine which elements to exclude, we systematically removed elements from set A and recorded those with the highest improvement on the accuracy of the networks when compared to the full set. The results indicated that Mo, and then La and Y, in that order, are the elements with the highest impact on the closest neighbour algorithm accuracy for both \textsc{fruity} and Monash ensembles. Therefore, we defined a new set, `set F' (`F' for final) by removing Mo, La and Y from set A. Removing more elements had marginal effects on the overall accuracy. Set F is, therefore, defined by: Fe, Rb, Sr, Zr, Ru, Nd, Ce, Sm, and Eu. 

While there are difficulties when measuring Sr related to blending with CN molecules, we have include this element because \citet{Roriz21_heavy} carefully selected Sr-lines free from contamination of the other elements/molecules. Furthermore, NLTE effects on Sr at solar metallicity lead to a correction of approximately $\pm$0.1 dex \citep{2012bergemann}, which is less than its typical $\pm$0.3 dex error bar.

\begin{figure}
    \centering
    \includegraphics[width=\linewidth]{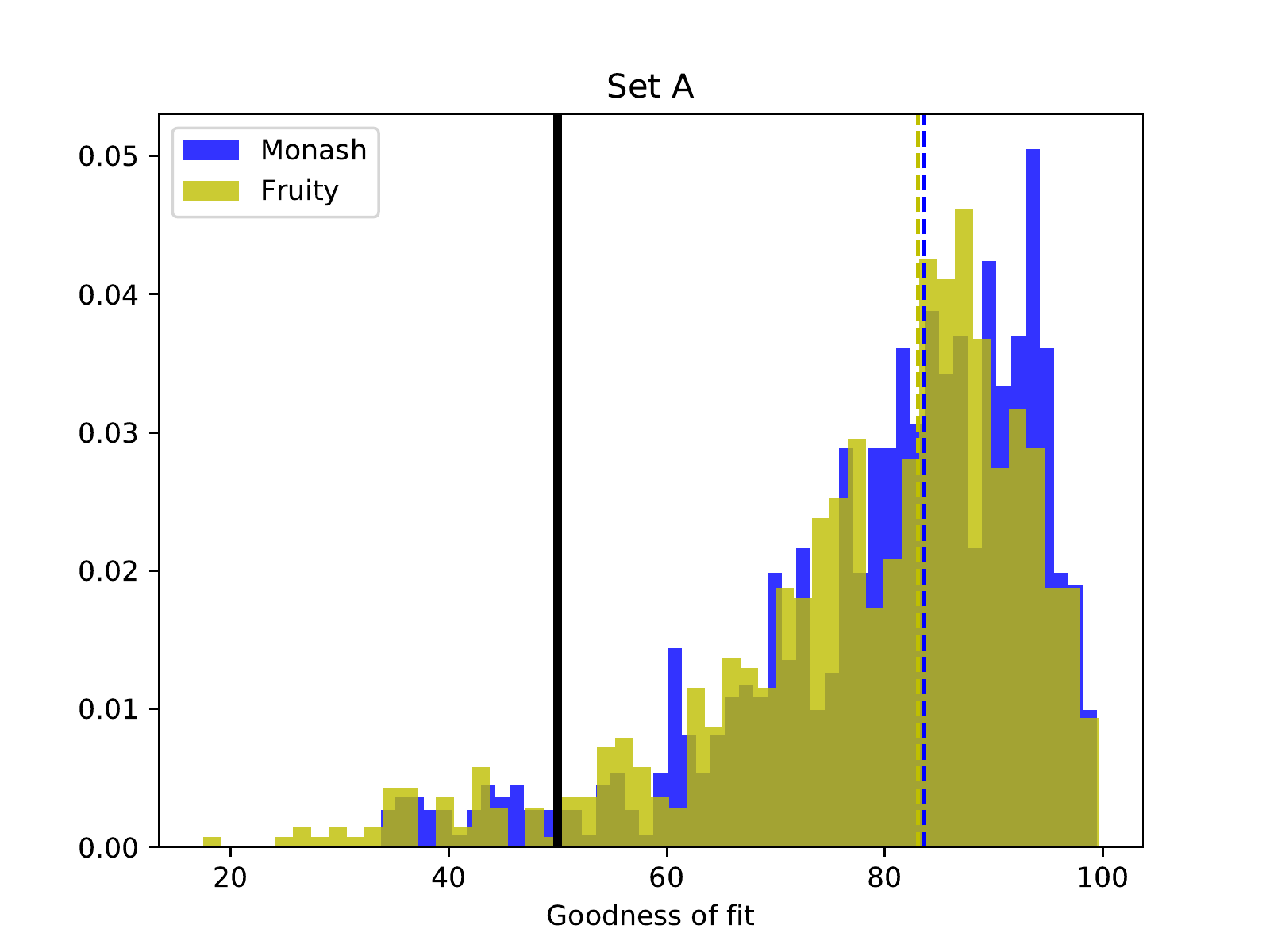}
    \includegraphics[width=\linewidth]{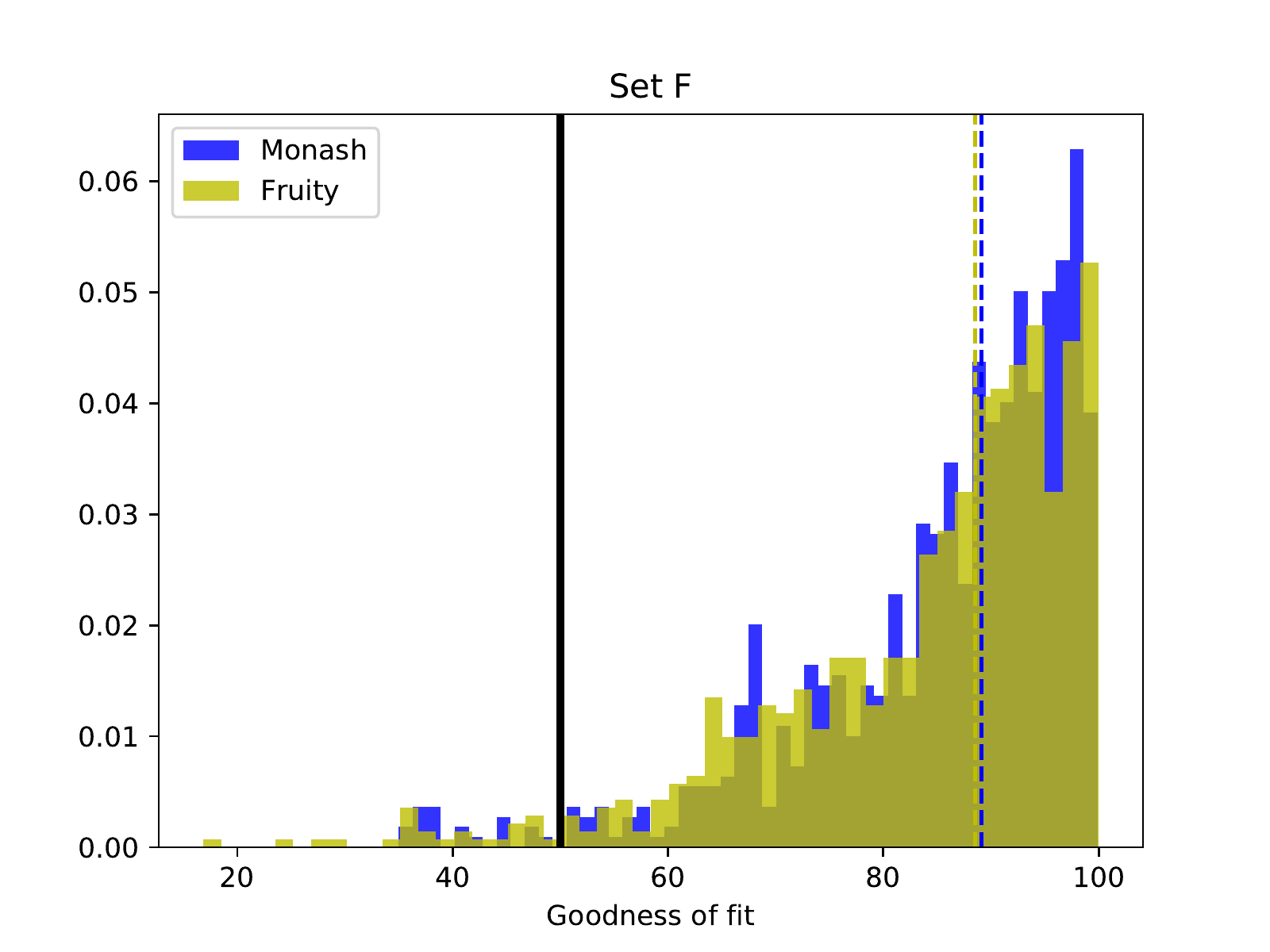}    
    \caption{Histograms of GoF distributions for Set A (top panel) and Set F (bottom panel). The results are shown for the output of the Nearest-neighbour algorithm, for the AGB final surface abundances of \textsc{fruity} (yellow) and Monash (blue). The solid black line marks the 50\% GoF, and the dashed lines are the median values of the two sets of AGB final surface abundances.}
    \label{fig:setAF}
\end{figure}

In Fig. \ref{fig:setAF} we show the distribution of GoF values for the classifications using the nearest-neighbour algorithm for each set of AGB final surface abundances. Differences between the distributions of Set A and Set F are clear: 1) the median values are higher when using Set F instead of Set A; 2) there are less identifications with a GoF below 50\% in Set F instead of Set A; and 3) the distribution peaks at a higher value in Set F compared to Set A. 

Finally, we note that our method of selecting the elements exploits the fact that a decrease in the GoF when a specific element is included is an indication that either 1) this element does not have a pure s-process origin, 2) the observations have issues (Section \ref{sec:obs_reasons}), or 3) the models are missing relevant physics affecting the abundance pattern (e.g., missing neutron fluxes, see discussion and models in Section \ref{sec:iproc}).
Our approach of changing the list of elements to maximize the GoF allows us to discover which of the observed elements do and which do not display the patterns predicted by the models. We remove element by element to not be affected by any previous bias (e.g., what is an s-process element and what is not, and known problematic lines to measure).

%-------------------------------------------------------------

\section{Classification of the 28 Ba stars considered in Paper 1}
\label{sec:28stars}

In this section we validate the ML algorithms by classifying the 28 stars previously analyzed in Paper 1. First, we present our classification based on the machine learning algorithms and we compare our results with paper 1. Second, we discuss the seven Ba stars where we find a different initial mass for the AGB companion compared to the analysis of \citet{jorissen19} (Groups 2 and 3 in Paper 1).

\subsection{Classification of the 28 stars with our algorithms}
\label{sec:ml28}

The final classifications of the subset of 28 stars are shown in Table \ref{tab:28stars}.  

When comparing the results of Paper 1 with our final classifications, for four stars we find a different mass range for the AGB star companion, independently from the two sets of stellar AGB abundances used: HD 18182, HD 84678, HD 107541, and HD 134698. These four stars were also listed among the seven stars reported by Paper 1 as difficult to identify, and we will discuss them in more detail in Section \ref{sec:28stars_prob}. For the other 24 stars we have at least one final classification which agrees with Paper 1. Therefore, we positively verified the capability of the ML algorithms implemented in the present analysis. 

Another method to check our algorithms and their results is to compare the metallicity ranges of the final classifications to the metallicity derived from observations, where the calculated [Fe/H] should be within the error bar of the observed values. This is the case for 27 (26) of the 28 final classifications using Monash (\textsc{Fruity}) predictions. The exceptions are HD84678 and HD18182, which are both part of the four stars that will be discussed in more detail in the following section. Thus, also from this comparison we can conclude that our algorithms are accurate classificators.

\begin{table*}
\centering
\caption{The final classification of the subset of 28 stars}
\vspace{-0.2cm}
\begin{tabular}{l|ll|ll|lll}
%\begin{longtable}{l|ll|ll|lll}
%\hline
 & \multicolumn{2}{|c}{Earlier works} & \multicolumn{2}{|c}{Paper 1} & \multicolumn{3}{|c}{This work}\\
Star & M$_{\rm{AGB}}$ (M$_{\odot}$)& [Fe/H] (dex)  & mass range & [Fe/H] values & mass range & [Fe/H] range & GoF \\ 
\hline
MONASH & & & & & & & \\ 
BD $-$14$^{\circ}$2678 	&3.5 & 0.01 $\pm$ 0.12 &  2.0 to 2.5 & 0.00 & 2.25 to 2.75 & -0.38 to 0.08 & 85 \\ 
CD $-$42$^{\circ}$2048& 3.1 & -0.23	$\pm$ 0.16  & 2.0 to 3.0 & -0.37, -0.24, -0.15 & 1.75 to 2.75 & -0.48 to 0.08 & 64 \\ 
CPD $-$64$^{\circ}$4333& 2.1 & -0.10 $\pm$ 0.18  & 2.0 to 2.5 & -0.24, -0.15 & 2.25 to 2.75 & -0.6 to -0.14 & 64 \\ 
HD 18182* & 1.9 & -0.17 $\pm$ 0.10  & 1.5 to 2.0\footnote{\label{onlyone} only one matching model} & -0.24, -0.15 & 2.25 to 2.75 & -0.23 to 0.23 & 78 \\
HD 20394 & 3.2 & -0.22 $\pm$ 0.12  & 2.0 to 3.0 & -0.24, -0.15 & 2.25 to 2.75 & -0.6 to -0.01 & 72 \\ 
HD 24035 & 1.8 & -0.23 $\pm$ 0.15 & 1.5 to 2.0 & -0.37 & 1.75 to 2.75 & -0.6 to -0.01 & 55 \\ 
HD 40430 & 2.8 & -0.23 $\pm$ 0.13 & 2.0 to 3.0 & -0.24, -0.15 & 1.75 to 2.75 & -0.48 to 0.08 & 93 \\ 
HD 49641 & 5.2 & -0.30 $\pm$ 0.17 & 2.0 to 3.0 & -0.37, -0.24, -0.15 & 2.75 to 4.25 & -0.6 to -0.14 & 83 \\
HD 53199 & 3.0 & -0.23 $\pm$ 0.13 & 2.0-3.0 & -0.24, -0.15 & 1.75 to 2.75 & -0.48 to 0.08 & 85 \\ 
HD 58121 & 3.1 & -0.01 $\pm$ 0.13 & 2.0 to 3.0 & -0.15, 0.00 & 1.75 to 2.25 & -0.23 to 0.08 & 79 \\ 
HD 58368 & 3.1 & 0.04  $\pm$ 0.14 & 2.0 to 2.5 & -0.15, 0.00 & 2.25 to 3.25 & -0.38 to 0.08 & 64 \\ 
HD 59852 & 3.0 & -0.22 $\pm$ 0.10 & 2.0 to 3.0 & -0.24, -0.15 & 1.75 to 2.75 & -0.48 to 0.08 & 94 \\ 
HD 84678* & 3.8 & -0.13 $\pm$ 0.16 & 2.5\textsuperscript{3} & -0.24 & 1.75 to 1.75 & -0.48 to -0.14 & 50 \\ 
HD 91208 & 2.8 & 0.05  $\pm$ 0.14 & 2.0 to 3.0 & -0.15, 0.00 & 1.75 to 2.75 & -0.38 to 0.08 & 71 \\ 
HD 92626 & 5.6 & -0.15 $\pm$ 0.22 & 2.0 to 2.5 & -0.37 & 2.25 to 2.75 & -0.6 to -0.14 & 57 \\ 
HD 95193 & 3.3 & 0.04  $\pm$ 0.12 & 2.0-3.0 & -0.15, 0.00 & 2.25 to 2.75 & -0.23 to 0.08 & 94 \\ 
HD 107541* & 1.4 & -0.63 $\pm$ 0.11 & 2.0\textsuperscript{3} & -0.67 & 2.25 to 2.75 & -0.6 to -0.14 & 62 \\ 
HD 119185 & 1.7 & -0.43 $\pm$ 0.10 & 1.3 to 2.0 & -0.37 & 1.75 to 2.75 & -0.48 to -0.14 & 65 \\ 
HD 134698* & 1.2 & -0.52 $\pm$ 0.12 & -\footnote{\label{nomatch}no match of the first peak} & -0.67 & 1.25 to 3.25 & -0.91 to -0.14 & 82 \\
HD 143899 & 3.0 & -0.27 $\pm$ 0.12 & 2.0 to 3.0 & -0.37, -0.24, -0.15 & 1.75 to 2.25 & -0.48 to 0.08 & 88 \\ 
HD 154430 & 3.6 & -0.36 $\pm$ 0.19 & 2.0 to 3.0 & -0.37, -0.24 & 1.25 to 1.75 & -0.48 to -0.01 & 96 \\ 
HD 180622 & 1.9 & 0.03  $\pm$ 0.12 & 1.5 to 2.0 & -0.15, 0.00 & 1.75 to 2.25 & -0.23 to 0.08 & 92 \\ 
HD 183915 & 1.9 & -0.39 $\pm$ 0.14 & 1.5 to 2.0 & -0.37 & 1.25 to 2.25 & -0.6 to -0.14 & 89 \\ 
HD 200063 & 2.4 & -0.34 $\pm$ 0.20 & 2.0 to 2.5 & -0.37, -0.24, -0.15 & 1.75 to 2.25 & -0.38 to -0.01 & 98 \\ 
HD 201657 & 3.0  & -0.34 $\pm$ 0.17& 2.0 to 3.0 & -0.37, -0.24 & 1.75 to 2.75 & -0.6 to -0.01 & 97 \\ 
HD 201824 & 2.8 & -0.33 $\pm$ 0.17 & 2.0 to 2.5 & -0.37, -0.24 & 1.75 to 2.25 & -0.48 to -0.01 & 96 \\ 
HD 210946 & 1.9 & -0.12 $\pm$ 0.13 & 1.5 to 2.0 & -0.24, -0.15, 0.00 & 2.25 to 2.75 & -0.38 to 0.08 & 83 \\ 
HD 211594 & 3.2 & -0.43 $\pm$ 0.14 & 2.0 to 2.5 & -0.37 & 2.25 to 2.75 & -0.48 to -0.01 & 87 \\ 
\hline
\textsc{Fruity} & & & & & & & \\ 
BD $-$14$^{\circ}$2678 & 3.5& 0.01 $\pm$ 0.12 & 2.5 to 3.5 & 0.00 & 1.75 to 2.5 & -0.23 to 0.23 & 98\\ 
CD $-$42$^{\circ}$2048 & 3.1 & -0.23 $\pm$ 0.16 & 2.0 to 3.0 & -0.30, -0.15 & 1.75 to 3.0 & -0.54 to 0.23 & 56\\ 
CPD $-$64$^{\circ}$4333 & 2.1 & -0.10	$\pm$ 0.18 & 2.0 & -0.15, 0.00 & 1.75 to 3.0 & -0.54 to 0.23 & 77 \\ 
HD 18182 & 1.9 & -0.17 $\pm$ 0.10 & 2.0\textsuperscript{3} & -0.15 & 2.5 to 3.5 & 0.1 to 0.23 & 68\\
HD 20394 & 3.2 & -0.22 $\pm$ 0.12 & 2.1 to 3.0 & -0.30, -0.15 & 1.75 to 2.25 & -0.38 to 0.08 & 82\\ 
HD 24035 & 1.8 & -0.23 $\pm$ 0.15 & 1.75 to 2.25 & -0.30 & 1.65 to 2.35 & -0.54 to -0.07 & 80\\
HD 40430 & 2.8 & -0.23 $\pm$ 0.13 & 2.5 to 3.25 & -0.30, -0.15 & 1.75 to 2.25 & -0.23 to 0.08 & 88\\ 
HD 49641 & 5.2 & -0.30 $\pm$ 0.17 & 2.1 to 4.0 & -0.30, -0.15 & 4.25 to 4.75 & -0.54 to -0.07 & 89\\
HD 53199 & 3.0 & -0.23 $\pm$ 0.13 & 2.5 to 3.5 & -0.30, -0.15 & 1.75 to 2.25 & -0.23 to 0.08 & 74\\ 
HD 58121 & 3.1 & -0.01 $\pm$ 0.13 & 2.75 to 3.5 & 0.00 & 1.75 to 2.75 & -0.23 to 0.23 & 80\\ 
HD 58368 & 3.1 & 0.04  $\pm$ 0.14 & 2.5 to 3.5 & 0.00 & - & - & - \\ 
HD 59852 & 3.0 & -0.22 $\pm$ 0.10 & 2.5 to 3.0 & -0.30, -0.15 & 2.25 to 3.25 & -0.23 to 0.08 & 91\\ 
HD 84678 & 3.8 & -0.13 $\pm$ 0.16 & -\textsuperscript{4} & -0.15, 0.00 & 1.5 to 1.75 & -0.23 to -0.07 & 52\\ 
HD 91208 & 2.8 & 0.05  $\pm$ 0.14 & 2.5 to 3.25 & 0.00 & 1.75 to 2.5 & -0.23 to 0.23 & 84\\ 
HD 92626 & 5.6 & -0.15 $\pm$ 0.22 & 2.1 to 2.75 & -0.30 & 1.75 to 2.5 & -0.54 to -0.07 & 95\\ 
HD 95193 & 3.3 & 0.04  $\pm$ 0.12 & 2.5-3.5 & 0.00 & 1.75 to 2.5 & -0.23 to 0.23 & 98\\ 
HD 107541 & 1.4 & -0.63 $\pm$ 0.11 & 1.5 to 2.0 & -0.70 & 2.25 to 2.75 & -0.54 to -0.47 & 88\\
HD 119185 & 1.7 & -0.43 $\pm$ 0.10 & 1.5 to 2.1 & -0.30 & 1.75 to 2.75 & -0.54 to -0.07 & 62\\
HD 134698 & 1.2 & -0.52 $\pm$ 0.12 & -\textsuperscript{4} & -0.70 & 3.25 to 3.75 & -0.23 to -0.07 & 83\\
HD 143899 & 3.0 & -0.27 $\pm$ 0.12 & 2.5 to 3.0 & -0.30, -0.15 & 1.75 to 2.25 & -0.23 to 0.08 & 86\\ 
HD 154430 & 3.6 & -0.36 $\pm$ 0.19 & 2.25 to 3.25 & -0.30 & 1.5 to 2.25 & -0.38 to -0.07 & 95\\ 
HD 180622 & 1.9 & 0.03  $\pm$ 0.12 & -\textsuperscript{4} & 0.00 & 1.75 to 3.0 & -0.23 to 0.23 & 91\\ 
HD 183915 & 1.9 & -0.39 $\pm$ 0.14 & 1.5 to 2.5 & -0.30 & 1.25 to 2.35 & -0.54 to -0.07 & 91\\
HD 200063 & 2.4 & -0.34 $\pm$ 0.20 & 1.9 to 2.75 & -0.30, -0.15 & 1.75 to 2.25 & -0.38 to -0.07 & 96\\ 
HD 201657 & 3.0 & -0.34 $\pm$ 0.17 & 2.5 to 3.25 & -0.30 & 1.75 to 3.0 & -0.54 to 0.08 & 96\\ 
HD 201824 & 2.8 & -0.33 $\pm$ 0.17 & 2.1 to 3.0 & -0.30 & 1.75 to 2.25 & -0.38 to -0.07 & 97\\ 
HD 210946 & 1.9 & -0.12 $\pm$ 0.13 & 1.5 to 2.25 & -0.15, 0.00 & 1.75 to 3.25 & -0.23 to 0.23 & 82\\ 
HD 211594 & 3.2 & -0.43 $\pm$ 0.14 & 2.0 to 3.5 & -0.30 & 2.5 to 3.0 & -0.54 to -0.07 & 83\\ 
%3=onlyone, 4=nomatch atm
\end{tabular}
    \begin{tablenotes}
      \item \small The first three columns of both tables show the name of the star and the mass and metallicity range inferred from observations (taken from \citet{deCastro}). The fourth and fifth columns list the masses and metallicities of the classifications from Paper 1, and the final three columns show the mass and metallicity ranges of our final classifications, and the minimum GoF of the classifications. Stars with a * are described in the text, $^3$ means there was only one match in Paper 1, and $^4$ means no matches were found in Paper 1.
    \end{tablenotes}
\label{tab:28stars}
\end{table*}

\subsection{Four challenging classifications: HD 18182, HD 84678, HD 107541, and HD 134698.}
\label{sec:28stars_prob}

\begin{figure*}
    \centering
    \includegraphics[width=\linewidth]{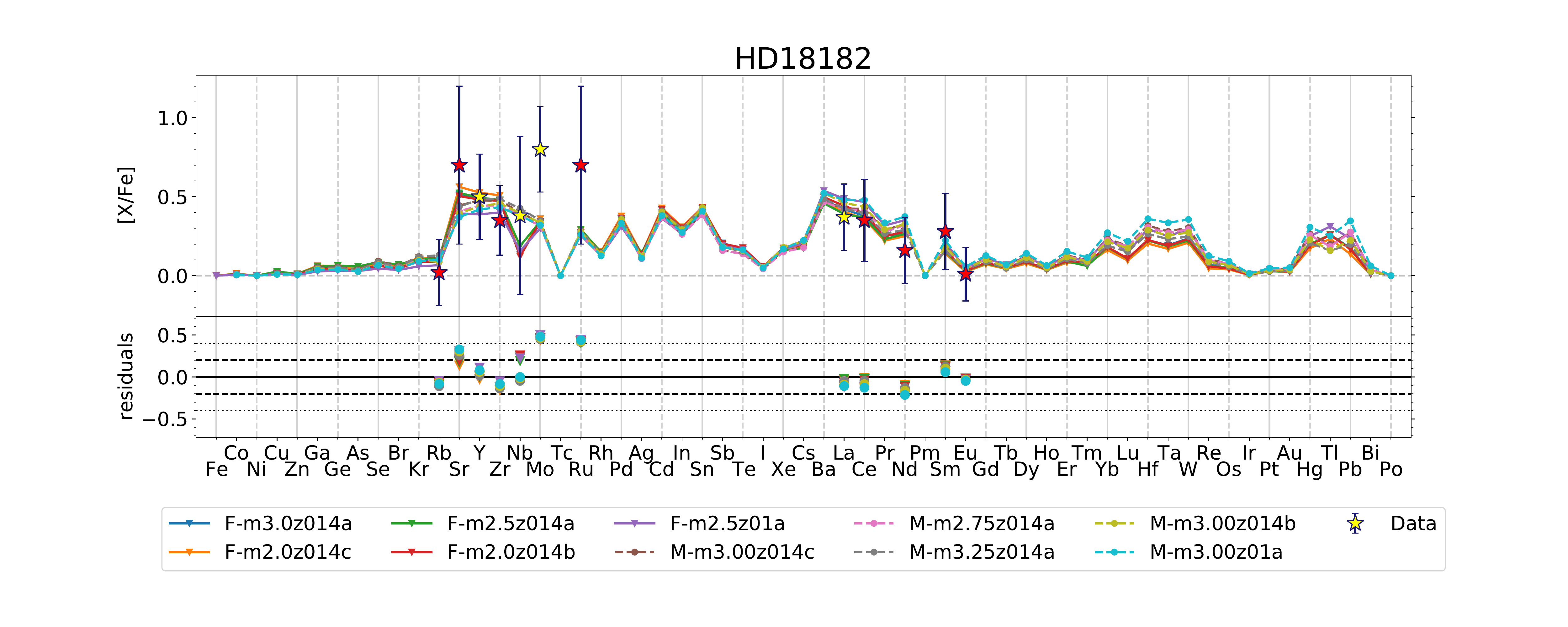}
    \caption{Top panel: comparison between the observed abundances (star symbols where red/yellow symbols are elements that are included/excluded in our classification), and the results (lines) of the nearest-neighbour algorithm. Bottom panel: residuals of the classification, i.e. the observed abundance minus the predicted abundance for each element. This star is difficult to classify, because the abundances in the first s-process peak are higher than predicted by the best-fit models.}
    \label{fig:HD18182}
\end{figure*}

%HD181812
The mass and metallicity ranges found for HD18182 in Paper 1 for both sets of AGB final surface abundances were similar to the observed values: a mass of 1.5 to 2 M$_{\odot}$ and a metallicity of [Fe/H] = $-$0.24 or $-$0.15, while the observation-derived values are 1.9 M$_{\odot}$ and a metallicity of $-$0.17 $\pm$ 0.10. Our final classification using the Monash (\textsc{Fruity}) set includes a mass range of 2.25 to 2.75 (2.5 to 3.5) M$_{\odot}$ and a metallicity range of $-$0.23 to 0.23 (0.1 to 0.23), thus the Monash values fit the observed values better. The minimum GoF of the final classification using the Monash set is 78\% and 68\% using the \textsc{Fruity} set, see Table \ref{tab:28stars} and Fig. \ref{fig:HD18182}. In the figure we only show the classifications of the nearest-neighbour algorithm, as for the ANN we grouped the predicted abundances with the same initial mass and metallicity into one label which thus does not correspond to a single set of abundances. We generate short model names to make our figures more readable and easily identify specific AGB models. Those are summarized with the relevant model parameters in Tables \ref{tab:names_fruity} and \ref{tab:names_monash}.
When comparing the models with observations [Mo/Fe] is not matched, affecting the overall fit of the observed abundance trend.

%HD84678
\begin{figure*}
    \centering    
    \includegraphics[width=\linewidth]{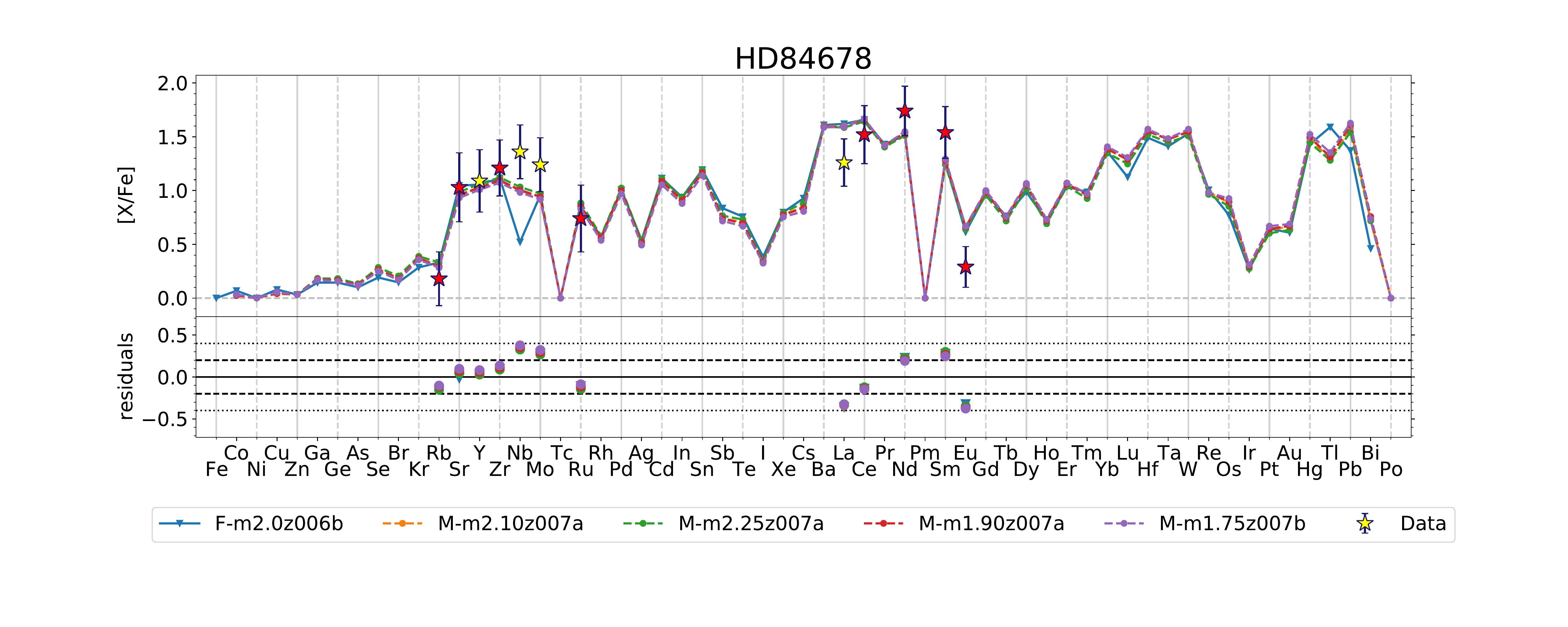}   
    \caption{Same figure format as in Fig. \ref{fig:HD18182}.
    This star was difficult to classify manually, because the binary-derived AGB mass is higher than the initial masses of the best-fit models (Group 3 stars in Paper 1). We are able to classify this star with our algorithms, indeed with models with initial masses lower than the binary-derived value, but the matching the observed values in the second $s$-process peak is still an issue.}
    \label{fig:HD84678}
\end{figure*}
The second challenging star is HD 84678 (see Fig. \ref{fig:HD84678}). The main reason is that the focus in the manual classification was on models with an initial mass close to the mass derived from binary parameters (M = 3.8 M$_{\odot}$). The conclusion in Paper 1 was that the initial mass should be below M = 2.5 M$_{\odot}$, although there was only one match found in Paper 1, with M = 2.5 M$_{\odot}$. Both our final classifications using both sets of AGB final surface abundances agree on an initial mass 1.75 M$_{\odot}$ and an initial metallicity range of [Fe/H] = $-$0.23 to $-$0.07, which overlaps with the observed metallicity range of $-$0.13 $\pm$ 0.16. The minimum GoF is only 50\% and 52\%, which is due to the high values of [Nd/Fe] and [Sm/Fe], in combination with the low value for [Eu/Fe]. The classification of the second $s$-process peak is thus still an issue.

\begin{figure*}
    \centering
    \includegraphics[width=\linewidth]{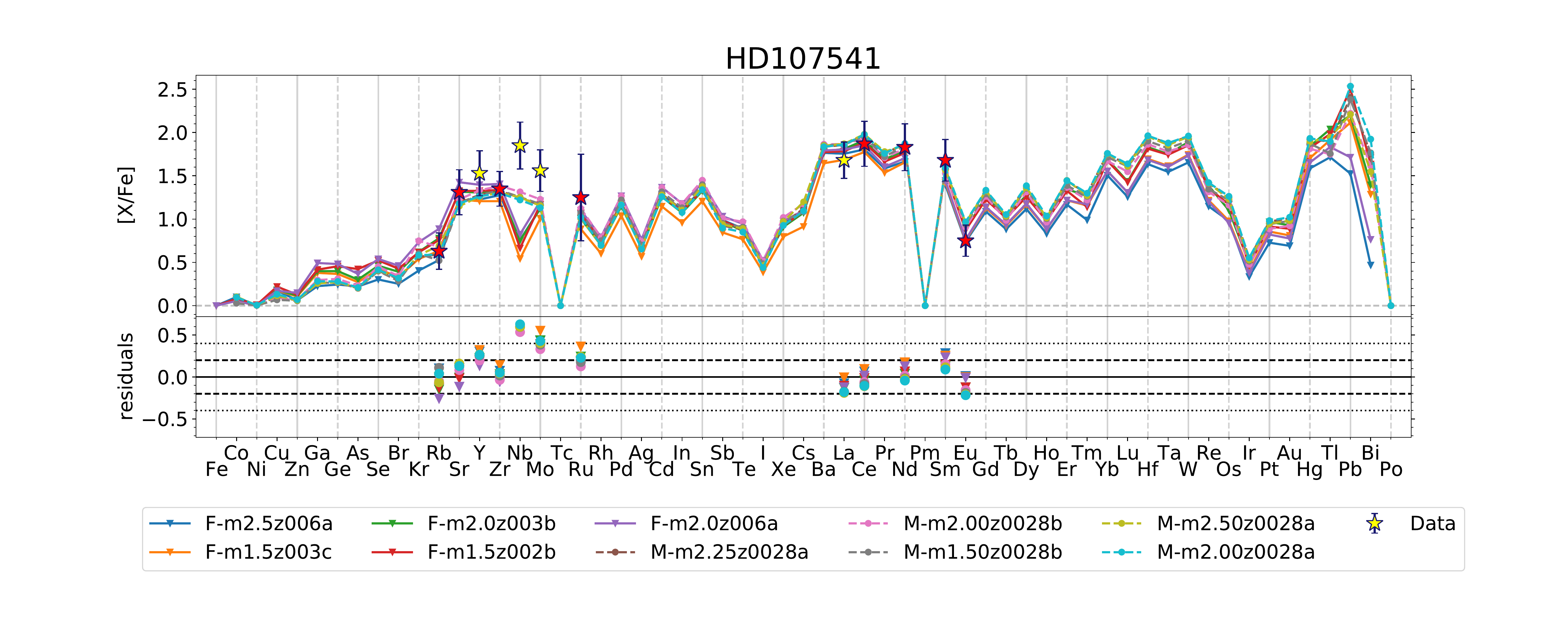} 
    \caption{Same figure format as in Fig. \ref{fig:HD18182}.
    This Ba star was difficult to classify manually because the abundances in the first s-process peak are higher than predicted by the best-fit models (Group 2 stars in Paper 1). We are able to classify it with models with slightly higher initial masses than their mass derived from binary parameters.}
    \label{fig:HD107541}
\end{figure*}

%HD107541
The third star to discuss is HD 107541 (see Fig. \ref{fig:HD107541}). Paper 1 did not find good agreement between models and the star, because the masses derived from binary parameters point at an initial mass lower than M = 2 M$_{\odot}$. Our algorithms however, agree on an final classification of M = 2.25 to 2.75 M$_{\odot}$ and [Fe/H] = $-$0.54 to $-$0.47, with a GoF of at least 62\%. Thus, the mass derived from binary parameters is likely too low. 

\begin{figure*} 
    \centering
    \includegraphics[width=\linewidth]{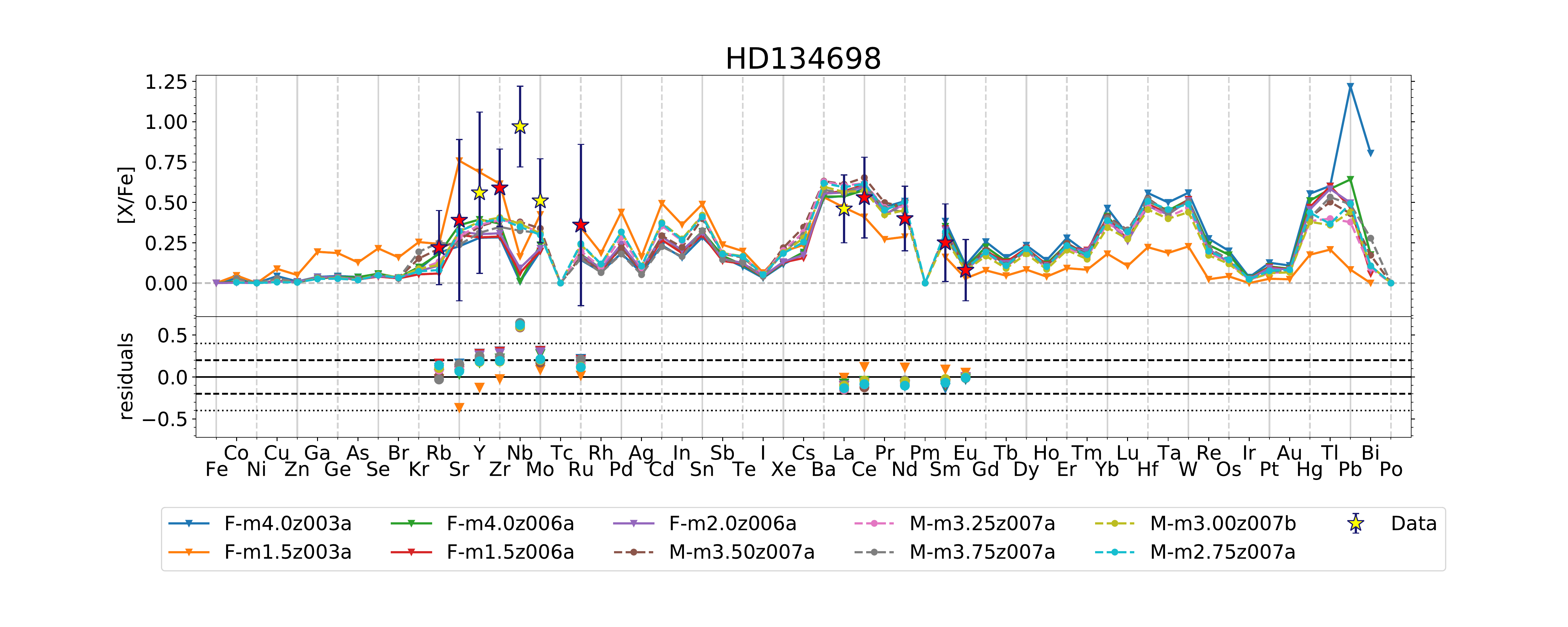}
    \caption{Same figure format as in Fig. \ref{fig:HD18182}.
    This star is one of the four Ba stars which was difficult to classify manually because the abundances in the first s-process peak are higher than predicted by the best-fit models (Group 2 stars in Paper 1). We found an final classification for this star with our algorithms. However, [Nb/Fe] is too high to be matched by our AGB final surface abundances. }
    \label{fig:HD134698}
\end{figure*}

%134698
The last star to discuss here is HD 134698 (see Fig \ref{fig:HD134698}), for which no classification were found in Paper 1. The mass derived from binary parameters is M = 1.2 M$_{\odot}$. However, current stellar models with such low initial masses do not develop TDUs during the AGB phase, and their envelope does not become s-process rich. The final classification using the Monash (\textsc{Fruity}) set is an initial mass range of M = 1.25 to 3.25 M$_{\odot}$ (3.25 to 3.75 M$_{\odot}$) and an initial metallicity range of [Fe/H] = $-$0.91 to $-$0.14 ($-$0.23 to $-$0.07). The mass and metallicity derived from binary parameters for this start are M=1.2 M$_{\odot}$ and [Fe/H] = $-$0.52 $\pm$ 0.12, these are thus both matched by the final classification using the Monash AGB final surface abundances.

In summary, we can classify these four stars by using our ML tools, but issues remains in fitting the element abundance patterns.

%----------------------------------------------------------------
\section{Classification of the full set of Ba stars}
\label{sec:full-set-classification}

Here we present the classifications of the full set of Ba stars and the statistics of these classifications. We present the Ba stars for which we have final classifications, one for the Monash AGB models and/or one for the \textsc{Fruity} AGB models, in Section \ref{sec:overlap}. 
In Section \ref{sec:nooverlap} we discuss the stars for which we do not have a final classification that is constructed of the overlap in mass range and metallicity of the outcomes of the ANNs and the nearest-neighbour algorithm.

\subsection{Ba stars for which classifications overlap}
\label{sec:overlap}

\begin{figure*}
    \centering
    \includegraphics[width=0.45\linewidth]{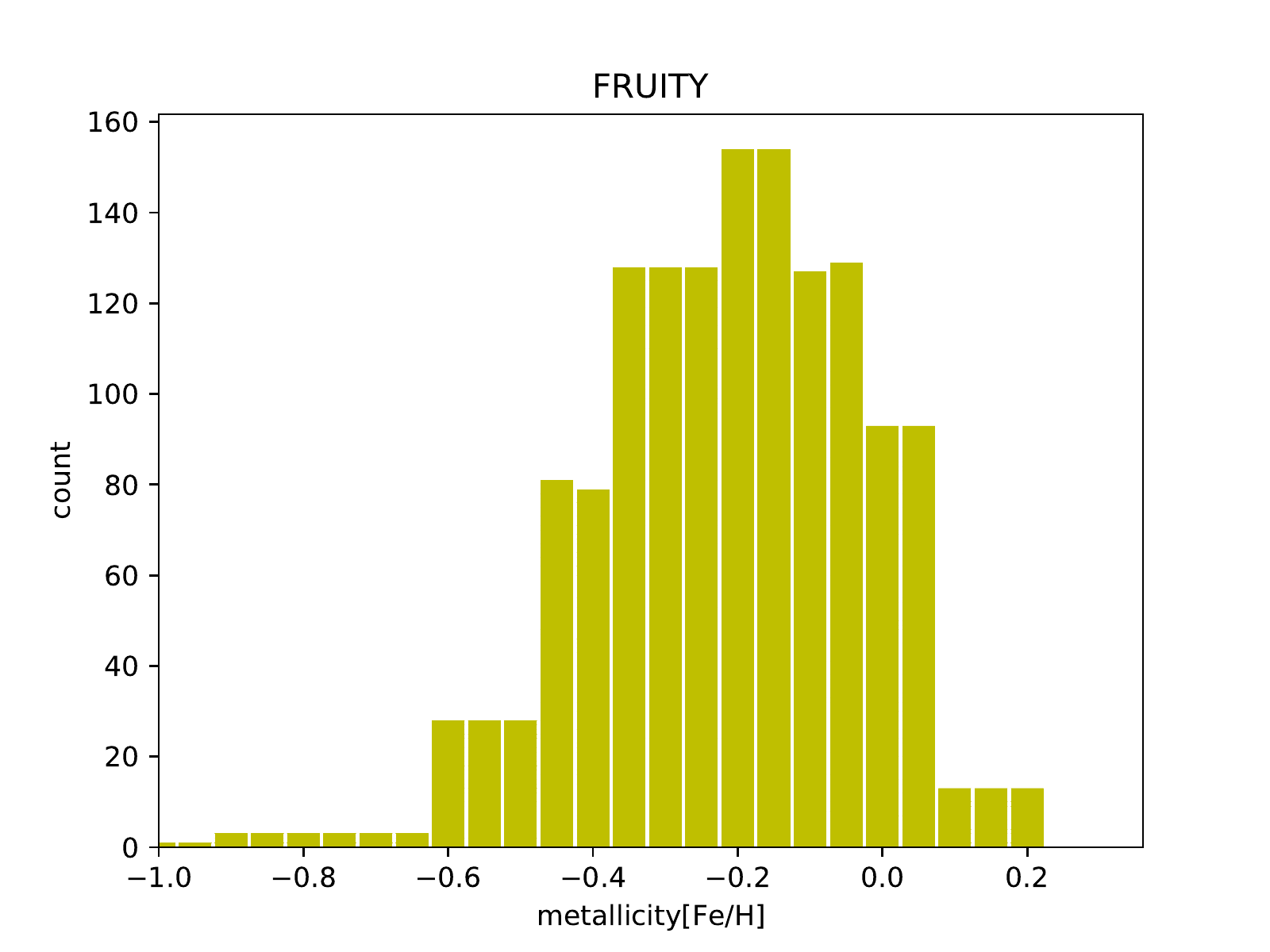}
    \includegraphics[width=0.45\linewidth]{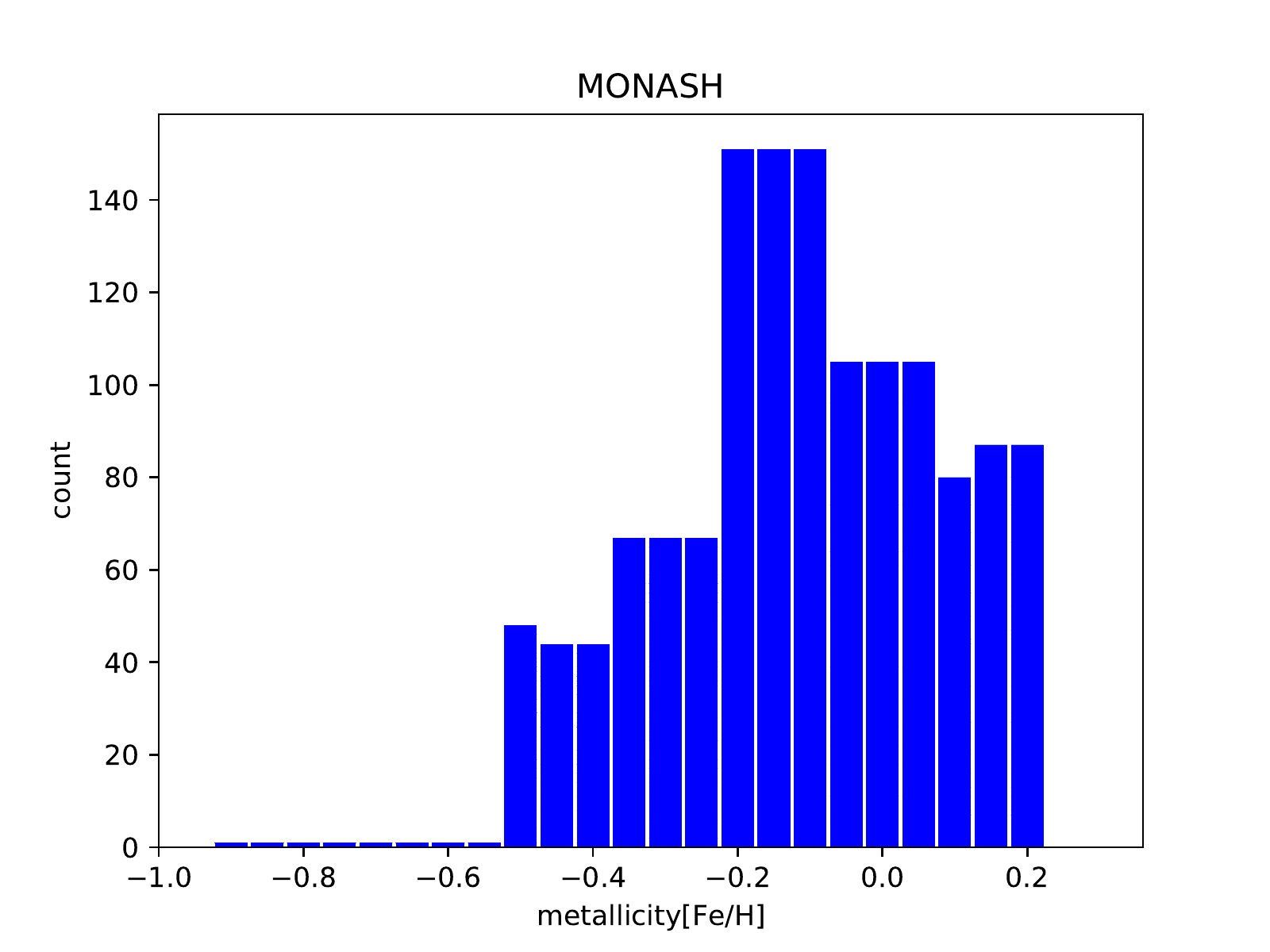}   
    \includegraphics[width=0.45\linewidth]{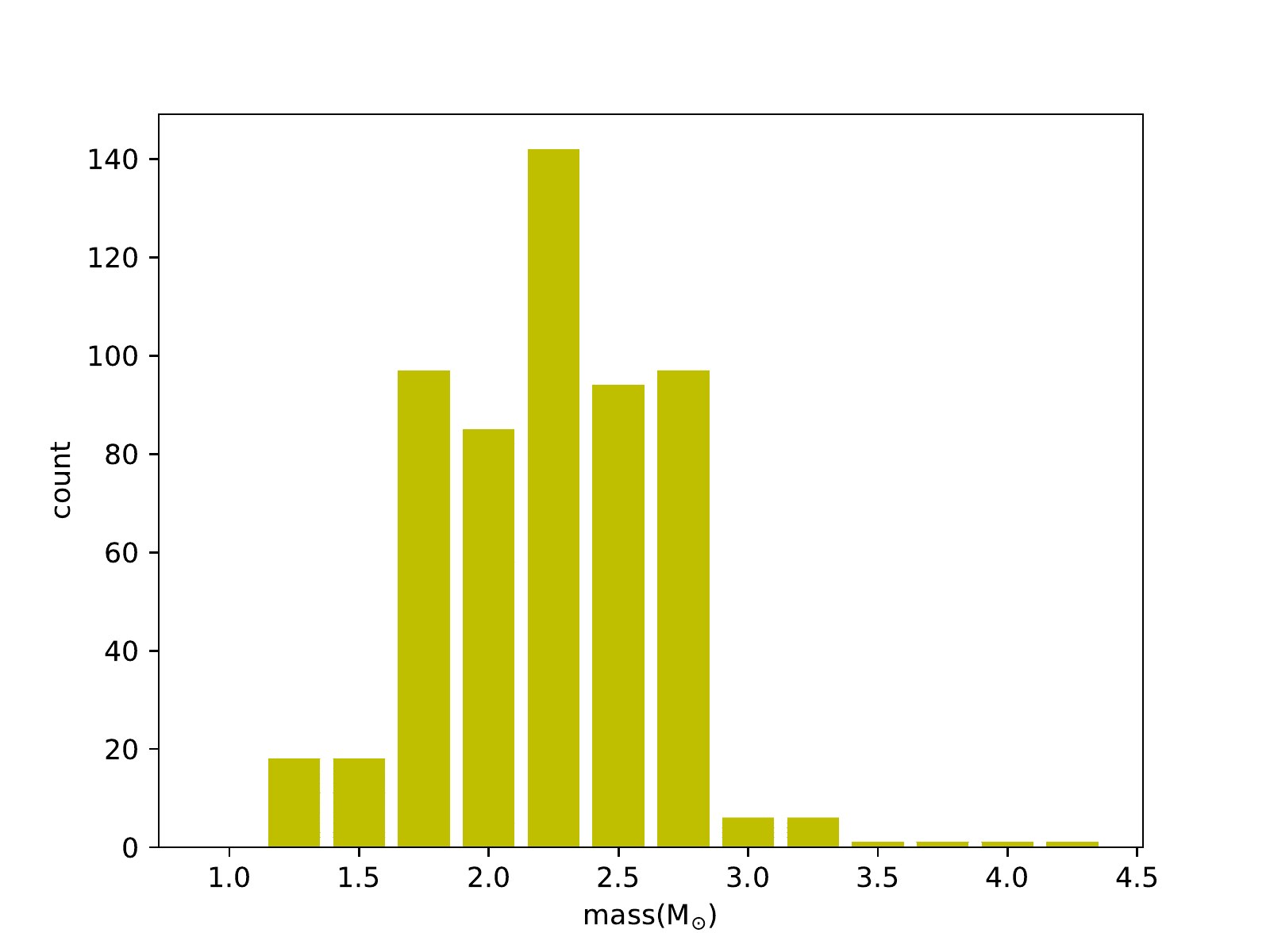}
    \includegraphics[width=0.45\linewidth]{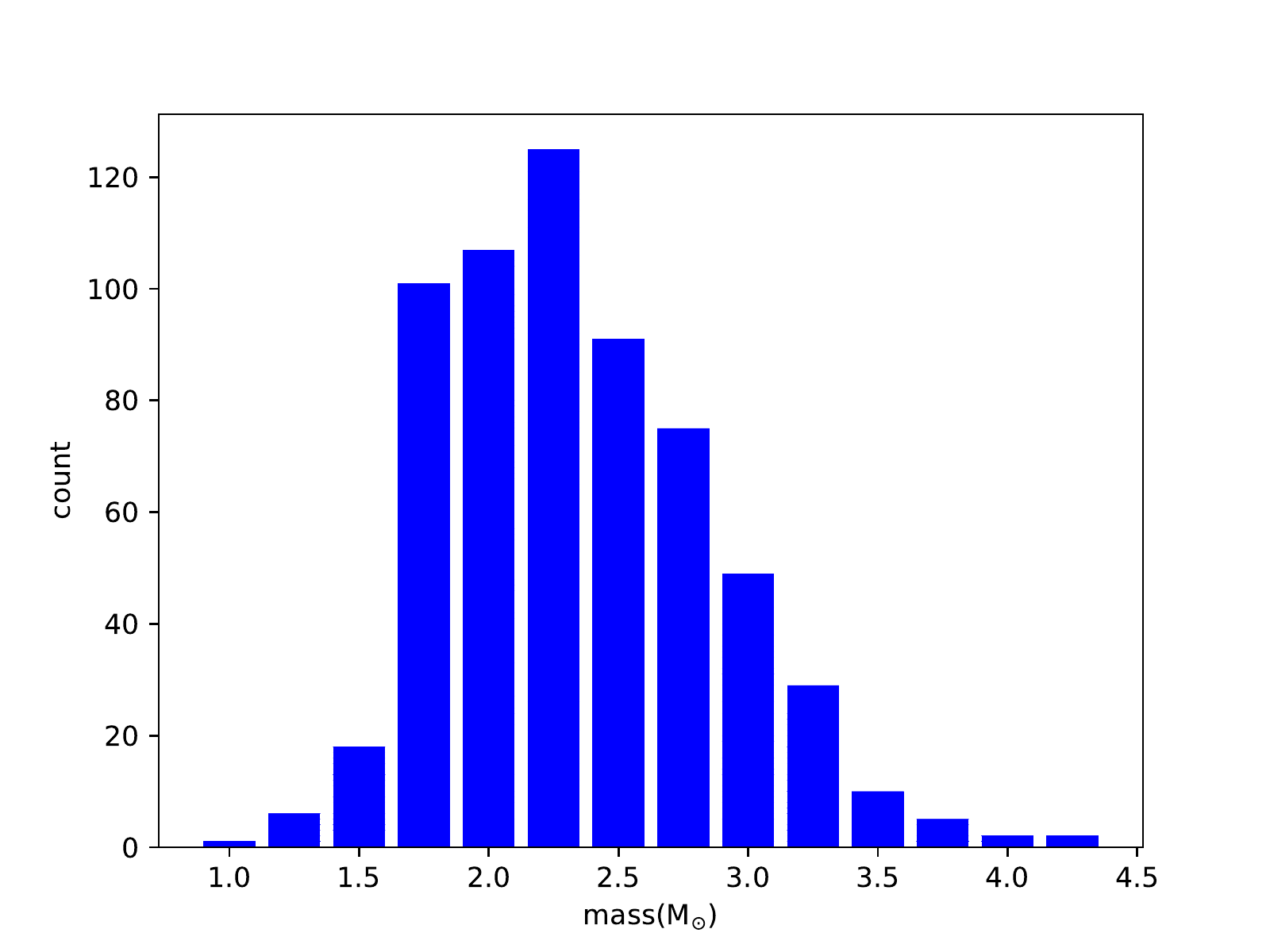}
    \includegraphics[width=0.45\linewidth]{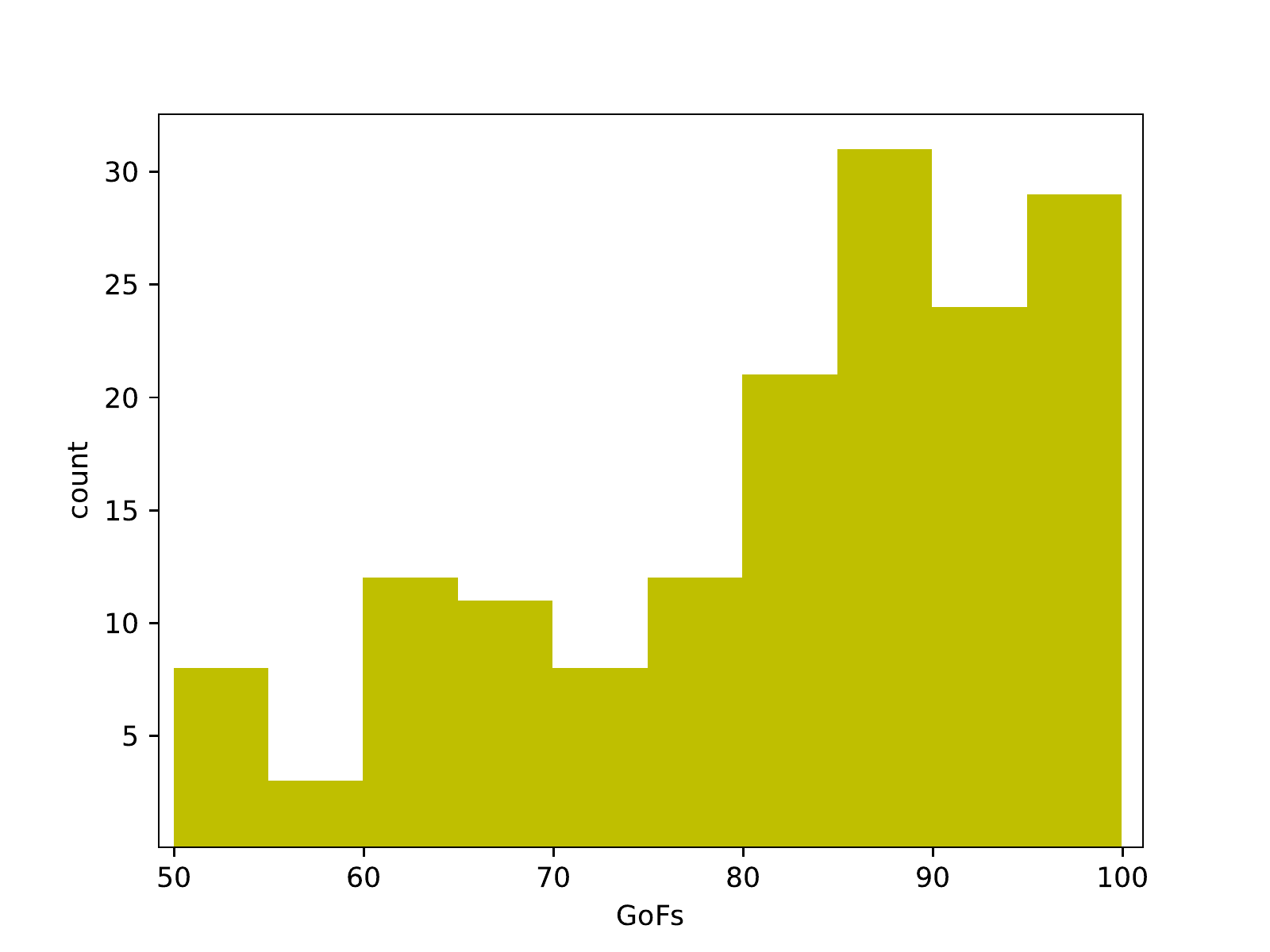}
    \includegraphics[width=0.45\linewidth]{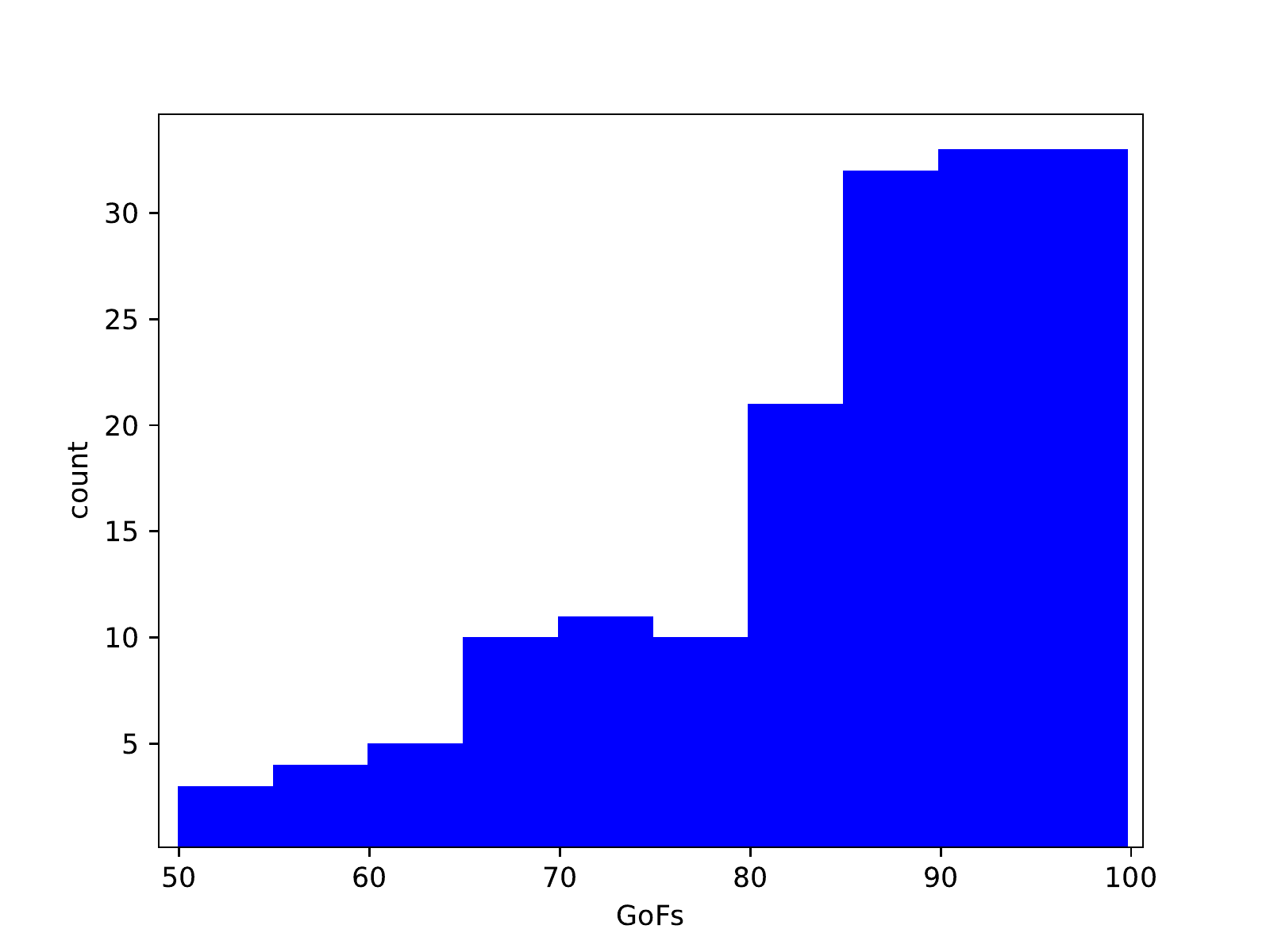}
    \caption{Distributions of the final classifications for both sets of AGB final surface abundances. From top to bottom we show the distributions for the 1) metallicities, 2) masses, and 3) minimum GoFs of the classifications as listed in the tables in Appendix \ref{sec:app_complete_tabs}. The left column shows the distribution for \textsc{fruity} and the right column Monash.}
    \label{fig:classes}
\end{figure*}

Our final classifications of both sets of AGB final surface abundances are listed in Tables \ref{tab:rest_fruity} and \ref{tab:rest_monash}. 
Using both \textsc{Fruity} and Monash AGB models, 85\% of the [Fe/H] observed values are consistent with our results within 0.02 dex. 
Since we do not have initial mass estimates for the full set of 169 Ba stars to compare with, we cannot perform the comparison between masses.

The distributions of the mass and metallicity ranges of the final classifications are shown in the top two rows of Fig. \ref{fig:classes}. In particular, we can see that the final classifications of \textsc{fruity} and Monash peak in the same bin for both the same mass and metallicity. The mean and standard deviation of the final classifications using the \textsc{Fruity} set are M = 2.23 and 0.44 M$_{\odot}$ for the mass and [Fe/H] = $-$0.21 and 0.18 for the metallicity. For the final classifications using the Monash set the mean and standard deviation are M = 2.34 and 0.52 M$_{\odot}$ for the mass and [Fe/H] = $-$0.11 and 0.19 for the metallicity. 
The distributions are not the same, as expected from Section \ref{sec:corr_el}.

The bottom row of Fig. \ref{fig:classes} shows the distributions for the minimum value of the GoFs for each stars. The mean and standard deviation for the \textsc{Fruity} classifications is 78\% and 23\%, while for the \textsc{Monash} classifications we obtain 81\% and 20\% respectively. 
In \citet{deCastro}, Gaussian distributions were fitted to the data. The mean and standard deviation of those distributions are [Fe/H] = $-$0.12 $\pm$ 0.49 and M$_{\rm{Ba\_star}}$ = 2.76 $\pm$ 0.84 M$_{\odot}$. The average [Fe/H] agrees well with our results within the uncertainties. Instead, we cannot compare directly our mass distribution derived for the former AGB star companions, since
\citet{deCastro} derived the current mass of the Ba stars observed after the mass transfer event. 

\subsection{Ba stars with no final classifications}
\label{sec:nooverlap}

In the following subsection we discuss the four stars for which our algorithms were not able to find final classifications with the AGB final surface abundances from either \textsc{Fruity} or Monash.

\begin{figure*}
    \centering
    \includegraphics[width=\linewidth]{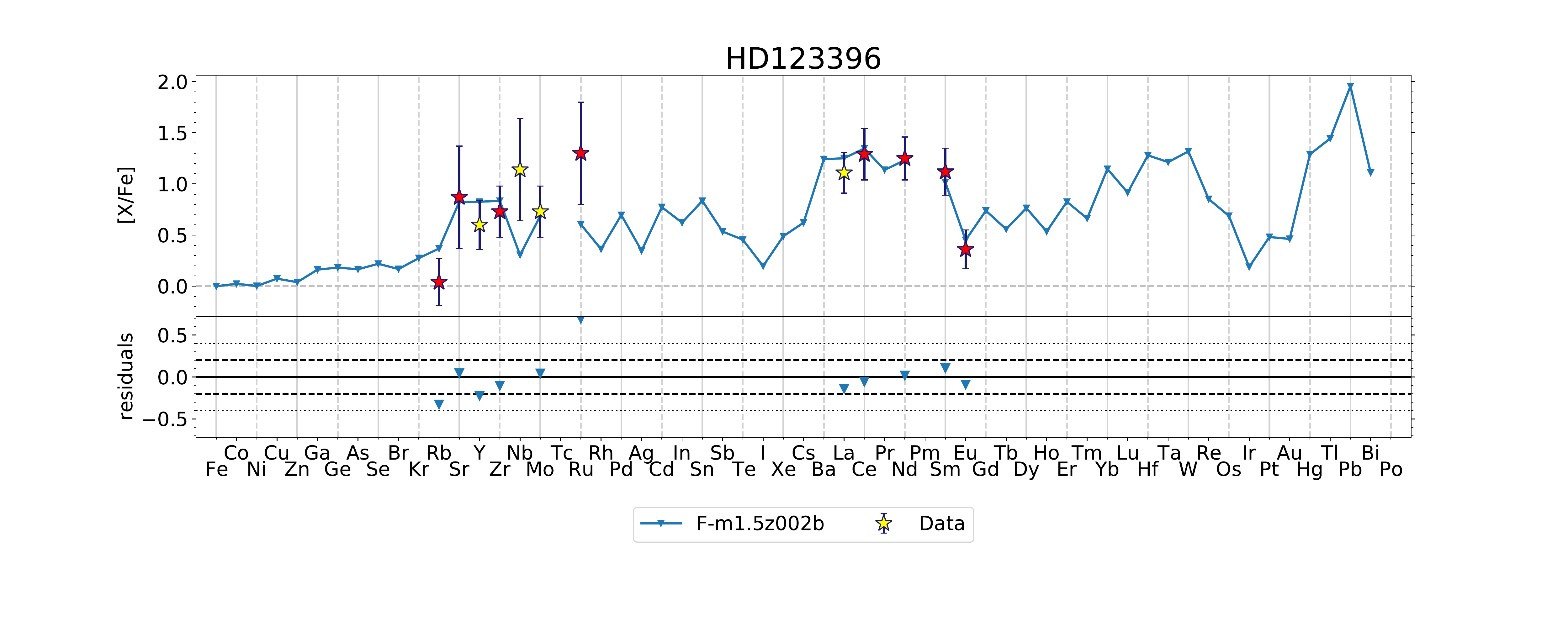}
    \includegraphics[width=\linewidth]{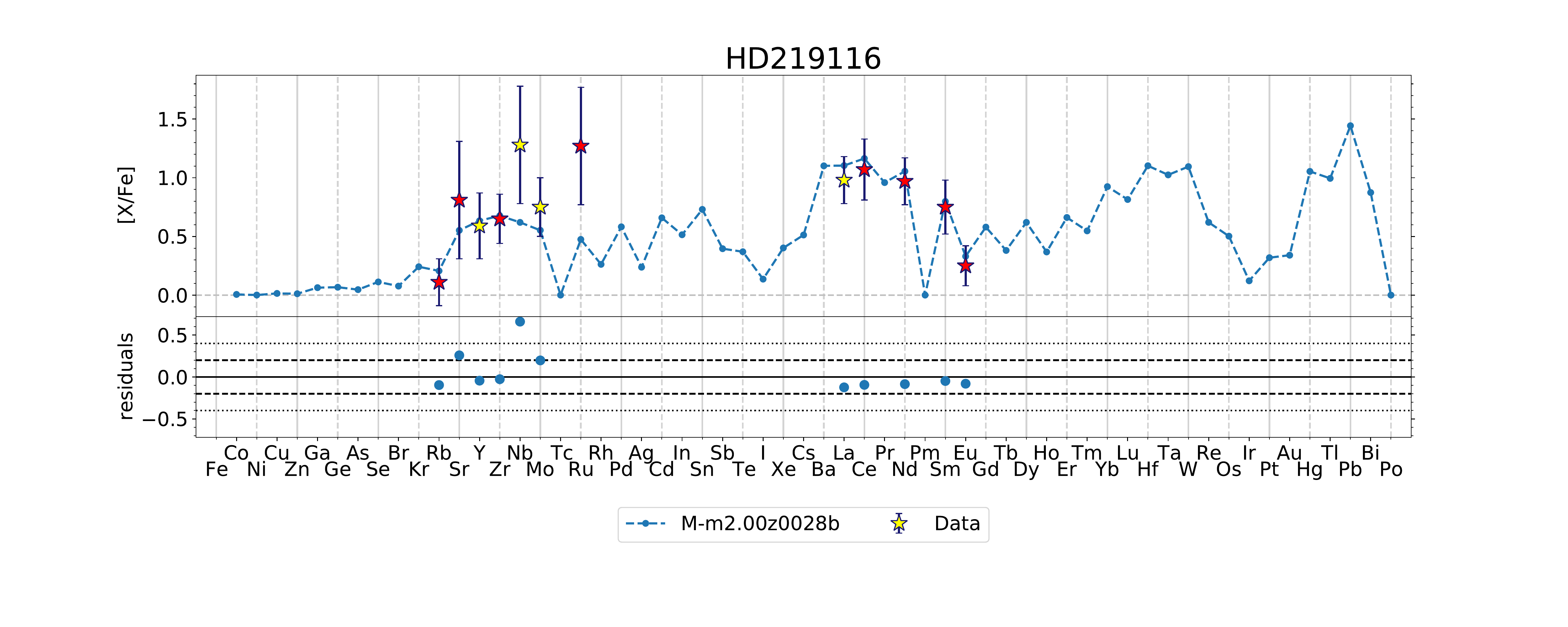}
    \caption{Same figure format as in Fig. \ref{fig:HD18182}.
    For both stars in this figure only one model was able to reach a GoF above 50\%, the reason for this is in both stars the unusually high [Ru/Fe] combined with low [Rb/Fe].}
    \label{fig:fig_odd_first_peak}
\end{figure*}

\subsubsection{Issue with first peak pattern: HD 123396, HD 219116 (Fig. \ref{fig:fig_odd_first_peak})}

We found only one diluted AGB s-process distribution for each of these two stars that matches the observed abundances with a GoF above 50\% (see Fig. \ref{fig:fig_odd_first_peak}). Their GoF are barely above that cut-off limit and both sit below 55\%. In both cases, the observed abundances in the second s-process peak are matched or almost matched by the model. The reason for the low GoFs is the first s-process peak. In HD 123396 the [Rb/Fe] and [Ru/Fe] are not reproduced, while in HD 219116 [Ru/Fe] is problematic. Furthermore, the trend of increasing abundance with increasing atomic number is 
missing in the stellar model predictions. 
In Section \ref{sec:disc}, we will speculate that the low number of models with a GoF > 50\% and the peculiar abundance pattern at the first s-process peak may be a signature of a nucleosynthesis contribution different from the s-process.

\begin{figure*}
    \centering    
    \includegraphics[width=\linewidth]{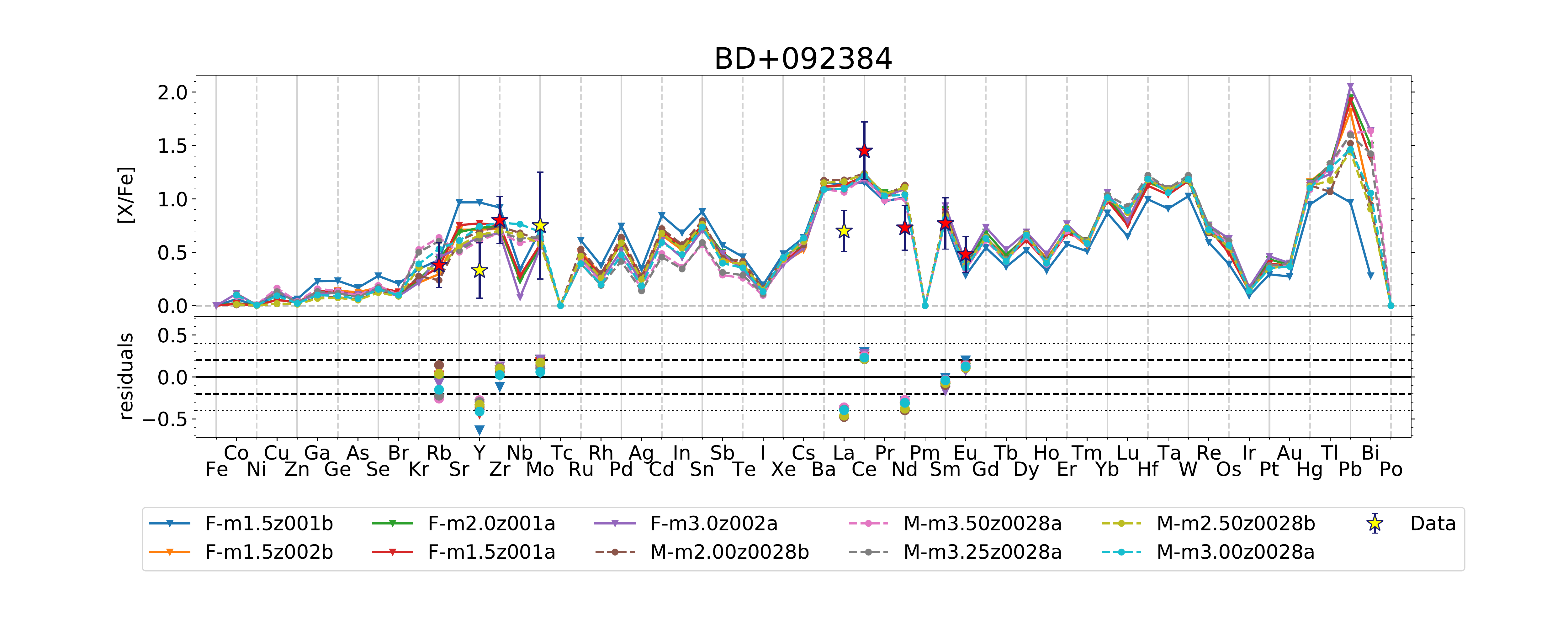}
    \caption{Same figure format as in Fig. \ref{fig:HD18182}.
    For this star no final classifications were found. The reasons for the issues with these two stars are likely the abundances in the second s-process peak, in particular Ce and Nd.}
    \label{fig:fig_2ndpeak}
\end{figure*}

\subsubsection{Issue with second peak: BD +09$^{\circ}$2384 (Fig. \ref{fig:fig_2ndpeak})}

For BD +09$^{\circ}$2384 we cannot find final classifications. There are two reasons for this: [Sr/Fe] and [Ru/Fe] are unknown and thus we miss crucial information to match the first s-process peak properly. Furthermore, the abundances in the second s-process peak show an unusually high ratio of [Ce/Fe] over [Nd/Fe] in combination with small error bars. These abundances would be even more difficult to match if also [La/Fe] was included in the classification. These are all signs that there might be another nucleosynthetic process responsible for this enrichment pattern than the s process, which we will discuss in Section \ref{sec:disc}.

\begin{figure*}
    \centering
    \includegraphics[width=\linewidth]{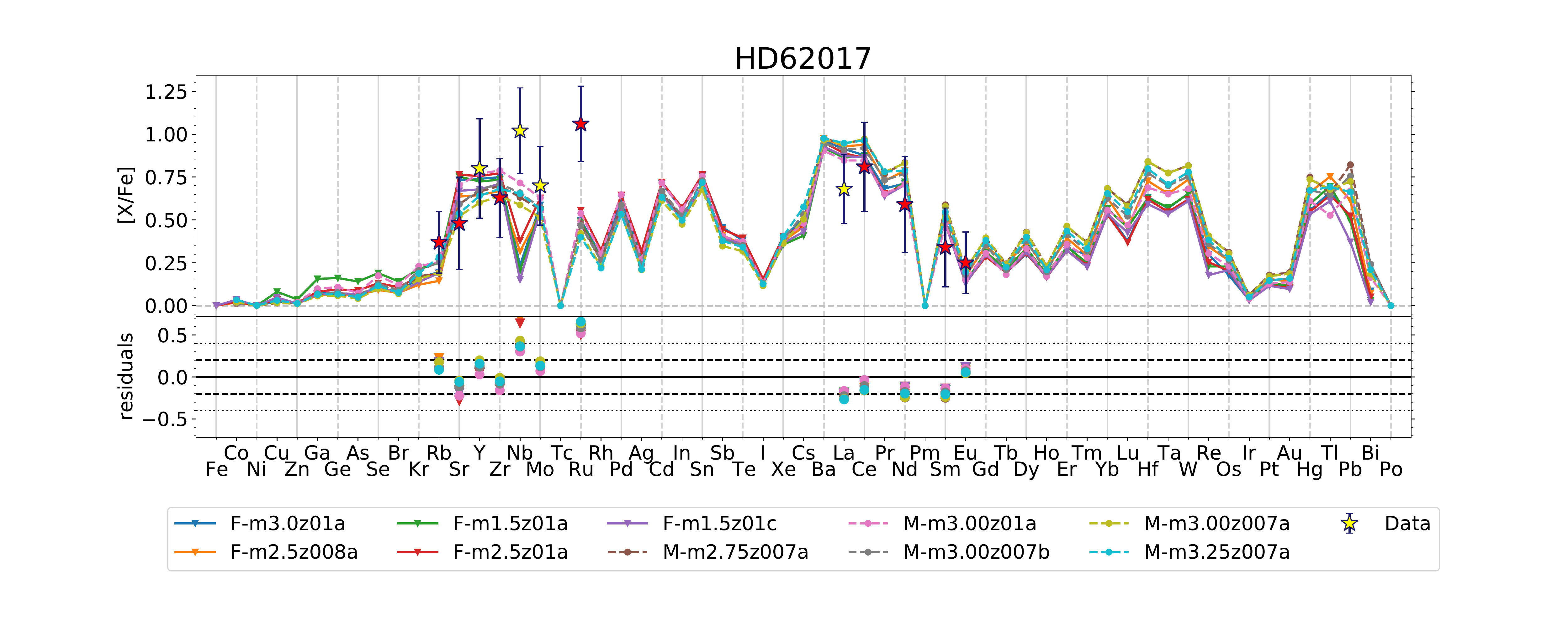}
    \caption{Same figure format as in Fig. \ref{fig:HD18182}. There are no classifications with a GoF above 50\% for this star due to the strong increase in the elemental abundances of the first s-process peak and the strong decrease in the elemental abundances of the second s-process peak.}
    \label{fig:only3elements}
\end{figure*}

\subsubsection{No classifications with a GoF above 50\%: HD 62017 (Fig. \ref{fig:only3elements})}

For HD 62017 all GoFs were found to be between 23\% and 34 \%, thus below our threshold of 50\%. However, the ANNs do agree with the classifications of the nearest-neighbour classificator. The final classification based on the low GoFs and the ANNs using \textsc{Fruity} models is M = 3.5 M$_{\odot}$ with [Fe/H] $\simeq$ $-$0.14 to 0. Using the Monash models we find M = 2.5 M$_{\odot}$ with [Fe/H] $\simeq$ $-$0.14 to 0. The two sets of AGB final surface abundances thus do not agree on the mass, but they do agree on the metallicity range. The observed metallicity are [Fe/H] = $-$0.32 $\pm$ 0.14, thus lower than the classifications.

The reason for the low GoFs is clear from Fig. \ref{fig:only3elements}: models struggle to match the high value of [Rb/Fe] at the same time as the low values for [Sr/Fe] and [Zr/Fe]. Furthermore, none of the models reach the high value for [Ru/Fe] or its error bar. Thus, we again reach the conclusion that there might be another nucleosynthetic process responsible for this enrichment pattern than the s process.

%----------------------------------------------------------------

\section{Discussion}
\label{sec:disc}

In this section we return to the elements that were excluded for the classifications and focus on understanding why the nearest-neighbour algorithm performs better by finding on average higher GoFs for the classifications (see Figure~\ref{fig:setAF}) when excluding these elements. We discuss whether there may be issues with the observations or specific astrophysics reasons causing the difficulties with the classifications.

\subsection{Different GoFs when using SET A}
\label{sec:setAsetF_comp}

We start by comparing the average GoFs of the classification of stars using Set A to the classifications using Set F. We label all the stars for which the difference between the two GoFs is larger than 10 percentual points, independently of whether the GoFs belong to the same AGB final surface abundance in both classifications or not. We found that in 43 Ba stars the difference between the average GoFs was bigger than this threshold. We list these 43 stars in Table \ref{tab:43stars}. 

\begin{figure*}
    \centering
    \includegraphics[width=0.3\linewidth]{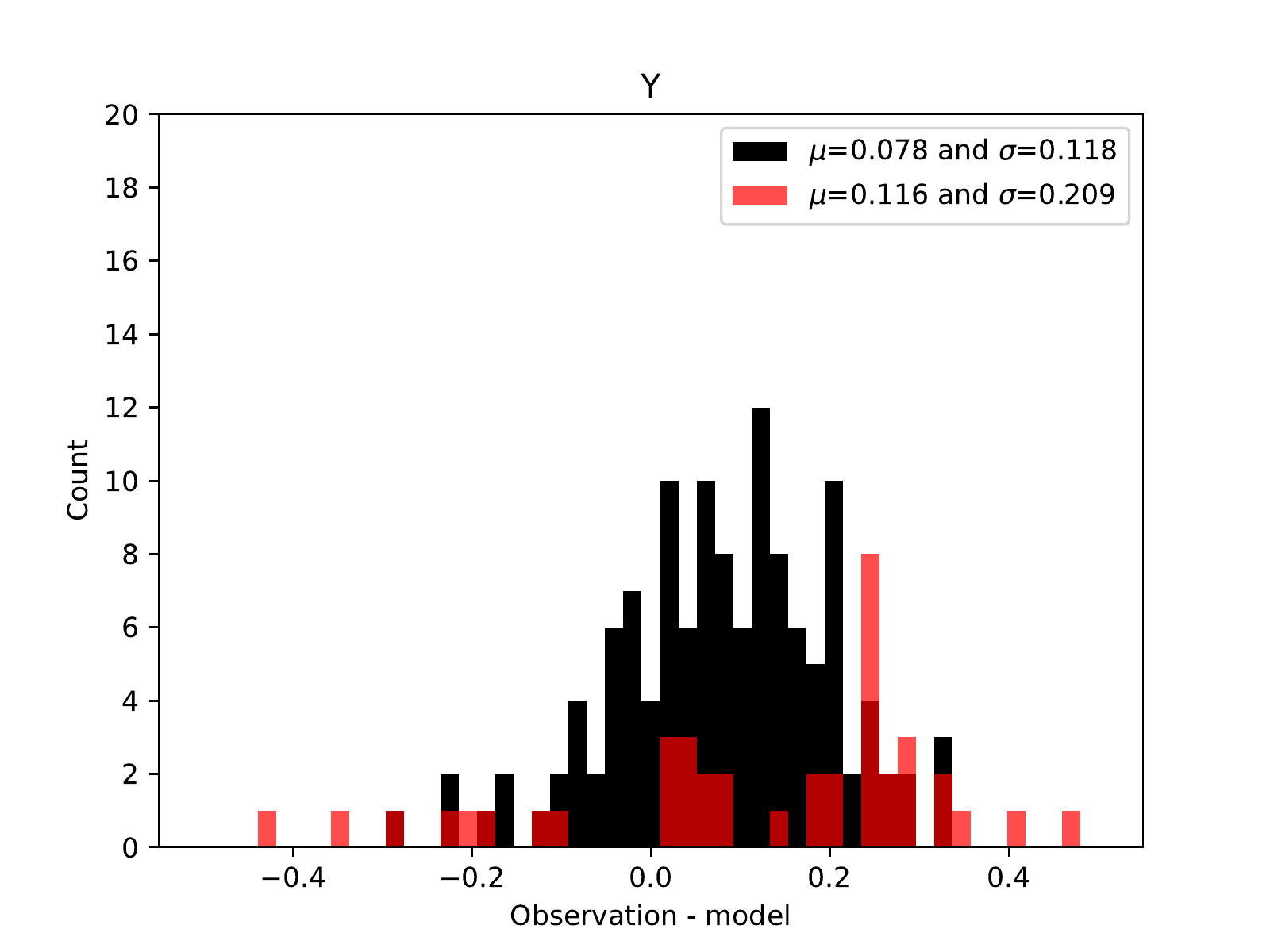}
    \includegraphics[width=0.3\linewidth]{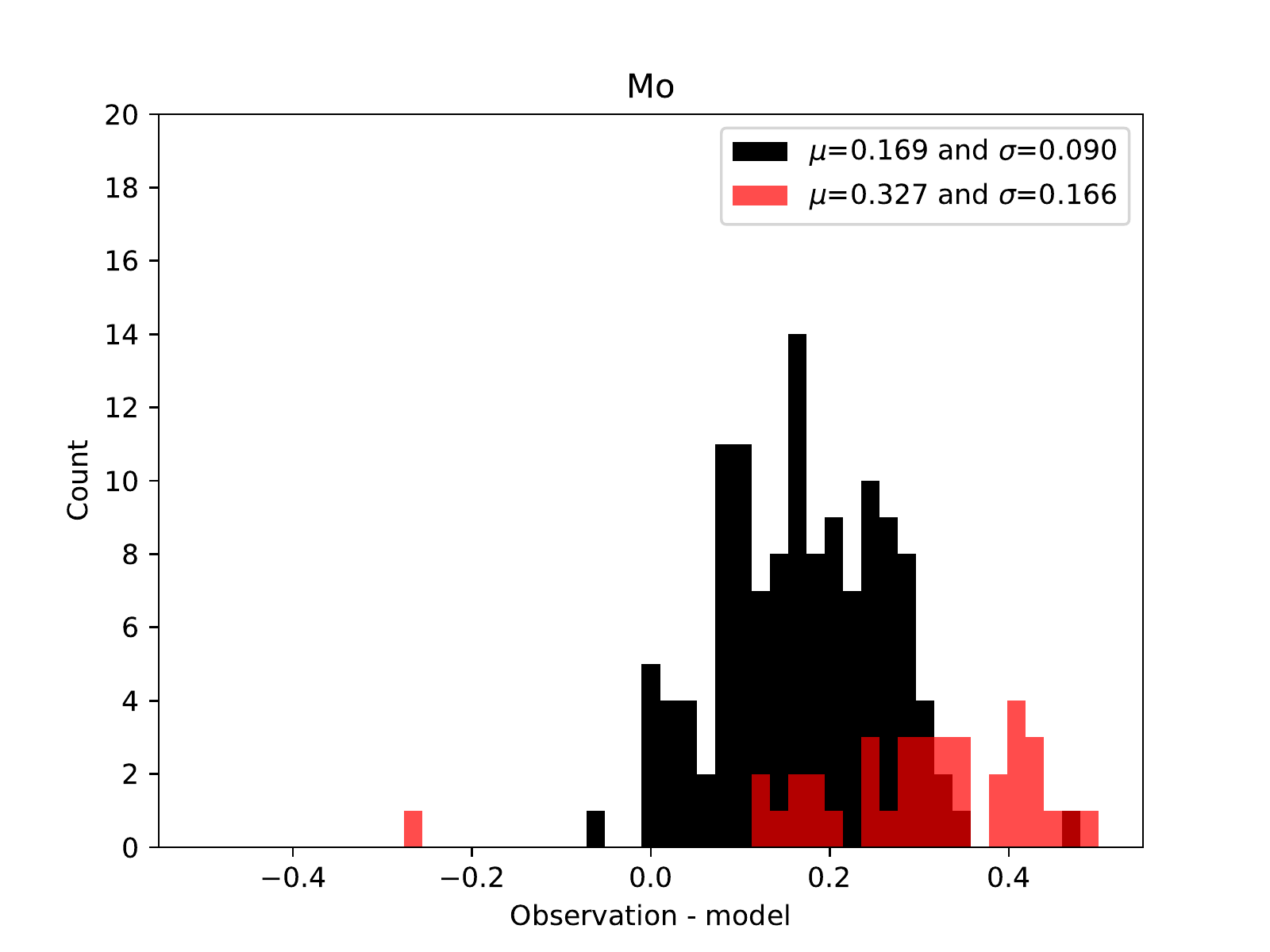}
    \includegraphics[width=0.3\linewidth]{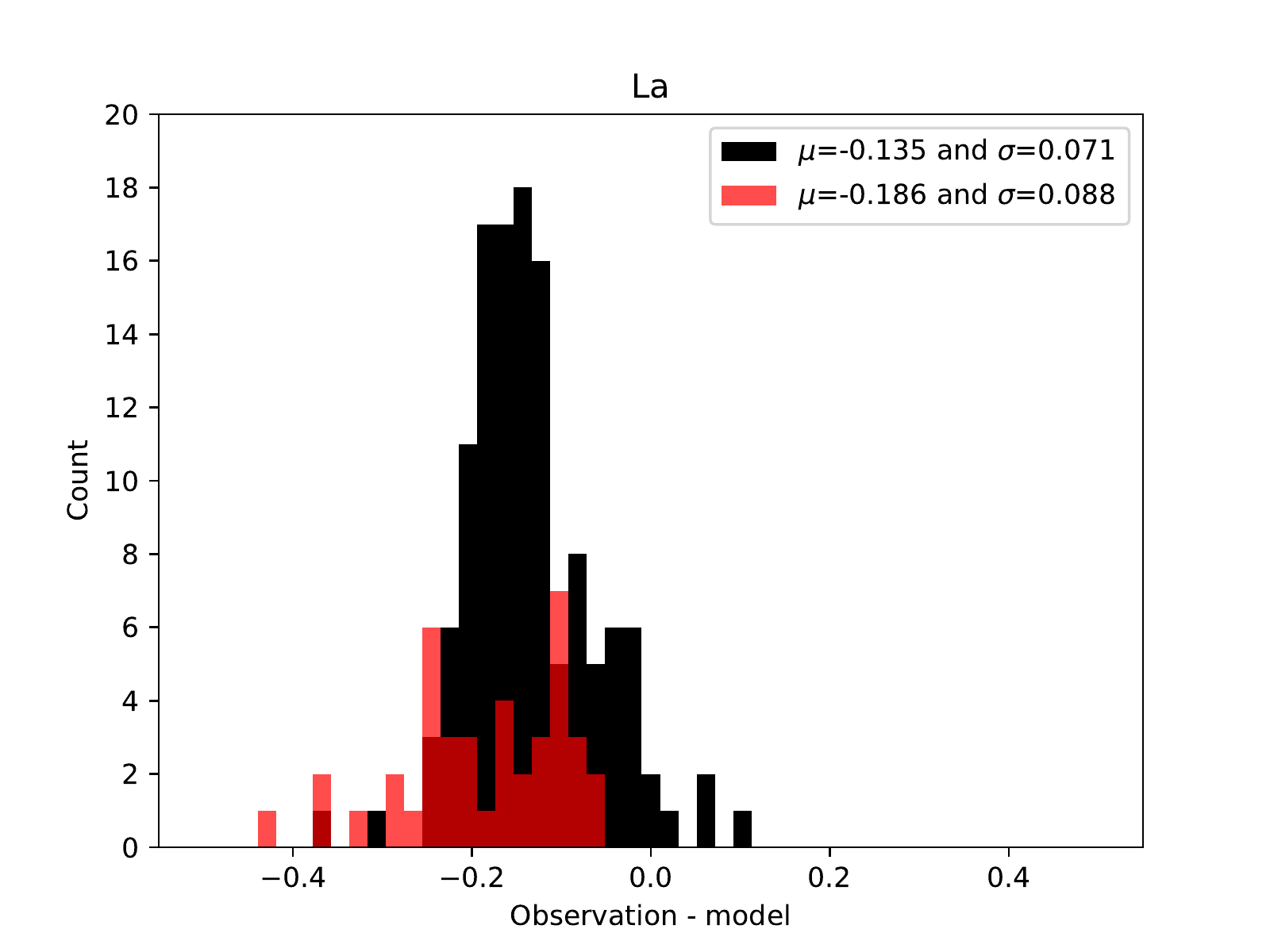}

    \includegraphics[width=0.3\linewidth]{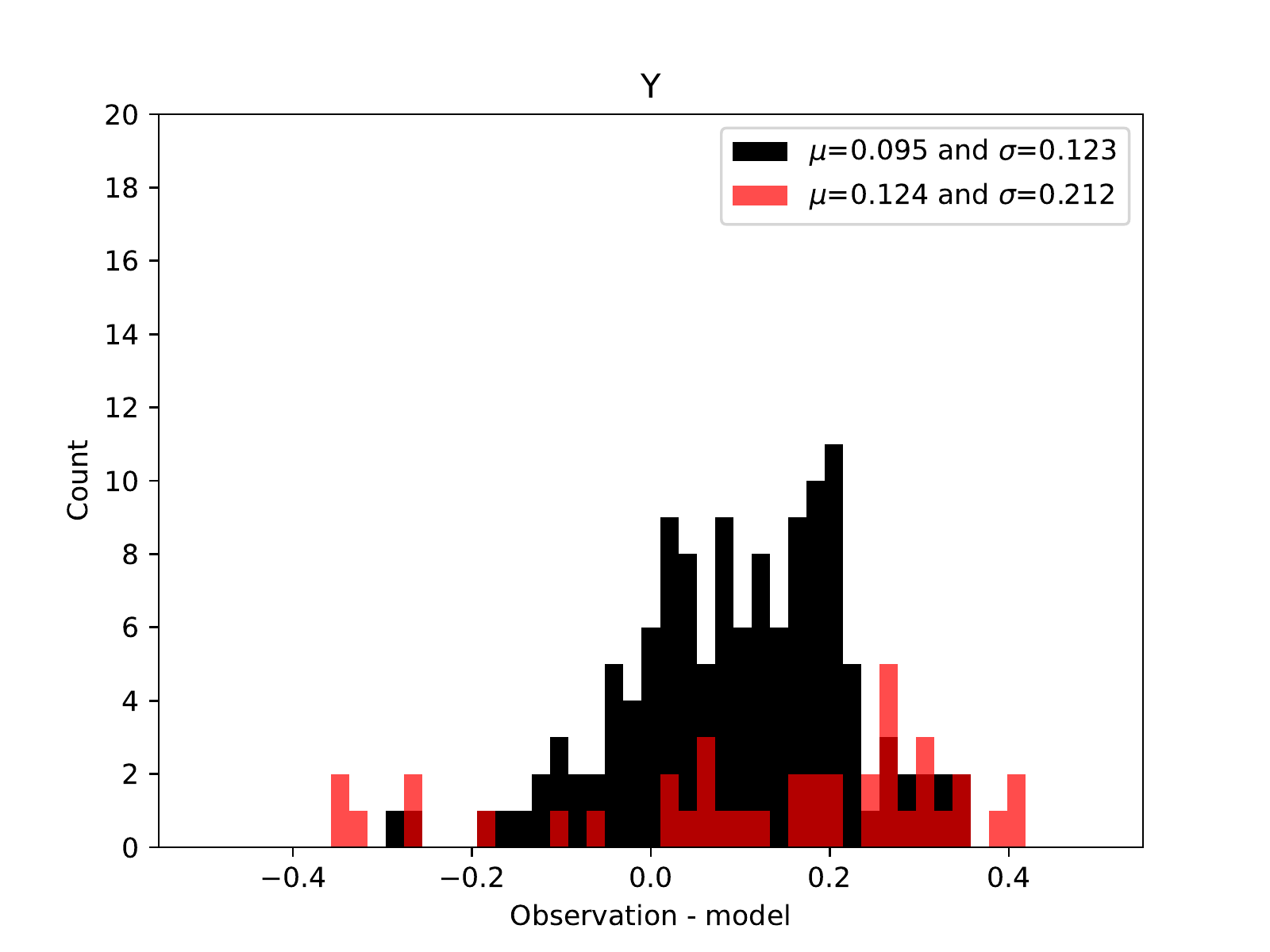}
    \includegraphics[width=0.3\linewidth]{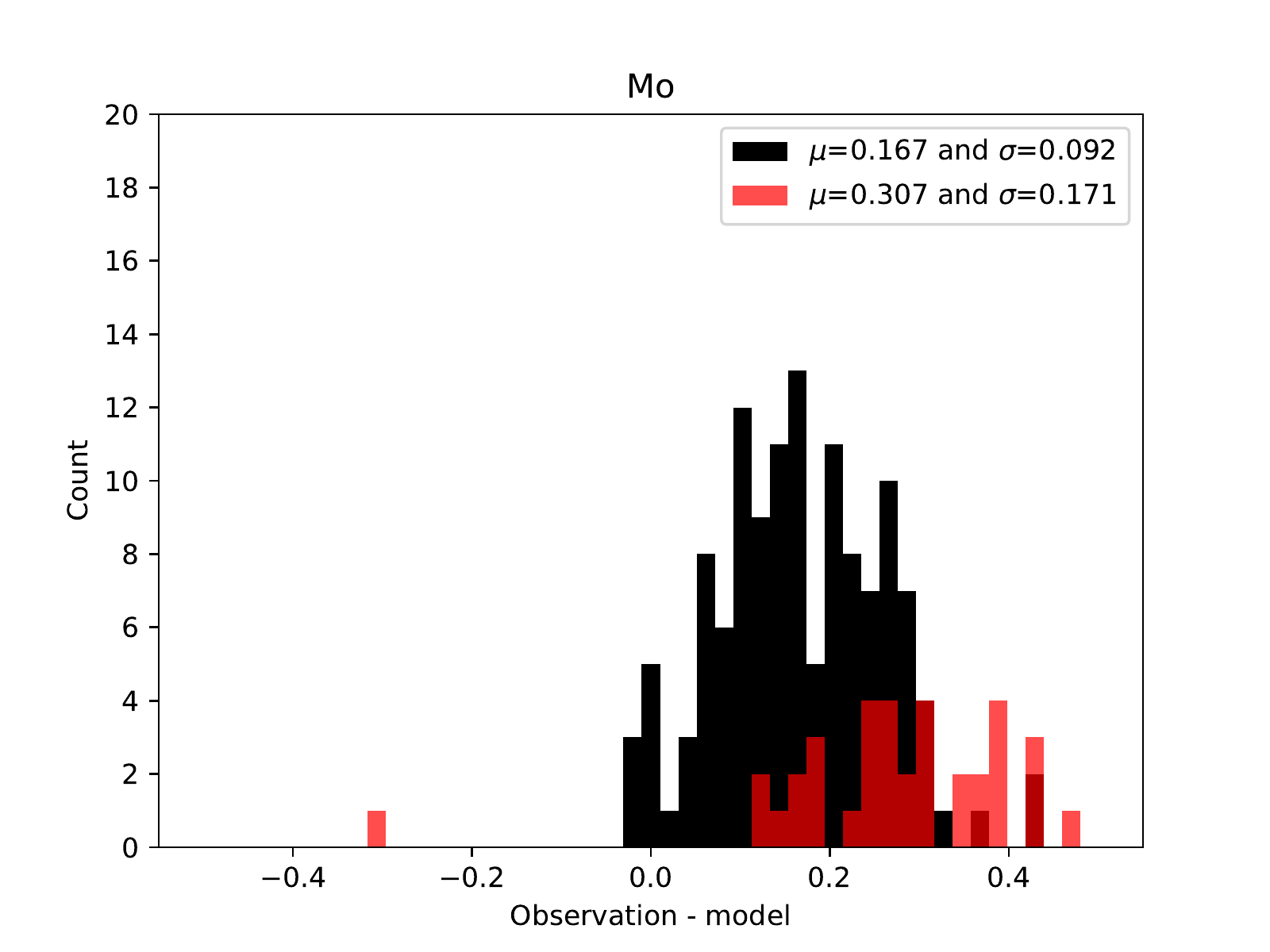}
    \includegraphics[width=0.3\linewidth]{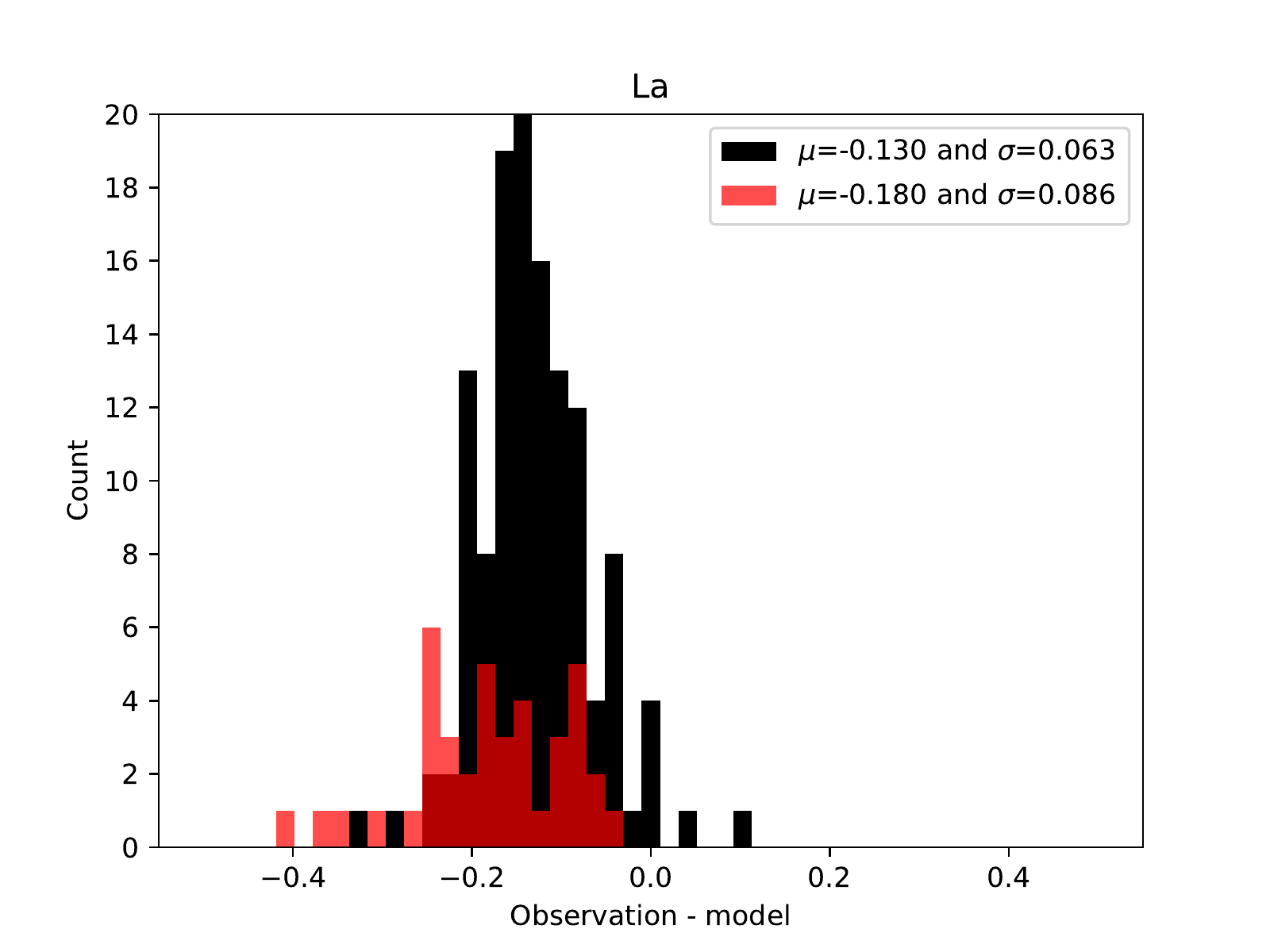}
    \caption{Histograms of the residuals between the observational data and the Set F nearest-neighbour classifications using the \textsc{Fruity} (top panel) and Monash (bottom panel) stellar sets for three elements excluded from the classification: Y, Mo and La. The residuals for the 43 stars listed in Table \ref{tab:43stars} are shown in red, and the other stars in black. The means and standard deviations of the two groups are shown in the legend.}
    \label{fig:hist_resi}
\end{figure*}

\begin{figure*}
    \centering
    \includegraphics[width=0.3\linewidth]{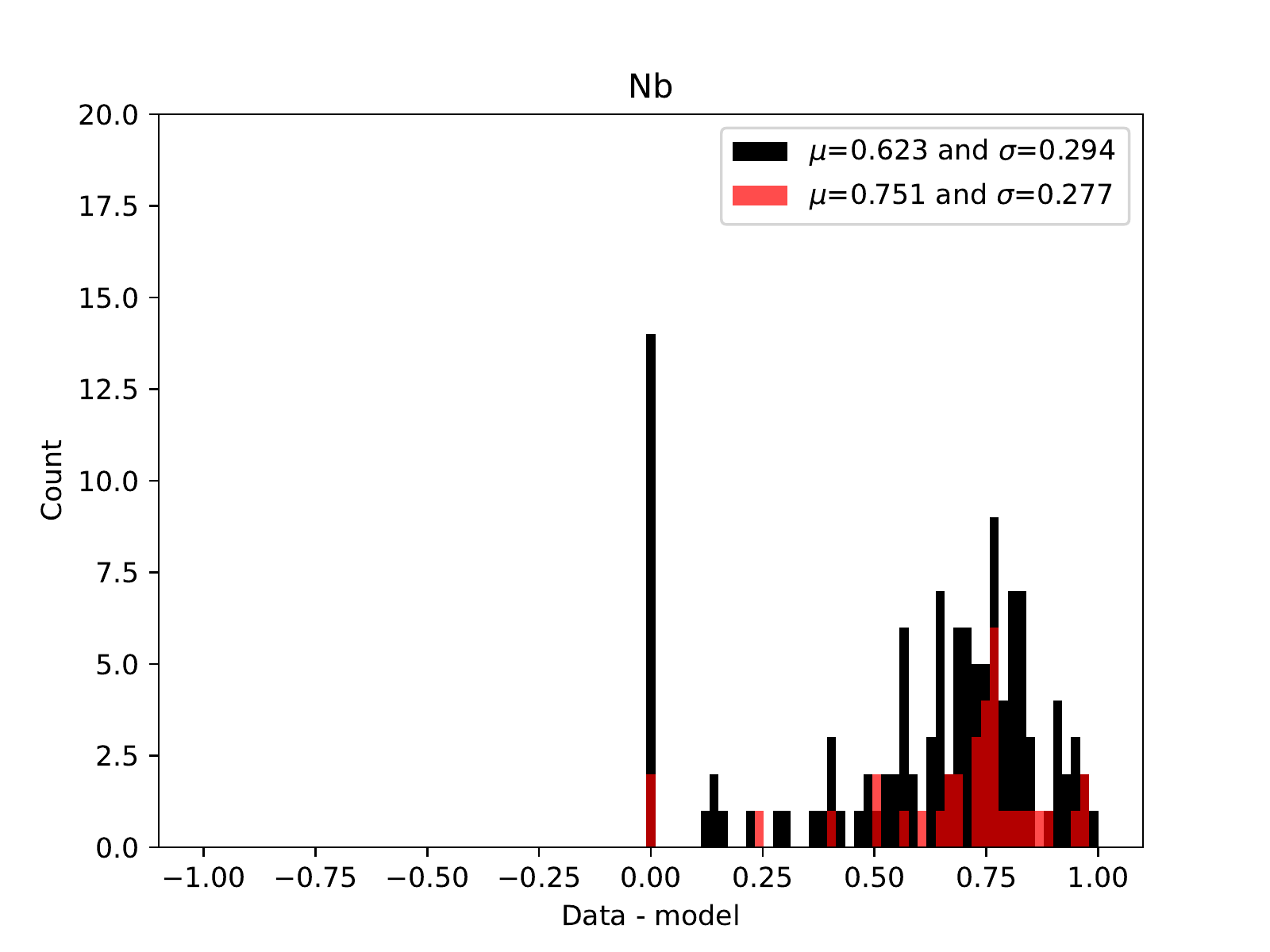}
    \includegraphics[width=0.3\linewidth]{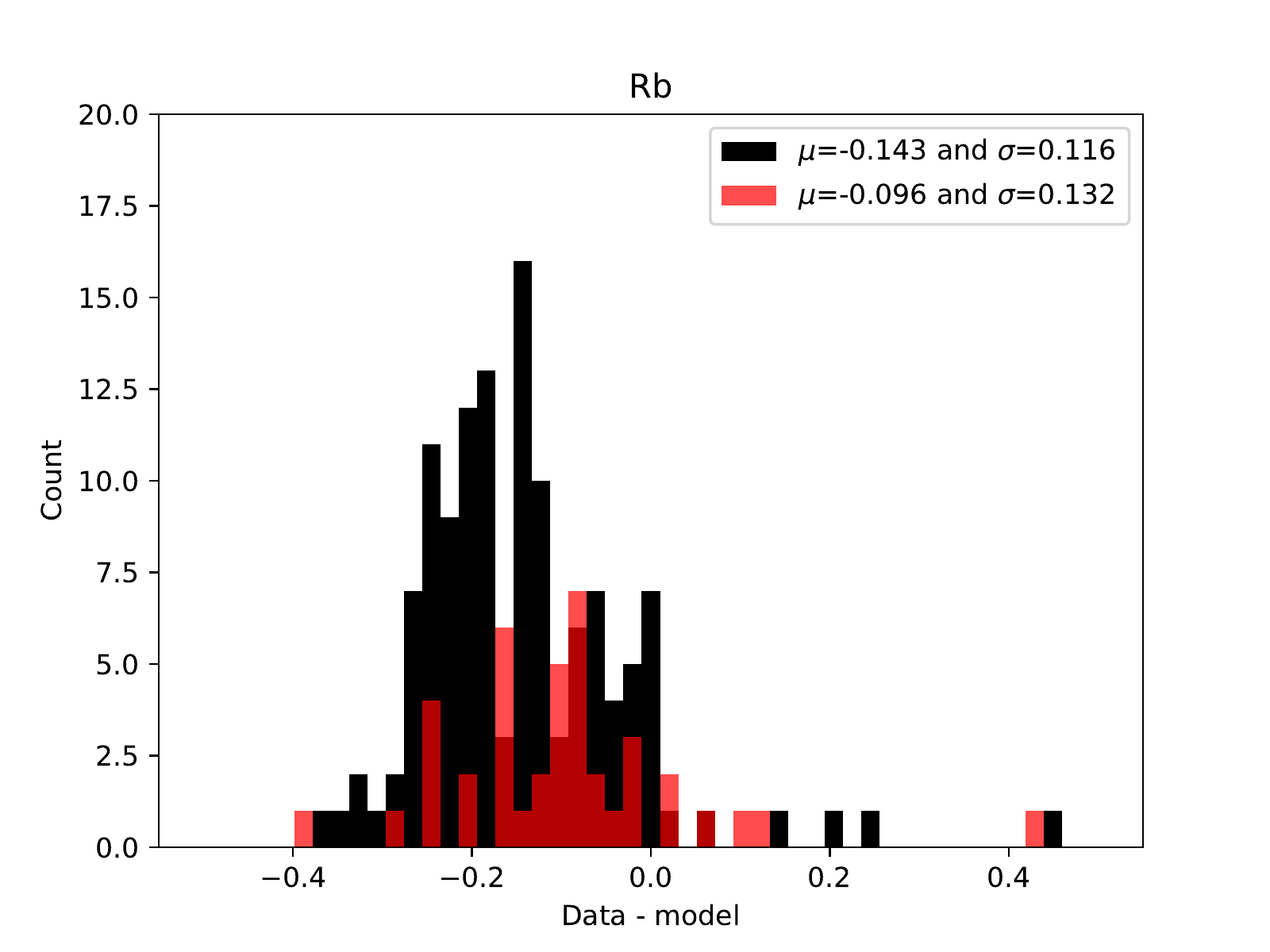}
    \includegraphics[width=0.3\linewidth]{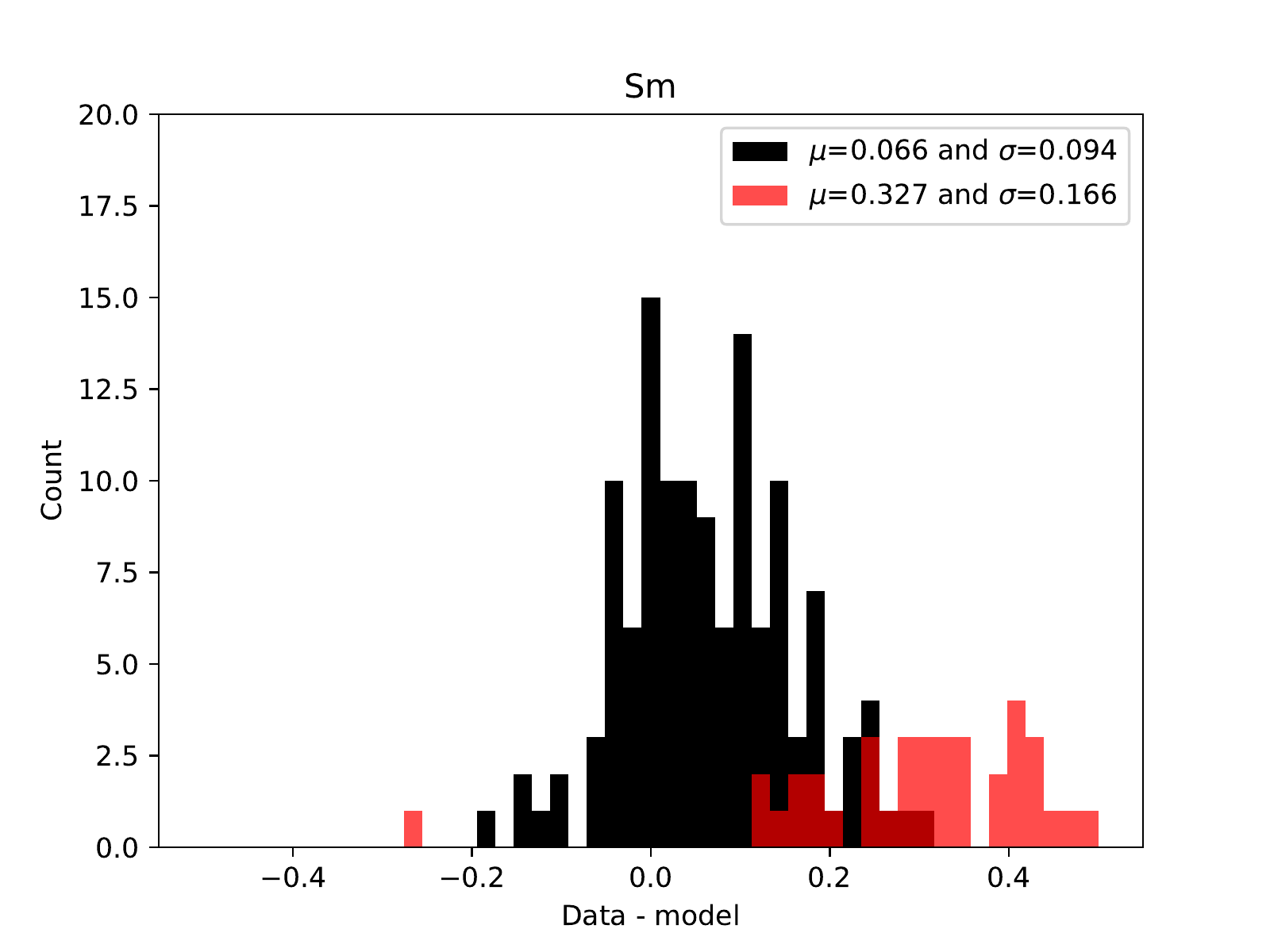}

    \includegraphics[width=0.3\linewidth]{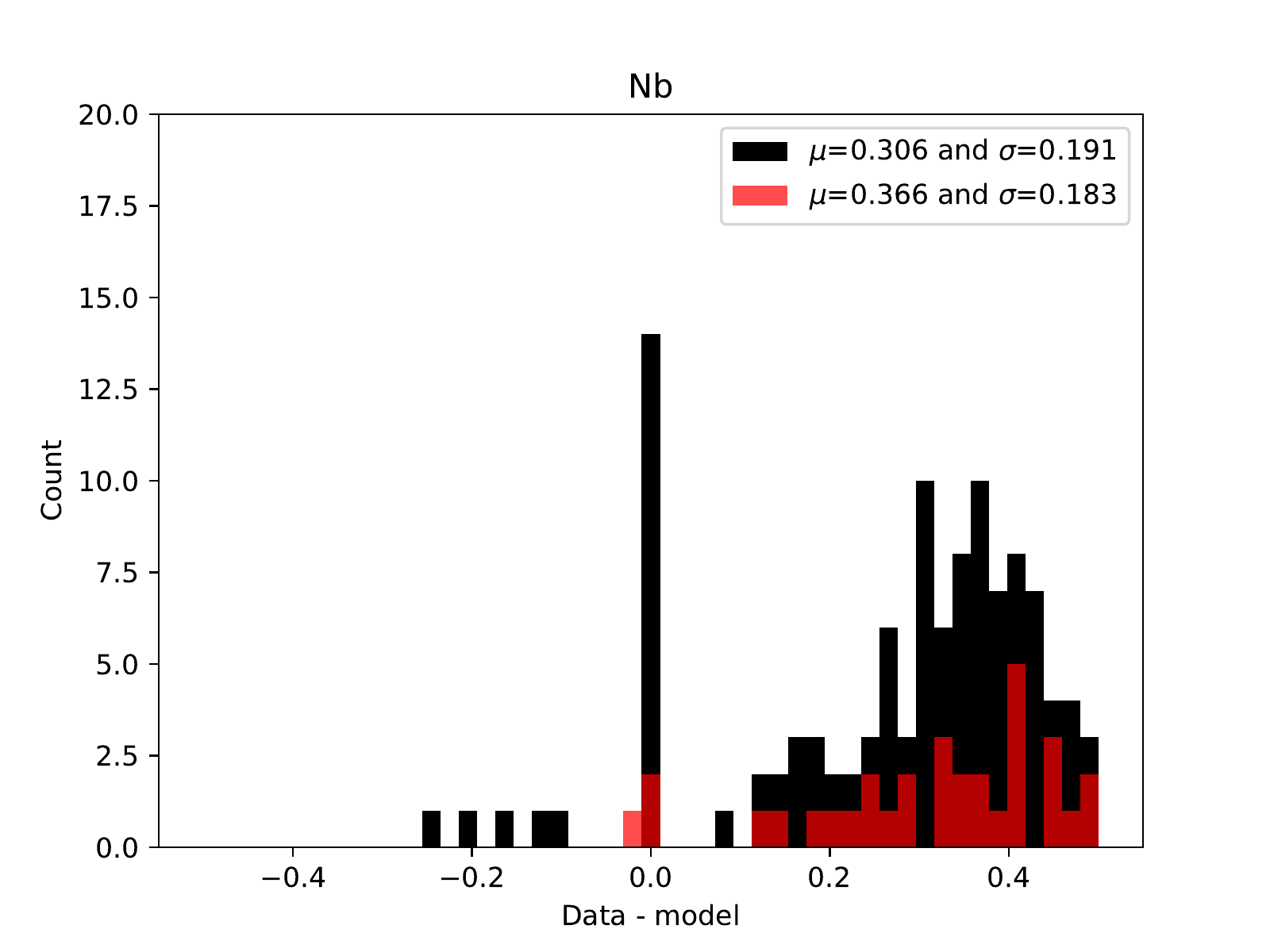}
    \includegraphics[width=0.3\linewidth]{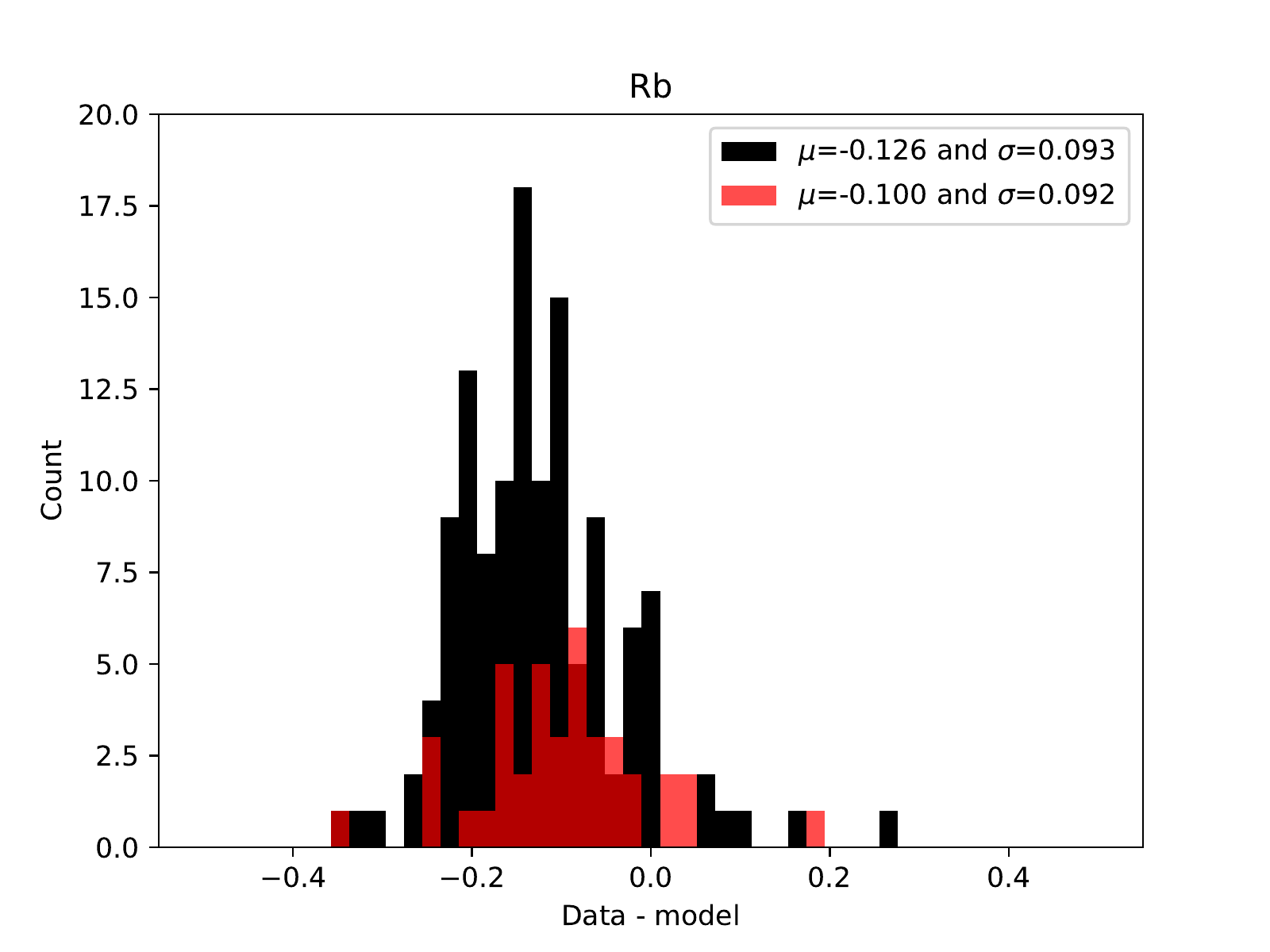}
    \includegraphics[width=0.3\linewidth]{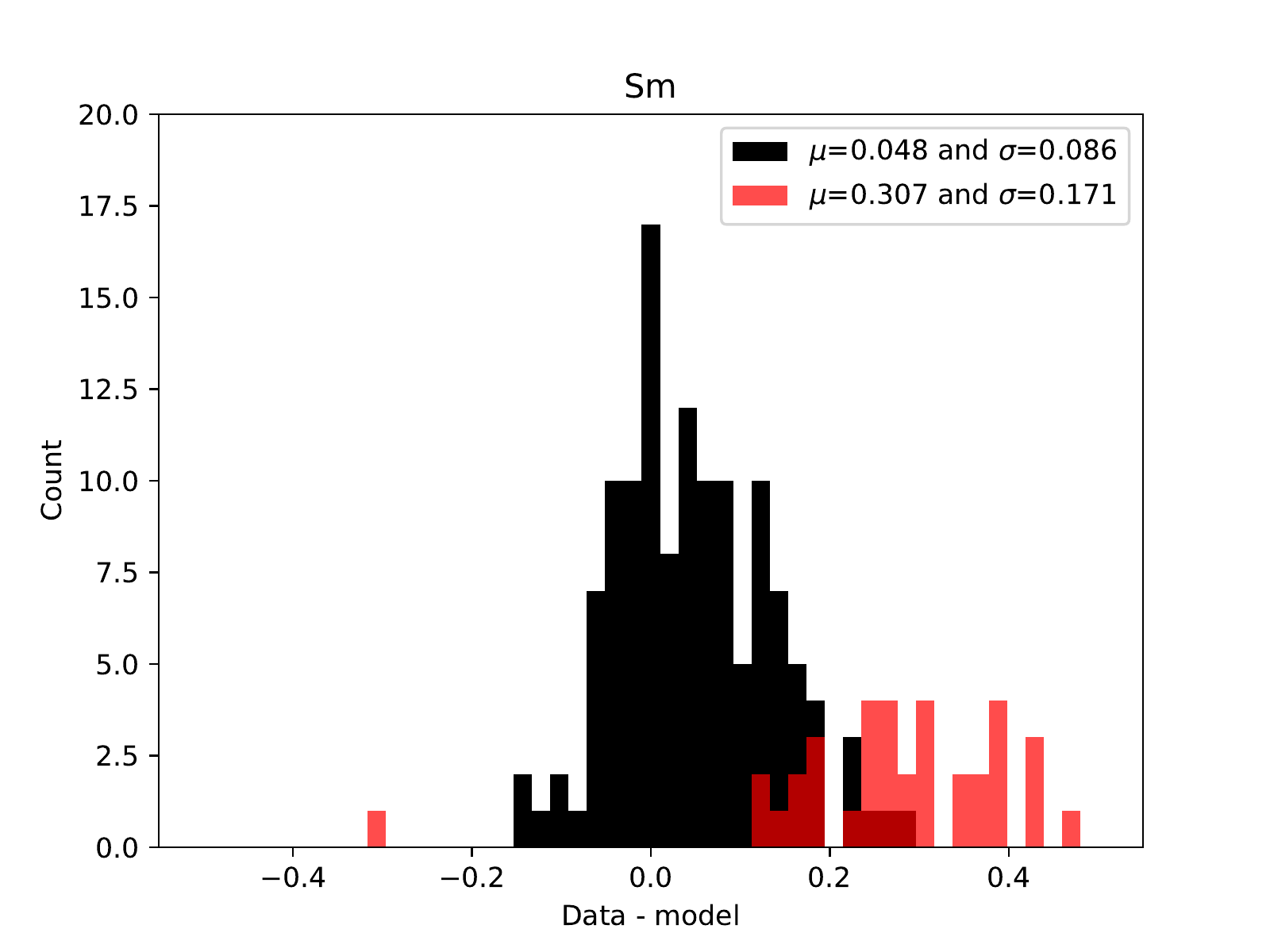}
    \caption{Like in Figure \ref{fig:hist_resi}, the histograms of the residuals (averaged per star) of the \textsc{Fruity} (top panel) and Monash (bottom panel) classifications are reported for Nb and the two extra elements Rb and Sm. }
    \label{fig:hist_res_RbNbSm}
\end{figure*}

To understand better where the difference comes from, we show in Fig.~\ref{fig:hist_resi} and \ref{fig:hist_res_RbNbSm} the residuals between the observational data and the Set F nearest-neighbour classifications for the elements that are excluded in Set F (Y, Mo, La, and Nb). The residuals for the \textsc{fruity} and Monash classifications are shown in the top and bottom panel, respectively. We did not include the error bars on the observed values in these calculations. We show the 43 stars listed in Table \ref{tab:43stars} in red and the other stars in black. In the legend we list the means, $\mu$ and standard deviations, $\sigma$, of both groups of stars. With these parameters of both groups we can perform an one-sides t-test to determine how likely it is that both groups could be samples of the same population. A t-test determines if there is a statistically significant difference between the mean of one group and the mean of another group. Our null hypothesis is thus $\mu_{\rm{43\_stars}}$ = $\mu_{\rm{rest}}$ and we set the confidence interval to 5\%. This means that if the probability p that the means are the same is smaller than 5\%, we reject the null hypothesis and the means are not the same. Our results for the t-tests are shown in Table \ref{tab:t-tests}. We are able to reject our null hypothesis for [Mo/Fe], [La/Fe] and [Nb/Fe] in the \textsc{fruity} and Monash classifications, but not for [Y/Fe]. Their t-statistics confirm our exclusion of these elements as described in Section \ref{sec:elements}: the biggest improvement in the classifications is made when [Mo/Fe] is excluded from the set, and the null hypothesis of $\mu_{\rm{43\_stars}}$ = $\mu_{\rm{rest}}$ is rejected with the highest certainty (highest p-value) for [Mo/Fe]. The mean of the [Mo/Fe] residuals is a larger positive value in the 43 stars than in the other stars. A smaller improvement is made when [La/Fe] is excluded, for which the null hypothesis is rejected but with a smaller certainty than for [Mo/Fe]. The mean of the [La/Fe] residuals is a larger negative value in the 43 stars than in the other stars. For [Y/Fe] the improvement on the classification was the smallest.
In the case of [Nb/Fe] and [Sm/Fe], we can instead reject the null hypothesis using both AGB stellar sets, where the 43 stars have a larger positive mean than the other stars (see Table \ref{tab:t-tests} and Figure \ref{fig:hist_res_RbNbSm}). 

We also checked with the same statistical test the other elemental abundances, and we found that the null hypothesis can be rejected for [Sm/Fe] and [Rb/Fe] as well. For [Sm/Fe] the 43 stars have a larger positive mean, while for [Rb/Fe] the 43 stars have a smaller negative mean residual than the other stars. However, for [Rb/Fe] we were only able to reject the null hypothesis for the \textsc{Fruity} classifications. 

In Paper 1 we found that Nb is often difficult to reach by the models: in 12 of the 28 stars considered in that paper Nb is the only element that is higher than the model predictions, and in several other stars a group of elements including Nb is high compared to the models. Thus, it would be interesting to investigate if the same issues with Nb as found in Paper 1 occur in our classifications of the 169 Ba stars, but we cannot do this with our classifications as shown in Table \ref{tab:rest_fruity} and \ref{tab:rest_monash} as Nb is not included in our algorithms. We thus performed another set of classifications including Nb. 
We focus here on the Ba stars for which the GoF differs by more than 10 percentual points when comparing with the classifications found with Set A with and without Nb, following a similar procedure as at the start of this section. We only did this test with the AGB final surface abundances of Monash. We find that for 43 of the 169 stars the difference is bigger than 10 percentual points. These 43 stars overlap with the ones in Paper 1 that were flagged for the high [Nb/Fe] value. Of these 43, however, only 4 are also included among the 43 stars included in Table \ref{tab:43stars}. We are thus dealing with a different subset of Ba stars. In conclusion, as also mentioned in Paper 1, there are still many open questions concerning the observed [Nb/Fe] values and more work is needed before drawing any conclusions.

In summary, the 43 stars have elemental abundances that are different from the other Ba stars and show statistically significant higher [Mo/Fe], lower [La/Fe], higher [Nb/Fe], and higher [Sm/Fe]. The [Rb/Fe] was found to be higher as well, but this is only found to be statistically significant in the \textsc{Fruity} classifications. 

\begin{table}
    \centering
    \caption{The names of the 43 stars for which the GoF of set A and set F differ by more than 10 percentage points.}
    \begin{tabular}{llll}
BD +09$^{\circ}$2384 	&HD 107270 &    HD 154430 &	HD 29370 \\
BD $-$09$^{\circ}$4337 	&HD 107541 &	HD 168214 &	HD 30554 \\
BD $-$14$^{\circ}$2678 	&HD 109061 &	HD 18182  &	HD 38488 \\
BD $-$18$^{\circ}$821	&HD 113291 &	HD 193530 &	HD 43389 \\
CD $-$27$^{\circ}$2233 	&HD 122687 &	HD 211173 &	HD 45483 \\
CD $-$30$^{\circ}$8774 	&HD 123949 &	HD 216809 &	HD 5825  \\
CD $-$34$^{\circ}$6139 	&HD 130255 &	HD 217143 & HD 62017 \\
CD $-$42$^{\circ}$2048	&HD 139266 &	HD 217447 &	HD 66291 \\
CD $-$53$^{\circ}$8144 	&HD 139409 &    HD 223617 &	HD 67036  \\
CPD $-$64$^{\circ}$4333 &HD 143899 &	HD 24035  & MFU 112 \\
HD 107264               &HD 148177 &    HD 273845 & \\
    \end{tabular}
    \label{tab:43stars}
\end{table}

\begin{table}
    \centering\caption{Results of one-sided t-tests}
    \begin{tabular}{c|c|c|c}
       \multicolumn{3}{c}{\textsc{Fruity}} \\
       Element  & t-statistic & p-value & $\mu_A$=$\mu_B$? \\
       \hline
       Y & -1.1 & 0.14 & possibly \\
       Mo & -5.7 & 3.3e-07 & no\\
       La & 3.3 & 7.8e-04 & no\\
       \hline
       Rb & -2.0 & 0.025 & no\\
       Nb & -2.5 & 7.1e-03 & no\\
       Sm & -9.4 & 7.0e-13 & no\\
       \hline
       \multicolumn{3}{c}{Monash}\\%  & &\\
       Element  & t-statistic & p-value & $\mu_A$=$\mu_B$? \\
       \hline
       Y  & -0.8 & 0.22 & possibly\\
       Mo & -4.9 & 6.1e-06 & no\\
       La & 3.4 & 6.6e-04 & no\\
       \hline
       Nb & -1.8 & 0.041 & no\\      
       Rb & -1.5 & 0.065 & possibly\\
       Sm & -9.1 & 4.2e-12 & no\\
    \end{tabular}
     \begin{tablenotes}
      \item We performed an one-sided t-test to test our null hypothesis that the means of the elemental abundances are the same for the two groups of Ba stars and
    list the outcome, which is the t-statistic, and certainty level, which is the p-value for the \textsc{fruity} and Monash classifications. If the p-value is below our confidence interval of 0.05 then we can reject the null hypothesis, which is indicated in the most right column of this table.
    \end{tablenotes}   
    \label{tab:t-tests}
\end{table}

\subsection{Possible observational problems}
\label{sec:obs_reasons}

For none of the three elemental abundances discussed in Section \ref{sec:setAsetF_comp} we have been able to identify observational issues, except for Y. Ionised lines were used for the analysis of Y while neutral lines were used for Sr and Zr, which are the other two elements in the first s-process peak. As noted by \citet{AB2006}, the derivation of Sr and Zr abundances from neutral lines results in lower abundances than using ionised lines. 
This difference makes Sr and Zr abundances lower than Y, and can cause a discrepancy in the pattern of the first s-process peak when compared to AGB stellar models. As discussed in Paper 1, the differences between ionised and neutral lines are attributed to departures from the LTE approximation, which was assumed to derive the abundances from our stellar sample \citep[][]{deCastro}. Therefore, the first peak could be easier to classify with a high GoF value if Y is removed.

Overall, the error bars of the second peak elements are smaller than those at the first peak. This is related to the fact that more lines with lower standard deviation of the derived abundances were used in the analysis for the second peak elements. The fact that fewer lines were available for the determination of some elements (e.g., Sr) is also reflected in the lack of error bars since no error bars are available in the original study of \citet{deCastro} and \citet{Roriz21_heavy} for elements with less than three lines were used to derive the abundance.

As a result, it is more challenging to match both the observed [La/Fe] and [Ce/Fe], because of the small error bars reported. However, we note that [La/Fe] and [Ce/Fe] were determined in separate studies: La together with Sr, Nb, Mo, Ru, Sm and Eu \citep{Roriz21_heavy}, while Ce was determined by \citet{deCastro} together with Y, Zr, Nd. This might potentially introduce additional errors in the equivalent width measurements, well beyond the reported errors.

Observational issues for Rb and Nb are also possibly relevant in our analysis presented here. In some of the sample stars [Rb/Fe] is negative, which is impossible to match with AGB models. 
Rb abundances used in this study were derived from only one Rb I line, the $D_{2}$ resonance line at 7800.2 \AA, assuming LTE approximation and taking also into account the hyperfine structure of Rb. Recently \cite{Korotin20} calculated NLTE corrections in cool stars with metallicities in the range of our sample stars for the same Rb line. This author has shown that the errors, when neglecting the NLTE effects, can be as high as 0.3 dex, depending on temperature, metallicity, log$\,g$, and also the Rb abundance itself. However, this effect is weak, around 0.1 dex, in the case of our sample stars with mean temperature of 4800 $\pm$ 260 K and log$\,g$ of 2.3 $\pm$ 0.48 \citep[see Figs. 4 and 5 in][]{Korotin20}. On the other hand, \citet{Korotin20} also found that departures from NLTE strengthens the Rb resonance lines in the solar spectrum and found NLTE absolute Rb abundance\footnote{A(E) = 12 + log$_{10}$(n(E)/n(H))} A(Rb) = 2.35 $\pm$ 0.05. This is in agreement with the meteoritic value of A(Rb) = 2.36 $\pm$ 0.03 from \cite{lodders19}, and much lower than the solar value used to derive the Rb abundances in this study, 2.60 $\pm$ 0.15 \citep{grevesse98}. In contrast, \cite{Takeda21} found a lower NLTE correction for the solar value of $-$0.05 dex. This could have a much larger impact on the observed [Rb/Fe] abundances, but a more detailed analysis is required. In any case, applying the lower value instead of the higher value would result in higher Rb values for our sample stars, and therefore it would imply non-negative [Rb/Fe] abundances for all of our sample stars.

Due to blending with other lines, Nb abundances were calculated from maximum 3 lines using the spectral synthesis technique. We note that almost half of the sample stars have artificial error bars for Nb in this study, since there were less than 3 lines used for the abundance derivation. The consideration of the hyperfine splitting in the calculation of the Nb abundances avoids the overestimation of Nb due to neglecting this factor. The fact that the Nb abundances in the 43 stars are higher than the models could therefore point to a process different from the $s$-process occurring in the polluter AGB stars and producing these high Nb values seen in about 25\% of our Ba stars sample.

\subsection{Signatures of i-process nucleosynthesis in Ba star spectra?} \label{sec:iproc}

From our sample of 169 Ba stars, the ML algorithms adopted in this work have identified 43 stars (equivalent to 25$\%$ of the total) showing anomalous abundance patterns, mainly at the first neutron magic peak above iron (N = 50). 
This could suggest that, along with the s-process, one or more unidentified nucleosynthesis component contributed to the production of heavy elements in this atomic mass region in the evolved AGB stellar companion. 

\begin{figure*}
    \centering
    \includegraphics[width=8cm]{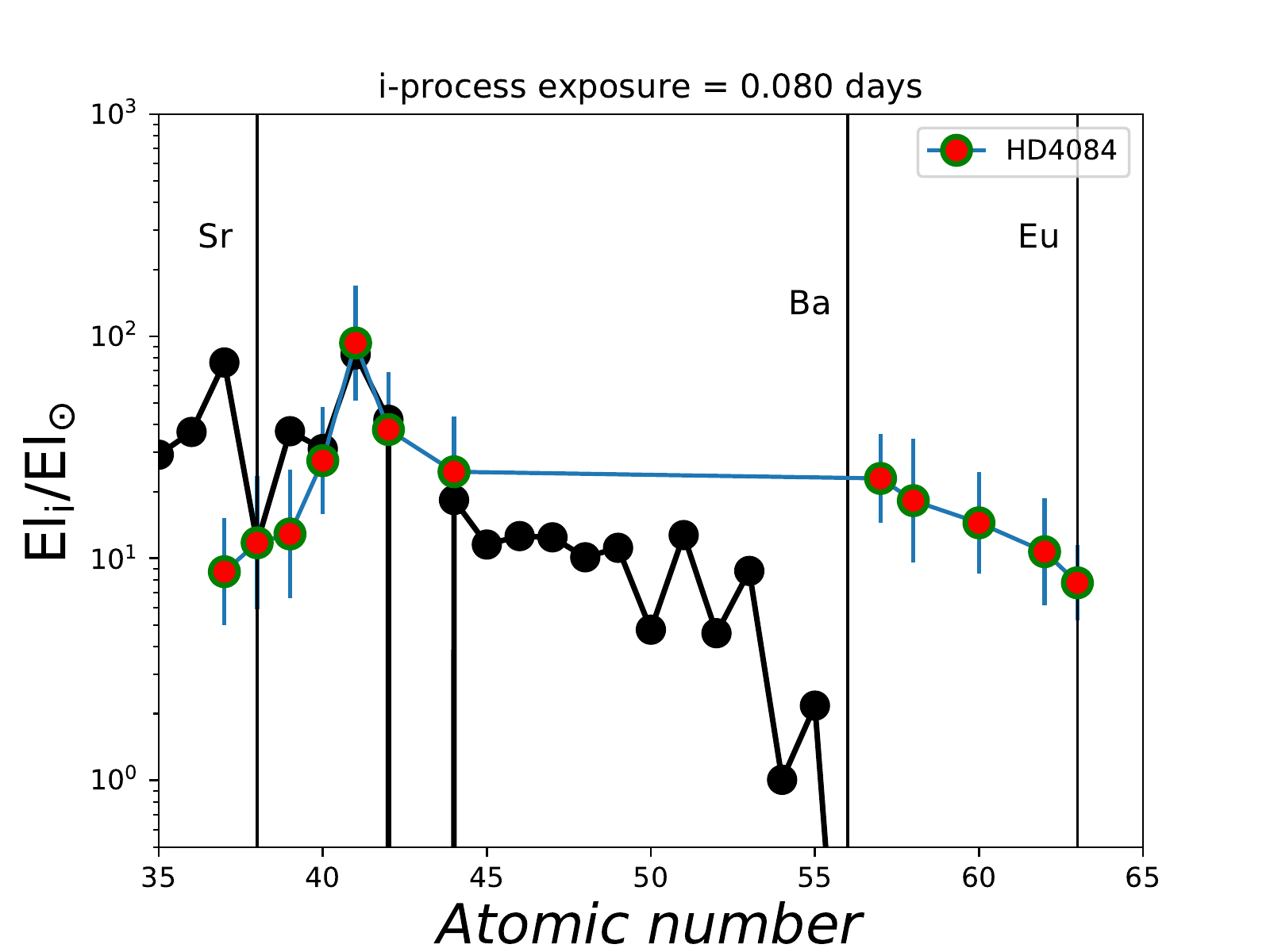}
    \includegraphics[width=8cm]{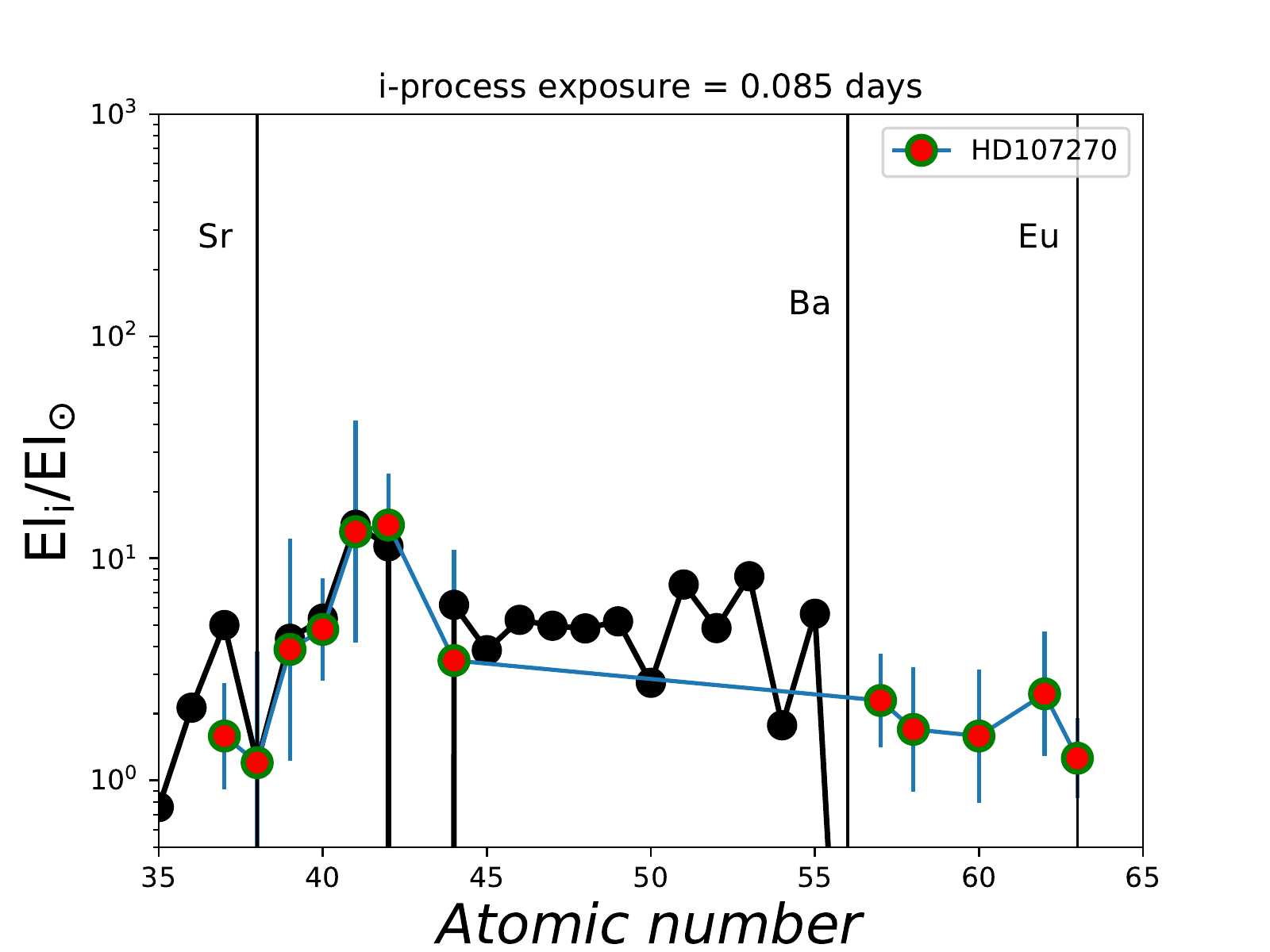}    
    \includegraphics[width=8cm]{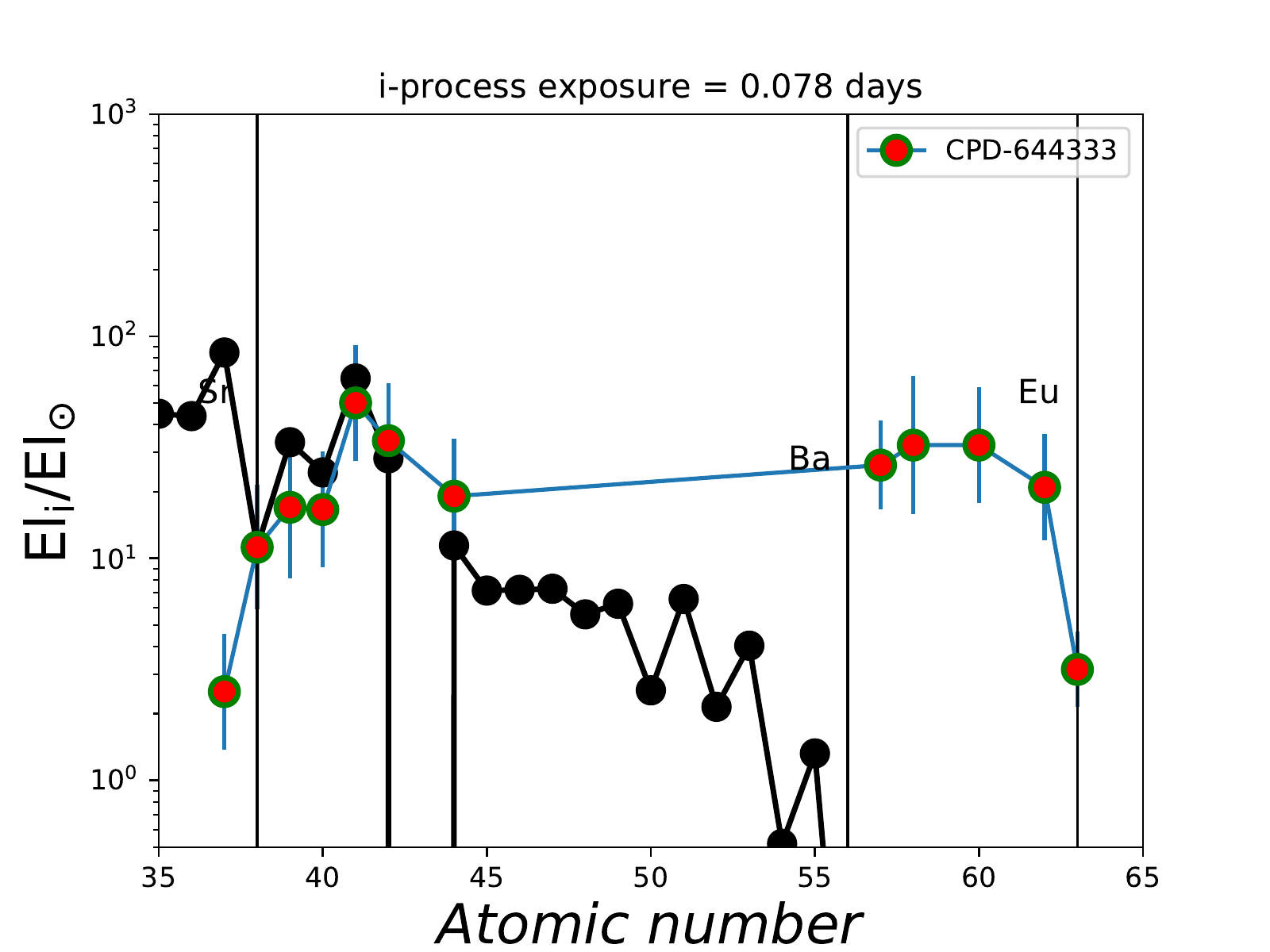}    
    \includegraphics[width=8cm]{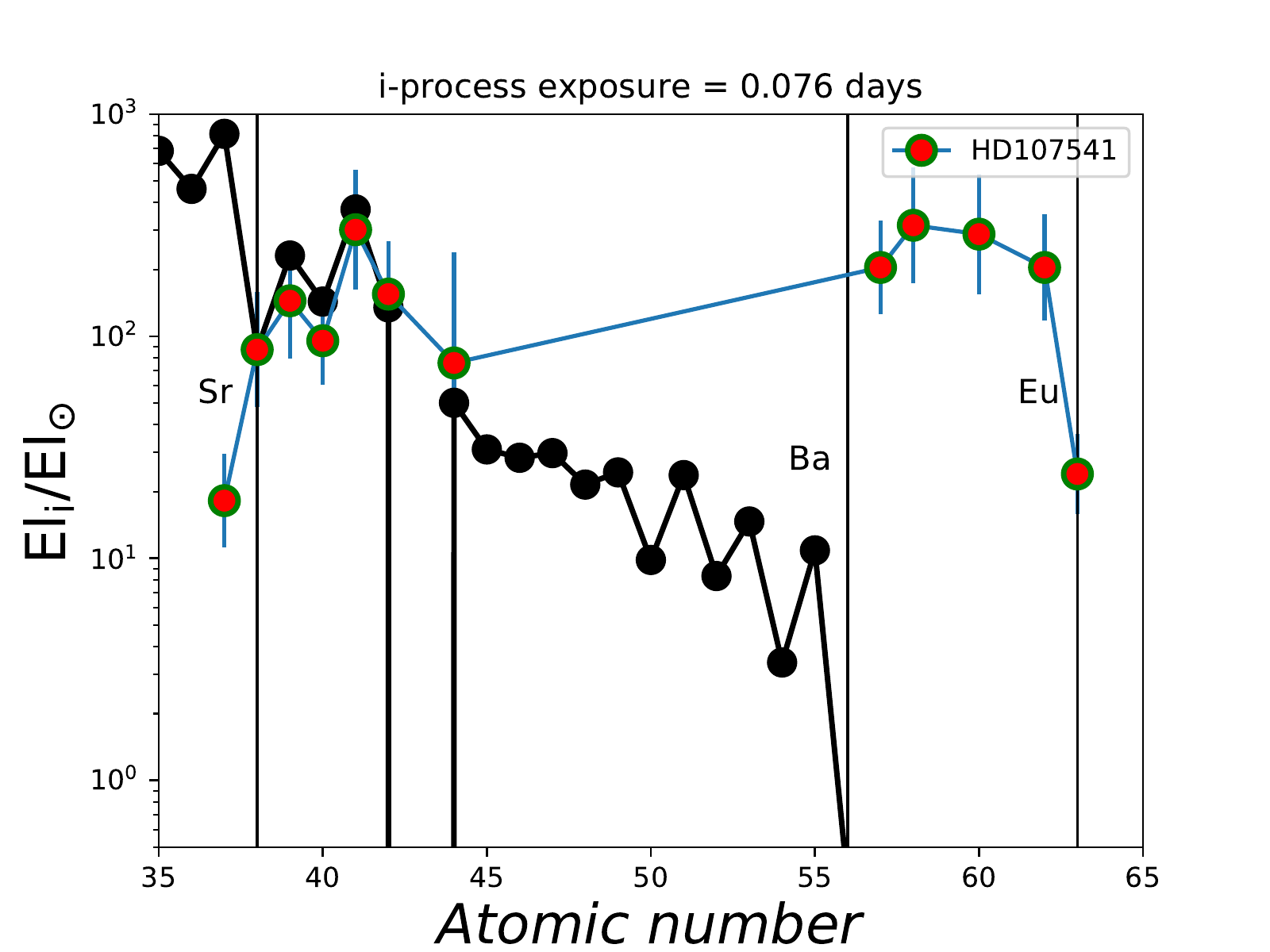}
    \caption{Abundance distributions normalized to solar for four stars are shown (HD 4084 at [Fe/H] = $-$0.42, HD 107270 at [Fe/H] = $+$0.05, CPD $-$64$^{\circ}$4333 at [Fe/H] = $-$0.1, and HD 107541 at [Fe/H] = $-$0.63), carrying a composition in the mass region Rb-Ru (at the light neutron magic peak, N = 50) anomalous with respect to the s-process in AGB models. Observations are compared with the i-process calculations at time of exposure between 0.076 and 0.085 days (see text for details). The theoretical abundances are normalized to the observations at Sr. In the plots vertical lines are added in correspondence of the atomic number of Sr, Ba and Eu.}
    \label{fig:comp_iprocess_sample}
\end{figure*}

Together with the classical s-process, an additional contribution from the intermediate neutron capture process \citep[i-process,][]{cowan:77} has been identified in a fraction of carbon-enhanced metal-poor stars (CEMP stars) carrying heavy element enrichment, the analogue of Ba stars at much lower metallicities \citep[e.g.,][]{dardelet:14,abate:15,cristallo:16,2016ApJHampel,2019Hampel, choplin:21}. Additionally, the i-process has been proposed to be responsible of some of the anomalous abundances observed in metal-poor post AGB stars \citep[e.g.,][]{lugaro:15,2016ApJHampel}. In the same metallicity range of the Ba stars discussed here the post-AGB star Sakurai's object was the first star with a clearly identified i-process nucleosynthesis signature \citep{herwig:11}.
Therefore, a reasonable hypothesis would be that the anomalies in the first s-process peak are indeed signature of an i-process activation in about 25$\%$ of the evolved companions of the Ba stars observed by \citet{deCastro}. In order to test this hypothesis, in Fig. \ref{fig:comp_iprocess_sample} we compare the heavy element abundances of four of the 43 anomalous Ba stars with i-process abundances. These stars are: HD 4048 and HD 107270, for which we only found one final classification, CPD $-$64$^{\circ}$4333, for which we found two final classifications with low GoFs, and HD 107541 which was also problematic to classify in Paper 1. The last three of these four stars are part of the 43 stars, see Table \ref{tab:43stars}. HD 4048 is instead not included in that stellar sample.

For the calculations, we used a simplified i-process trajectory, as adopted by, e.g., \cite{bertolli:13} and \cite{dardelet:14}. By using a solar scaled initial composition beyond Fe, we assumed that there is no previous contribution from other processes. Nucleosynthesis calculations are performed using the NuGrid codes \citep[e.g.,][]{herwig:11}. Starting from the Fe seeds, in these conditions it takes about a couple of hours of i-process exposure at typical neutron densities of 10$^{14-16}$ neutrons cm$^{-3}$ to build the elements at the first peak. Despite the simplicity of the simulation framework adopted here, in Figure \ref{fig:comp_iprocess_sample} the i-process seems to reproduce well the pattern between Sr and Ru observed in the four stars. However, in all four stars considered in Fig. \ref{fig:comp_iprocess_sample} the i-process seems to make too much Rb compared to observations, by at least a factor of three. It is unlikely that such a large discrepancy could be fully explained by the issue with the NLTE correction, as discussed in Section \ref{sec:obs_reasons}.

Nuclear uncertainties could affect the i-process production of Rb compared to other nearby elements, where neutron-capture rates of neutron-rich radioactive nuclei along the i-process nucleosynthesis path are only theoretical \citep[e.g.,][]{denissenkov:18}. We should also consider the possibility that the calculated i-process trajectory is not reproducing correctly Rb because of the simplicity of the approach. 
First, we are ignoring the impact of the complex stellar environments hosting the i-process on the neutron-capture nucleosynthesis itself: following an H ingestion event where multi-dimensional hydrodynamics effects can shape the local stellar structure \citep[e.g.,][]{stancliffe:11,herwig:14,woodward:15}. Second, in our specific case the same stellar hosts (i.e., the evolved companions of Ba stars) were carrying also the s-process component and the i-process may therefore be activated in conjunction with the s process. We are not considering either of these aspects in our calculations. 

Notice also that with the present approach we cannot capture minor effects on elements heavier than Ru. The i-process may be able to produce these effects on a previously enriched s-process abundance distribution. This may account for the offset in our 43 stars of La and Sm residuals as shown in Fig. \ref{fig:hist_res_RbNbSm}. Such effects could be captured by combined i-process and s-process stellar simulations, and provide additional valuable constraints on the neutron exposure of the i-process events.

For the theoretical point of view, stellar models are predicted to possibly have a significant H ingestion event early on in the AGB phase at low metallicities \citep[e.g.,][]{iwamoto:04, cristallo:09, choplin:21}, creating the conditions for the i-process to happen. However, these events are not expected to take place during the AGB phase within the metallicity range typical of Ba stars.  

However, some post AGB stars develop a late thermal pulse or a very late thermal pulse event before becoming white dwarfs \citep[see, e.g.,][]{1995A&ABloecker}, during which some degree of H ingestion and possibly i-process are occurring as demonstrated by Sakurai's object \citep[see, e.g.][]{herwig:11}. Because of this, about 20$\%$ of AGB stars are becoming H-deficient white dwarfs \citep[e.g.,][]{werner:06}. 
Therefore, a possibility could be that the anomalous Ba stars have been polluted again in the final evolutionary stages of the evolved post-AGB companion with i-process-rich material. The feasibility of this scenario (or of other possible pathways within these binary systems) will need to be investigated in the future, with the guidance of multi-dimensional hydrodynamics simulations \citep[e.g.,][]{denissenkov:19}. From the observational point of view, it would be interesting to check if anomalous Ba stars are indeed preferentially associated to H-deficient white dwarf companions. This would be consistent with the similar statistics between anomalous Ba stars and post-AGB stars observations.

%-----------------------------------------------------------------

\section{Conclusions}
\label{sec:conclusions}

%context
We have developed two algorithms to classify the 169 Ba stars of \citet{deCastro} based on their abundance patterns. These algorithms include machine learning techniques to minimise the human bias on the classification and to speed up the analysis. For our study we use AGB final surface abundances of \textsc{Fruity} or Monash stellar sets, and only include the s-process elements heavier than Fe in the classification. The dilution of the AGB surface abundances required to reproduce the s-process enrichment in Ba stars is a free parameter in each algorithm.

%methods
The first algorithm is a ANN ensemble classificator and the second algorithm is a nearest-neighbour classificator. For the first algorithm we use Tensorflow to train an ensemble of 20 networks. Neural networks need distinguishable labels, so we had to group the AGB final surface abundances. We grouped all AGB final surface abundances that result from models that have the same initial mass and metallicity. The final classification is found by taking the medians of all output nodes. For the nearest-neighbour algorithm we calculate the modified $\chi^2$ value for a Monte Carlo sample set based on the observed values, taking into consideration the error bars on the abundances as well as dependencies in the atmospheric parameters. We then calculate the $\chi^2$ value of the diluted set of AGB final surface abundances, find the model with the highest GoF and return its name and the corresponding dilution. Our final classification is the range of mass and metallicity that are included in the classifications coming from the two algorithms using the same set of AGB final surface abundances. We also extend these ranges to account for the discrete character of our two sets of AGB final surface abundances. To determine which elements to include in our classification, we systematically removed elements from the set of 12 s-process elements 
and we recorded the impact of the removal on the GoF distribution of the classification with the nearest-neighbour algorithm. We found that the set of elements leading to highest GoFs with our AGB final surface abundances is: Fe, Rb, Sr, Zr, Ru, Nd, Ce, Sm, and Eu. We thus excluded Y, Mo, La, and Nb and investigated their abundances in more detail. 

After validating the ML tools with a smaller sample of 28 Ba stars, we apply them to study the full stellar sample. We found one or two classifications using Monash and \textsc{Fruity} AGB nucleosynthesis predictions for 166 of the 169 Ba stars. For 85\% of the stars we find [Fe/H] consistent with observations. The average stellar mass and metallicity of the stellar sample is consistent with previous studies. 

%big group - exceptions
For three of the 169 Ba stars we were unable to find any classification. Two of these stars show an abundance pattern in the first s-process peak that is different from typical AGB s-process abundances. The third star shows an unusual pattern in the second s-process peak. 

%discussion - GoF test
We want to better understand why our nearest-neighbour algorithm performs better when we exclude Y, Mo, and La from our classification, and therefore we compare our classification results to the classifications we would get when we include those three elements. We find that for 43 stars the average GoFs differ for more than 10 percentage points. We performed statistical tests to find out if that group of 43 stars is different from the other stars and found that this is indeed the case for Nb, Mo, La, and Sm. The 43 stars have statistically significant higher [Nb/Fe], [Mo/Fe], [Sm/Fe], and lower [La/Fe]. 

%discussion i-process
Inspired by the CEMP-stars in which i-process nucleosynthesis signatures are found, we suspect that at least part of these 43 stars might be enriched by an i-process component. We performed i-process nucleosynthesis calculations to test this hypothesis. The resulting nucleosynthesis pattern seems to well reproduce the observed pattern between Sr and Ru of these four stars.
In a future work we will study these 43 stars in more detail, to determine whether an extra i-process component is indeed the reason why they are different from the other Ba stars. 

% Pd107
Using ML tools, we may have found evidence that in the surface abundances of about $\sim$20\% of the AGB stars, that are in a binary system with Ba stars, another nucleosynthetic component could be present. These AGB stars are also part of Galactic Chemical Evolution (GCE), and they may have contributed to the radioactive signature of the early solar system. We note that \citet{2022ApJtrueman} found that $^{107}$Pd was missing in GCE calculations, it could be that both findings are connected and that the radioactive signatures are the key in solving this puzzle.

\begin{acknowledgements}
This work is supported by the ERC Consolidator Grant (Hungary) programme (RADIOSTAR, G.A. n. 724560). We thank the ChETEC COST Action (CA16117), supported by the European Cooperation in Science and Technology, and the ChETEC-INFRA project funded by the European Union’s Horizon 2020 research and innovation program under grant agreement No 101008324. MP acknowledges the financial support of NuGrid/JINA-CEE (NSF Grant PHY-1430152) and STFC (through the University of Hull’s Consolidated Grant ST/R000840/1), and ongoing access to {\tt viper}, the University of Hull High Performance Computing Facility. MP thanks the Joint Institute for Nuclear Astrophysics - Center for the Evolution of the Elements, USA and the the US IReNA Accelnet network (Grant No. OISE-1927130). ML and MP acknowledges the support from the "Lend{\"u}let-2014" Programme of the Hungarian Academy of Sciences (Hungary). The work of AYL was supported by the US Department of Energy through the Los Alamos National Laboratory. Los Alamos National Laboratory is operated by Triad National Security, LLC, for the National Nuclear Security Administration of U.S.\ Department of Energy (Contract No.\ 89233218CNA000001).
The authors would like to thank A. Jorissen for providing data. B.Cs. and M.L. acknowledge the support of the Hungarian National Research, Development and Innovation Office (NKFI), grant KH\_18 130405. B.V. is supported by the \'UNKP-21-1 New National Excellence Program of the Ministry for Innovation and Technology from the source of the National Research, Development and Innovation Fund.
M.P.R. acknowledges financial support by Coordena\c c\~ao de Aperfei\c coamento de Pessoal de N\'ivel Superior (CAPES). N.A.D. acknowledges financial support by Russian Foundation
for Basic Research (RFBR) according to the research projects 18\_02-00554 and 18-52-06004.
Software: Numpy \citep{numpy}, matplotlib \citep{matplotlib}, and Tensorflow \citep{tensorflow}.

\end{acknowledgements}

\bibliographystyle{aa}
\bibliography{references}

\onecolumn
\begin{appendix} 

\section{Table with complete classifications - \textsc{fruity}}
\label{sec:app_complete_tabs}

\begin{longtable}[c]{l|llll}
\caption{This table lists the 141 Ba stars and their matched FRUITY models (without the 28 Ba stars that were already included in Table \ref{tab:28stars}).}\label{tab:rest_fruity}\\ 
Star & Mass range (M$_{\odot}$) & Metallicity range ([Fe/H]) & min GoF(\%) \\ 
\hline
BD $-$08$^{\circ}$3194 & 2.25 to 2.75 & -0.6 to -0.14 & 52 \\ 
BD $-$09$^{\circ}$4337 & 2.25 to 2.25 & -0.6 to -0.14 & 50 \\ 
CD $-$27$^{\circ}$2233 & 1.75 to 2.75 & -0.48 to 0.08 & 85 \\ 
CD $-$29$^{\circ}$8822 & 1.75 to 2.75 & -0.48 to 0.08 & 69 \\ 
CD $-$30$^{\circ}$8774& 1.75 to 2.25 & -0.38 to 0.08 & 83 \\ 
CD $-$38$^{\circ}$585 & 1.75 to 2.25 & -0.48 to -0.14 & 69 \\ 
CD $-$53$^{\circ}$8144 & 2.25 to 2.75 & -0.38 to 0.08 & 90 \\ 
CD $-$61$^{\circ}$1941 & 2.25 to 2.75 & -0.6 to -0.01 & 58 \\ 
CPD $-$62$^{\circ}$1013 & 2.25 to 2.75 & -0.38 to 0.23 & 87 \\ 
HD 5424 & 2.25 to 2.75 & -0.6 to -0.14 & 53 \\ 
HD 5825 & 2.25 to 2.75 & -0.6 to -0.14 & 89 \\ 
HD 15589 & 2.25 to 2.75 & -0.6 to -0.14 & 60 \\ 
HD 21989 & 1.75 to 2.25 & -0.48 to -0.01 & 88 \\ 
HD 22285 & 2.25 to 2.75 & -0.6 to -0.14 & 86 \\ 
HD 22772 & 2.25 to 2.75 & -0.38 to 0.08 & 88 \\ 
HD 29370 & 2.25 to 2.75 & -0.48 to 0.08 & 90 \\ 
HD 29685 & 1.75 to 2.25 & -0.38 to 0.08 & 90 \\ 
HD 30240 & 1.75 to 2.75 & -0.38 to 0.08 & 86 \\ 
HD 30554 & 2.25 to 2.75 & -0.48 to 0.08 & 84 \\ 
HD 32712 & 1.25 to 1.55 & -0.6 to -0.44 & 75 \\ 
HD 32901 & 1.05 to 1.55 & -0.91 to -0.44 & 84 \\ 
HD 35993 & 1.75 to 2.75 & -0.48 to 0.08 & 80 \\ 
HD 36650 & 2.25 to 2.75 & -0.48 to 0.08 & 73 \\ 
HD 38488 & 1.75 to 2.25 & -0.48 to -0.01 & 53 \\ 
HD 43389 & 2.25 to 2.75 & -0.38 to -0.01 & 68 \\ 
HD 51959 & 2.25 to 2.75 & -0.38 to 0.08 & 95 \\ 
HD 61332 & 1.75 to 2.25 & -0.38 to 0.08 & 74 \\ 
HD 64425 & 1.75 to 2.75 & -0.38 to 0.08 & 77 \\ 
HD 67036 & 1.75 to 2.25 & -0.48 to -0.01 & 92 \\ 
HD 71458 & 1.75 to 2.25 & -0.48 to -0.01 & 73 \\ 
HD 74950 & 1.75 to 2.25 & -0.48 to -0.01 & 89 \\ 
HD 82221 & 1.75 to 2.75 & -0.38 to 0.08 & 99 \\ 
HD 83548 & 2.25 to 2.75 & -0.38 to 0.08 & 71 \\ 
HD 84610 & 1.75 to 2.75 & -0.38 to 0.08 & 84 \\ 
HD 88035 & 1.75 to 2.75 & -0.48 to 0.08 & 65 \\ 
HD 88562 & 1.25 to 2.75 & -0.6 to -0.14 & 85 \\ 
HD 89175 & 1.75 to 2.25 & -0.48 to -0.14 & 68 \\ 
HD 91979 & 1.75 to 2.25 & -0.38 to 0.08 & 96 \\ 
HD 105902 & 1.75 to 2.75 & -0.48 to 0.08 & 89 \\ 
HD 107264 & 2.25 to 2.75 & -0.38 to 0.08 & 99 \\ 
HD 110483 & 2.25 to 2.75 & -0.48 to 0.08 & 92 \\ 
HD 110591 & 2.25 to 2.75 & -0.6 to -0.14 & 83 \\ 
HD 111315 & 1.75 to 2.75 & -0.23 to 0.08 & 81 \\ 
HD 113291 & 1.75 to 2.25 & -0.48 to -0.01 & 77 \\ 
HD 116869 & 1.75 to 2.25 & -0.48 to -0.01 & 93 \\ 
HD 120571 & 1.75 to 2.25 & -0.48 to -0.14 & 99 \\ 
HD 120620 & 1.75 to 2.75 & -0.6 to 0.08 & 50 \\ 
HD 122687 & 1.75 to 2.25 & -0.38 to 0.08 & 91 \\ 
HD 123701 & 2.25 to 2.75 & -0.48 to -0.01 & 67 \\ 
HD 123949 & 1.75 to 1.75 & -0.48 to -0.01 & 82 \\ 
HD 126313 & 1.75 to 2.75 & -0.48 to 0.08 & 84 \\ 
HD 130255 & 2.75 to 3.25 & -1.38 to -0.92 & 89 \\ 
HD 131670 & 2.25 to 2.75 & -0.38 to 0.08 & 89 \\ 
HD 136636 & 1.75 to 2.25 & -0.48 to -0.01 & 89 \\ 
HD 142571 & 1.75 to 2.25 & -0.38 to 0.08 & 90 \\ 
HD 147884 & 2.25 to 2.75 & -0.38 to 0.08 & 94 \\ 
HD 148177 & 1.25 to 1.75 & -0.23 to 0.08 & 97 \\ 
HD 162806 & 1.75 to 2.25 & -0.38 to 0.08 & 99 \\ 
HD 168214 & 2.25 to 2.75 & -0.38 to 0.23 & 90 \\ 
HD 168560 & 1.75 to 2.25 & -0.38 to 0.08 & 91 \\ 
HD 168791 & 2.25 to 2.75 & -0.23 to 0.08 & 98 \\ 
HD 176105 & 1.75 to 2.25 & -0.38 to 0.08 & 93 \\ 
HD 180996 & 1.75 to 2.75 & -0.23 to 0.23 & 97 \\ 
HD 182300 & 1.75 to 2.25 & -0.38 to 0.08 & 87 \\ 
HD 187308 & 1.75 to 2.75 & -0.48 to 0.08 & 56 \\ 
HD 193530 & 1.25 to 1.75 & -0.48 to 0.08 & 96 \\ 
HD 196445 & 1.75 to 2.75 & -0.48 to 0.08 & 98 \\ 
HD 199435 & 1.75 to 2.75 & -0.6 to -0.01 & 70 \\ 
HD 200995 & 1.75 to 2.25 & -0.38 to -0.01 & 95 \\ 
HD 207277 & 1.25 to 1.75 & -0.6 to -0.14 & 68 \\ 
HD 210709 & 1.25 to 1.75 & -0.48 to -0.14 & 84 \\ 
HD 211173 & 1.75 to 2.25 & -0.38 to -0.01 & 91 \\ 
HD 211954 & 1.25 to 1.75 & -0.6 to -0.14 & 65 \\ 
HD 214579 & 1.25 to 3.25 & -0.48 to -0.01 & 97 \\ 
HD 217143 & 1.75 to 2.75 & -0.6 to -0.01 & 98 \\ 
HD 217447 & 1.75 to 2.75 & -0.48 to 0.08 & 75 \\ 
HD 223586 & 2.25 to 2.75 & -0.38 to 0.08 & 92 \\ 
HD 223617 & 2.25 to 2.75 & -0.38 to 0.08 & 92 \\ 
HD 252117 & 2.25 to 2.75 & -0.38 to 0.08 & 96 \\ 
HD 273845 & 1.75 to 2.25 & -0.48 to -0.01 & 81 \\ 
HD 288174 & 2.25 to 2.75 & -0.38 to 0.08 & 86 \\ 
BD $-$18$^{\circ}$821 & 1.75 to 2.25 & -0.48 to -0.14 & 62 \\ 
CD $-$26$^{\circ}$7844 & 2.25 to 2.75 & -0.23 to 0.23 & 100 \\ 
CD $-$30$^{\circ}$9005 & 1.75 to 2.75 & -0.38 to 0.08 & 80 \\ 
CD $-$34$^{\circ}$6139 & 2.25 to 2.75 & -0.38 to 0.23 & 62 \\ 
CD $-$34$^{\circ}$7430 & 1.75 to 2.25 & -0.23 to 0.08 & 98 \\ 
CD $-$46$^{\circ}$3977 & 1.75 to 2.75 & -0.38 to 0.08 & 85 \\ 
HD 18361 & 2.25 to 2.75 & -0.23 to 0.23 & 90 \\ 
HD 26886 & 2.25 to 2.75 & -0.48 to 0.08 & 82 \\ 
HD 31812 & 2.25 to 2.75 & -0.23 to 0.23 & 81 \\ 
HD 33709 & 2.25 to 2.75 & -0.23 to 0.08 & 77 \\ 
HD 39778 & 2.25 to 2.75 & -0.48 to 0.08 & 89 \\ 
HD 41701 & 1.75 to 2.75 & -0.23 to 0.08 & 98 \\ 
HD 45483 & 2.25 to 2.75 & -0.38 to 0.08 & 98 \\ 
HD 48814 & 1.75 to 2.25 & -0.38 to 0.08 & 98 \\ 
HD 49017 & 1.75 to 2.25 & -0.38 to -0.01 & 67 \\ 
HD 49778 & 1.25 to 2.75 & -0.48 to -0.01 & 64 \\ 
HD 50075 & 1.75 to 2.75 & -0.48 to 0.08 & 54 \\ 
HD 50843 & 1.75 to 2.25 & -0.38 to -0.14 & 95 \\ 
HD 88495 & 1.75 to 2.25 & -0.23 to 0.08 & 86 \\ 
HD 90167 & 2.25 to 2.75 & -0.23 to 0.08 & 84 \\ 
HD 109061 & 1.25 to 1.75 & -0.91 to -0.14 & 84 \\ 
HD 113195 & 2.25 to 2.75 & -0.38 to 0.08 & 90 \\ 
HD 115277 & 1.75 to 2.25 & -0.23 to 0.23 & 98 \\ 
HD 119650 & 1.75 to 2.25 & -0.38 to 0.08 & 90 \\ 
HD 139266 & 1.75 to 2.25 & -0.48 to -0.01 & 98 \\ 
HD 139409 & 2.25 to 2.75 & -0.23 to -0.01 & 65 \\ 
HD 169106 & 1.75 to 2.25 & -0.23 to 0.08 & 91 \\ 
HD 184001 & 2.25 to 2.75 & -0.38 to 0.08 & 92 \\ 
HD 204886 & 1.25 to 1.75 & -0.38 to -0.01 & 90 \\ 
HD 213084 & 1.75 to 2.75 & -0.48 to 0.08 & 89 \\ 
HD 223938 & 1.75 to 2.75 & -0.48 to -0.01 & 76 \\ 
MFU 214 & 1.75 to 2.25 & -0.23 to 0.08 & 94 \\ 
MFU 229 & 2.25 to 2.75 & -0.23 to 0.08 & 96 \\ 
HD 12392 & 1.25 to 2.25 & -0.6 to -0.01 & 66 \\ 
HD 17067 & 1.75 to 2.75 & -0.48 to -0.14 & 90 \\ 
HD 90127 & 2.25 to 2.75 & -0.08 to 0.08 & 64 \\ 
HD 102762 & 1.25 to 1.75 & -0.48 to -0.01 & 94 \\ 
HD 114678 & 1.75 to 2.25 & -0.38 to -0.01 & 87 \\ 
HD 210030 & 1.75 to 2.25 & -0.23 to 0.08 & 77 \\ 
HD 214889 & 2.25 to 2.75 & -0.38 to 0.08 & 99 \\ 
HD 215555 & 2.25 to 2.75 & -0.23 to 0.23 & 84 \\ 
HD 216809 & 2.75 to 3.25 & -0.08 to 0.23 & 74 \\ 
HD 221879 & 1.75 to 1.75 & -0.23 to 0.23 & 91 \\ 
HD 749 & 2.25 to 2.75 & -0.48 to 0.08 & 63 \\ 
HD 88927 & 1.75 to 2.25 & -0.23 to 0.08 & 83 \\ 
HD 89638 & 1.75 to 2.75 & -0.48 to 0.08 & 88 \\ 
HD 187762 & 1.75 to 2.25 & -0.48 to -0.01 & 87 \\ 
HD 4084 & - & - & - \\ 
HD 49641 & 2.75 to 4.25 & -0.6 to -0.14 & 83 \\
HD 66291 & - & - & -\\ 
HD 123396 & - & - & -\\ 
HD 177192 & - & - & - \\ 
HD 204075 & 2.25 to 2.75 & -0.23 to 0.23 & 97 \\ 
HD 219116 & - & - & - \\
MFU 112 & - & - & - \\
HD 21682 & 2.25 to 2.75 & -0.6 to -0.14 & 76\\ 
HD 49661 & 1.25 to 2.25 & -0.38 to -0.01 & 72 \\ 
HD 62017 & - & - & - \\
HD 107270 & - & - & - \\
HD 148892 & - & - & - \\ 
BD +09$^{\circ}$2384 & - & - & - \\
\end{longtable}

\newpage
\section{Table with complete classifications - Monash}
\begin{longtable}{l|llll}
\caption{This table lists the 141 Ba stars and their matched MONASH models (without the 28 Ba stars that were already included in Table \ref{tab:28stars}).}\label{tab:rest_monash}\\ 
Star & Mass range (M$_{\odot}$) & Metallicity range ([Fe/H]) & min GoF(\%) \\ 
\hline
BD $-$08$^{\circ}$3194 & 1.75 to 2.75 & -0.54 to 0.08 & 68 \\
BD $-$09$^{\circ}$4337 & 2.75 to 3.5 & -0.38 to 0.08 & 56 \\
CD $-$27$^{\circ}$2233 & 1.75 to 2.25 & -0.23 to 0.08 & 83 \\
CD $-$29$^{\circ}$8822 & 1.75 to 2.5 & -0.23 to 0.23 & 88 \\
CD $-$30$^{\circ}$8774 & 2.0 to 2.75 & -0.23 to 0.23 & 80 \\
CD $-$38$^{\circ}$585 & 2.25 to 2.75 & -0.54 to -0.07 & 67 \\
CD $-$53$^{\circ}$8144 & 1.75 to 3.0 & -0.54 to 0.23 & 87 \\
CD $-$61$^{\circ}$1941 & 1.75 to 2.25 & -0.23 to 0.08 & 67 \\
CPD $-$62$^{\circ}$1013 & 2.25 to 3.25 & -0.23 to 0.23 & 89 \\
HD 4084 & 3.25 to 3.25 & -0.23 to -0.07 & 51 \\
HD 5424 & 1.75 to 2.5 & -0.54 to -0.07 & 65 \\
HD 5825 & 2.5 to 2.75 & -0.54 to -0.07 & 84 \\
HD 15589 & 1.75 to 2.75 & -0.54 to -0.07 & 82 \\
HD 21989 & 1.75 to 2.25 & -0.38 to 0.08 & 86 \\
HD 22285 & 2.5 to 2.75 & -0.54 to -0.47 & 78 \\
HD 22772 & 1.75 to 3.0 & -0.54 to 0.23 & 86 \\
HD 29370 & 1.75 to 3.0 & -0.54 to 0.08 & 85 \\
HD 29685 & 1.5 to 2.25 & -0.23 to 0.23 & 92 \\
HD 30240 & 1.75 to 2.5 & -0.23 to 0.23 & 94 \\
HD 30554 & 1.5 to 3.0 & -0.54 to 0.23 & 83 \\
HD 32712 & 1.25 to 2.15 & -0.54 to -0.07 & 92 \\
HD 32901 & 1.65 to 2.35 & -0.54 to -0.07 & 86 \\
HD 35993 & 1.75 to 2.25 & -0.23 to 0.23 & 88 \\
HD 36650 & 2.25 to 3.25 & -0.23 to 0.08 & 63 \\
HD 38488 & 1.75 to 2.25 & -0.23 to 0.23 & 76 \\
HD 43389 & 2.5 to 3.0 & -0.54 to -0.07 & 74 \\
HD 51959 & 1.75 to 2.25 & -0.23 to 0.23 & 97 \\
HD 61332 & 1.75 to 2.25 & -0.23 to 0.23 & 89 \\
HD 64425 & 1.75 to 2.25 & -0.23 to 0.23 & 96 \\
HD 66291 & 2.75 to 3.5 & -0.54 to -0.07 & 94 \\
HD 67036 & 1.75 to 3.0 & -0.54 to -0.07 & 88 \\
HD 71458 & 1.75 to 2.25 & -0.38 to 0.23 & 85 \\
HD 74950 & 1.75 to 2.25 & -0.38 to 0.08 & 88 \\
HD 82221 & 1.75 to 2.5 & -0.23 to 0.23 & 96 \\
HD 83548 & 1.75 to 2.25 & -0.23 to 0.23 & 91 \\
HD 84610 & 2.0 to 2.75 & -0.23 to 0.23 & 91 \\
HD 88035 & 1.75 to 2.25 & -0.23 to 0.23 & 66 \\
HD 88562 & 1.75 to 2.25 & -0.38 to -0.07 & 91 \\
HD 89175 & 1.85 to 2.5 & -0.54 to -0.07 & 76 \\
HD 91979 & 1.75 to 2.5 & -0.23 to 0.23 & 94 \\
HD 105902 & 1.75 to 2.5 & -0.23 to 0.23 & 89 \\
HD 107264 & 2.0 to 3.25 & -0.54 to 0.23 & 96 \\
HD 110483 & 1.75 to 2.25 & -0.38 to 0.23 & 95 \\
HD 110591 & 2.5 to 2.75 & -0.54 to -0.47 & 84 \\
HD 111315 & 1.75 to 2.75 & -0.23 to 0.23 & 92 \\
HD 113291 & 1.75 to 2.25 & -0.38 to 0.08 & 86 \\
HD 116869 & 1.75 to 2.75 & -0.54 to -0.07 & 95 \\
HD 120571 & 2.25 to 2.5 & -0.54 to -0.07 & 98 \\
HD 120620 & 3.5 to 4.75 & -0.54 to 0.23 & 91 \\
HD 122687 & 1.75 to 2.25 & -0.38 to 0.23 & 92 \\
HD 123701 & 2.25 to 3.0 & -0.54 to -0.07 & 53 \\
HD 123949 & 1.5 to 1.75 & -0.23 to 0.08 & 93 \\
HD 126313 & 1.75 to 2.25 & -0.23 to 0.23 & 86 \\
HD 131670 & 2.25 to 3.25 & -0.23 to 0.23 & 92 \\
HD 136636 & 1.75 to 2.25 & -0.38 to 0.23 & 94 \\
HD 142571 & 1.75 to 2.25 & -0.23 to 0.23 & 87 \\
HD 147884 & 1.75 to 2.75 & -0.23 to 0.23 & 96 \\
HD 148177 & 2.0 to 2.75 & -0.23 to 0.23 & 94 \\
HD 162806 & 1.75 to 2.25 & -0.23 to 0.08 & 98 \\
HD 168214 & 2.25 to 3.25 & -0.23 to 0.23 & 87 \\
HD 168560 & 1.75 to 2.25 & -0.38 to 0.08 & 88 \\
HD 168791 & 2.75 to 3.25 & -0.38 to 0.23 & 92 \\
HD 176105 & 2.0 to 3.0 & -0.23 to 0.23 & 84 \\
HD 180996 & 2.25 to 2.75 & 0.1 to 0.23 & 96 \\
HD 182300 & 1.5 to 2.25 & -0.23 to 0.23 & 96 \\
HD 187308 & 1.75 to 2.5 & -0.23 to 0.23 & 62 \\
HD 193530 & 1.5 to 2.25 & -0.23 to 0.23 & 93 \\
HD 196445 & 1.75 to 2.25 & -0.23 to 0.23 & 95 \\
HD 199435 & 1.75 to 2.75 & -0.54 to -0.07 & 71 \\
HD 200995 & 1.75 to 2.25 & -0.23 to 0.23 & 98 \\
HD 207277 & 1.5 to 2.25 & -0.38 to 0.08 & 74 \\
HD 210709 & 1.25 to 1.75 & -0.23 to 0.08 & 91 \\
HD 211173 & 2.25 to 3.25 & -0.38 to -0.07 & 88 \\
HD 211954 & 1.5 to 2.15 & -0.54 to -0.07 & 74 \\
HD 214579 & 2.25 to 3.0 & -0.54 to -0.07 & 99 \\
HD 217143 & 1.75 to 2.75 & -0.54 to -0.07 & 99 \\
HD 217447 & 1.75 to 2.5 & -0.23 to 0.23 & 63 \\
HD 223586 & 1.75 to 3.0 & -0.38 to 0.23 & 94 \\
HD 223617 & 2.75 to 3.25 & -0.23 to 0.23 & 88 \\
HD 252117 & 1.75 to 2.25 & -0.23 to 0.23 & 96 \\
HD 273845 & 1.25 to 2.25 & -0.38 to -0.07 & 92 \\
HD 288174 & 1.75 to 2.25 & -0.23 to 0.23 & 93 \\
MFU 112 & 2.75 to 3.25 & -0.54 to -0.07 & 93 \\
BD $-$18$^{\circ}$821 & 1.65 to 2.35 & -0.54 to -0.07 & 79 \\
CD $-$26$^{\circ}$7844 & 2.5 to 3.5 & -0.23 to 0.23 & 100 \\
CD $-$30$^{\circ}$9005 & 1.75 to 2.5 & -0.23 to 0.23 & 95 \\
CD $-$34$^{\circ}$6139 & 2.25 to 3.25 & -0.23 to 0.23 & 68 \\
CD $-$34$^{\circ}$7430 & 2.0 to 2.75 & -0.23 to 0.23 & 99 \\
CD $-$46$^{\circ}$3977 & 1.75 to 2.75 & -0.23 to 0.23 & 89 \\
HD 18361 & 2.25 to 2.75 & 0.1 to 0.23 & 91 \\
HD 26886 & 2.75 to 3.25 & -0.38 to -0.07 & 82 \\
HD 31812 & 2.25 to 2.75 & 0.1 to 0.23 & 77 \\
HD 39778 & 1.75 to 2.25 & -0.23 to 0.23 & 88 \\
HD 41701 & 1.75 to 2.5 & -0.23 to 0.23 & 100 \\
HD 45483 & 2.0 to 2.75 & -0.23 to 0.23 & 95 \\
HD 48814 & 1.75 to 2.5 & -0.23 to 0.23 & 98 \\
HD 49017 & 1.75 to 2.25 & -0.23 to 0.23 & 76 \\
HD 49778 & 1.75 to 2.5 & -0.54 to -0.07 & 68 \\
HD 50075 & 1.75 to 2.25 & -0.23 to 0.08 & 58 \\
HD 50843 & 1.75 to 2.25 & -0.38 to -0.07 & 96 \\
HD 88495 & 2.5 to 3.0 & -0.23 to 0.23 & 74 \\
HD 90167 & 2.25 to 2.75 & 0.1 to 0.23 & 84 \\
HD 107270 & 3.25 to 4.0 & -0.23 to 0.23 & 79 \\
HD 109061 & 1.0 to 2.5 & -0.93 to -0.47 & 88 \\
HD 113195 & 2.25 to 3.25 & -0.23 to 0.23 & 83 \\
HD 115277 & 2.25 to 3.0 & -0.23 to 0.23 & 95 \\
HD 119650 & 1.5 to 2.25 & -0.23 to 0.23 & 86 \\
HD 139266 & 1.75 to 2.25 & -0.38 to 0.08 & 98 \\
HD 139409 & 3.0 to 3.75 & -0.23 to -0.07 & 57 \\
HD 169106 & 1.75 to 2.5 & -0.23 to 0.23 & 96 \\
HD 184001 & 2.25 to 3.25 & -0.23 to 0.23 & 84 \\
HD 204886 & 1.25 to 1.75 & -0.23 to 0.23 & 98 \\
HD 213084 & 1.75 to 2.25 & -0.23 to 0.23 & 93 \\
HD 223938 & 2.5 to 3.0 & -0.54 to -0.07 & 71 \\
MFU 214 & 2.5 to 3.0 & -0.23 to 0.23 & 92 \\
MFU 229 & 2.0 to 2.75 & -0.23 to 0.23 & 98 \\
HD 12392 & 1.5 to 2.35 & -0.54 to -0.07 & 85 \\
HD 17067 & 2.5 to 3.0 & -0.54 to -0.07 & 88 \\
HD 90127 & 3.0 to 3.75 & -0.23 to -0.07 & 70 \\
HD 102762 & 1.5 to 2.25 & -0.38 to 0.08 & 98 \\
HD 114678 & 2.5 to 3.0 & -0.54 to -0.07 & 82 \\
HD 210030 & 1.75 to 3.0 & -0.23 to 0.23 & 74 \\
HD 214889 & 2.0 to 2.75 & -0.23 to 0.23 & 93 \\
HD 215555 & 2.5 to 3.5 & -0.23 to 0.23 & 81 \\
HD 216809 & 2.75 to 3.25 & 0.1 to 0.23 & 67 \\
HD 221879 & 2.5 to 3.25 & -0.23 to 0.23 & 89 \\
HD 749 & 1.75 to 3.25 & -0.54 to 0.08 & 60 \\
HD 88927 & 2.25 to 3.0 & -0.23 to 0.23 & 78 \\
HD 89638 & 1.75 to 3.25 & -0.23 to 0.08 & 81 \\
HD 187762 & 1.75 to 2.5 & -0.54 to -0.07 & 89 \\ 
HD 49641 & 4.25 to 4.75 & -0.54 to -0.07 & 89 \\
HD 123396 & - & - & - \\ 
HD 130255 & - & - & - \\ 
HD 177192 & 2.5 to 2.75 & -0.23 to 0.23 & 74 \\ 
HD 204075 & 2.5 to 2.75 & 0.1 to 0.23 & 98 \\ 
HD 219116 & - & - & - \\ 
HD 21682 & 1.75 to 2.5 & -0.54 to -0.07 & 78 \\ 
HD 33709 & - & - & - \\ 
HD 49661 & 1.75 to 2.5 & -0.23 to 0.23 & 67 \\ 
HD 62017 & - & - & - \\ 
HD 148892 & 4.75 to 4.75 & -0.54 to -0.07 & 71 \\ 
BD +09$^{\circ}$2384 & - & - & - \\
\end{longtable}

\newpage
\section{Table with parameter values associated with the labels - \textsc{fruity}}
\begin{table}[H]
\centering
\caption{Parameter values associated with the labels appearing in figures such as Fig.~\ref{fig:HD18182} for the \textsc{fruity} models.  } \label{tab:names_fruity}
\begin{tabular}{lcccr}
Name & mass (M$_\odot$)& metallicity (z)& $^{13}$C-pocket type & IRV (km/s)\\
\hline\\
F-m1.5z001a & 1.5 & 0.001 & Extended & 0\\
F-m1.5z001b & 1.5 & 0.001 & Standard & 60\\
F-m1.5z002b & 1.5 & 0.002 & Extended & 0\\
F-m1.5z003a & 1.5 & 0.003 & Standard & 60\\
F-m1.5z003c & 1.5 & 0.003 & Extended & 0\\
F-m1.5z006a & 1.5 & 0.006 & Extended & 0\\
F-m1.5z01a  & 1.5 & 0.01 & Extended & 0\\
F-m1.5z01c  & 1.5 & 0.01 & Extended & 60\\

F-m2.0z001a & 2.0 & 0.001 & Extended & 0\\
F-m2.0z003b & 2.0 & 0.003 & Extended & 0\\
F-m2.0z006a & 2.0 & 0.006 & Extended & 0\\
F-m2.0z006b & 2.0 & 0.006 & Standard & 0\\
F-m2.0z014b & 2.0 & 0.014 & Standard & 0\\
F-m2.0z014c & 2.0 & 0.014 & Standard & 10\\

F-m2.5z006a & 2.5 & 0.006 & Standard & 0\\
F-m2.5z008a & 2.5 & 0.008 & Standard & 0\\
F-m2.5z01a  & 2.5 & 0.01 & Standard & 0\\
F-m2.5z014a & 2.5 & 0.014 & Standard & 0\\

F-m3.0z002a & 3.0 & 0.002 & Standard & 0\\
F-m3.0z01a  & 3.0 & 0.01 & Standard & 0\\
F-m3.0z014a & 3.0 & 0.014 & Standard & 0\\

F-m4.0z003a & 4.0 & 0.003 & Standard & 0\\
F-m4.0z006a & 4.0 & 0.006 & Standard & 0\\
\hline
\end{tabular}
    \begin{tablenotes}
      \item The pocket type is listed as either standard or extended \protect\citep{2015cristallo}, and the initial rotational velocity (IRV) is given in km/s..
    \end{tablenotes}
\end{table}

\newpage
\section{Table with parameter values associated with the labels - Monash}
\begin{center}
\begin{table}[H]
\centering
\caption{Same as Table~\ref{tab:names_fruity} for the Monash models.  } \label{tab:names_monash}
\begin{tabular}{lcccr}
Name & mass (M$_\odot$)& metallicity (z)& PMZ (M$_\odot$) & N$_\text{ov}$\\
\hline\\
M-m1.5z0028b & 1.5 & 0.0028 & 6e-3 & 0\\
M-m1.75z007b & 1.75 & 0.007 &  2e-3 & 1\\
M-m1.9z007a  & 1.9 & 0.007 & 2e-3 & 0\\
M-m2.0z0028a & 2.0 & 0.0028 & 2e-3 & 0\\
M-m2.0z0028b & 2.0 & 0.0028 & 6e-3 & 0\\
M-m2.1z007a  & 2.1 & 0.007 & 2e-3 & 0\\
M-m2.25z0028a& 2.25 & 0.0028 & 2e-3 & 0\\
M-m2.25z007a & 2.25 & 0.007 & 2e-3 & 0\\
M-m2.5z0028a & 2.5 & 0.0028 & 2e-3 & 0\\
M-m2.5z0028b & 2.5 & 0.0028 & 4e-3 & 0\\
M-m2.75z007a & 2.75 & 0.007 & 2e-3 & 0\\
M-m2.75z014a & 2.75 & 0.014 & 2e-3 & 0\\
M-m3.0z0028a & 3.0 & 0.0028 & 1e-3 & 0\\
M-m3.0z007a  & 3.0 & 0.007 & 1e-3 & 0\\
M-m3.0z007b  & 3.0 & 0.007 & 2e-3 & 0\\
M-m3.0z01a   & 3.0 & 0.01 & 2e-3 & 0\\
M-m3.0z014b  & 3.0 & 0.014 & 1e-3 & 0\\
M-m3.0z014c  & 3.0 & 0.014 & 2e-3 & 0\\
M-m3.25z0028a& 3.25 & 0.0028  & 1e-3 & 0\\
M-m3.25z007a & 3.25 & 0.007 & 1e-3 & 0\\
M-m3.25z014a & 3.25 & 0.014 & 1e-3 & 0\\
M-m3.5z0028a & 3.5 & 0.0028 & 1e-3 & 0\\
M-m3.5z007a  & 3.5 & 0.007 & 1e-3 & 0\\
M-m3.75z007a & 3.75 & 0.007 & 1e-3 & 0\\
\hline
\end{tabular}
    \begin{tablenotes}
      \item The partial mixing zone (PMZ) is listed in M$_\odot$. N$_\text{ov}$ indicates the number of pressure scale heights by which the convective envelope is extended during the TDU \protect\citep{2016karakas}.
    \end{tablenotes}
\end{table}    
\end{center}

\end{appendix}

\end{document}